\documentclass[floats,floatfix,showpacs,amssymb,prd,twocolumn,superscriptaddress,nofootinbib,nolongbibliography,reprint]{revtex4-1}

\usepackage{amssymb,amsmath,verbatim,mathtools,needspace,enumitem,etoolbox,graphicx,physics,microtype,afterpage,xspace,tabularx,lmodern,multirow}
\usepackage{gensymb}

\usepackage[dvipsnames, usenames]{xcolor}

\definecolor{linkcolor}{rgb}{0.0,0.3,0.5}
\usepackage[unicode, colorlinks=true, linkcolor=linkcolor, citecolor=linkcolor, filecolor=linkcolor,urlcolor=linkcolor, pdfusetitle]{hyperref}
\usepackage[all]{hypcap}
\usepackage[T1]{fontenc}
\usepackage[utf8]{inputenc}
\usepackage[usenames,dvipsnames]{xcolor}

\definecolor{romared}{RGB}{142,0,28}
\hypersetup{colorlinks=true,
            citecolor=romared,
            linkcolor=romared,
            urlcolor=romared}

\newcommand{\nn}{\nonumber}

\newcommand{\be}{\begin{equation}}
\newcommand{\ee}{\end{equation}}

\def\be{\begin{equation}}
\def\ee{\end{equation}}
\newcommand{\beq}{\begin{eqnarray}}
\newcommand{\eeq}{\end{eqnarray}}

\begin{document}

\title{Quasinormal Modes and Stability of Firewalls}

\author{Ryan McManus}
\email{rmcmanu3@jhu.edu}
\affiliation{Department of Physics and Astronomy, Johns Hopkins University, 3400 N. Charles Street, Baltimore, MD 21218, US}

\author{Emanuele Berti}
\email{berti@jhu.edu}
\affiliation{Department of Physics and Astronomy, Johns Hopkins University, 3400 N. Charles Street, Baltimore, MD 21218, US}

\author{David E. Kaplan}
\email{david.kaplan@jhu.edu}
\affiliation{Department of Physics and Astronomy, Johns Hopkins University, 3400 N. Charles Street, Baltimore, MD 21218, US}

\author{Surjeet Rajendran}
\email{srajend4@jhu.edu}
\affiliation{Department of Physics and Astronomy, Johns Hopkins University, 3400 N. Charles Street, Baltimore, MD 21218, US}

\begin{abstract}
A solution to the black hole information problem requires propagation of information from the interior of the black hole to the exterior. Such propagation violates general relativity and could conceivably be accomplished through ``firewall'' models. Based on the existence of similar firewalls at the inner horizons of charged and rotating black holes, a model of a firewall was recently constructed where the exterior spacetime reduces to that of the Schwarzschild metric but with a dramatically different interior.  We investigate the radial and nonradial polar stability of these objects.  We first study the dynamics of the shell under spherically symmetric perturbations,  and impose constraints on the firewall model parameters by requiring a subluminal speed of sound on the firewall. We show that the demands of stability and subluminality impose significant constraints on the internal parameters of the firewall, narrowing down the range of objects that could be used to create such a structure. 
\end{abstract}

\preprint{RUP-19-17}

\date{\today}
\maketitle

\section{Introduction}
\label{sec:intro}

It is widely accepted that black holes (BHs) describe massive, highly compact and dark astronomical objects.
As vacuum solutions of the field equations of general relativity (GR), BHs are purely geometrical, and so tests of their nature are tests of the spacetime outside of their event horizons.
The properties of BH spacetimes are in the process of being rigorously examined (see e.g. the articles collected in Ref.~\cite{Berti:2019tcy}).
For example, gravitational waves probe the dynamics of binary BHs and the nonlinear regime of GR, with the inspiral testing the weak field regime, while the merger and ringdown test the strong field regime \cite{Berti:2015itd,Abbott:2016,Berti:2018a,Berti:2018b}.
Further, the gravitational shadow of BHs tests the geodesic structure of the spacetime, and in particular the region known as the ``light ring''~\cite{Cunha:2018}. 
However, these tests are most sensitive to the spacetime far from the event horizon of the BHs, while the spacetime near the horizon remains hard to probe.

There are important reasons to probe the near-horizon nature of BHs. The formation of a BH inevitably leads to Planckian matter densities as the matter crunches towards a singularity. Once such a region has formed, its future evolution need not be governed by GR, since the theory breaks down at these densities.  The conventional belief that the entire BH geometry is described by GR is an extrapolation, requiring the applicability of the theory in predicting the future of an object whose physics is not described by the theory. This extrapolation is particularly problematic since it is well known that the GR description of BHs runs afoul of the laws of quantum mechanics, giving rise to the BH information problem. It has been long recognized that the BH information problem cannot be solved as long as the BH has an empty event horizon \cite{Mathur:2009hf}. 

Whether motivated by the desire to rigorously test the BH spacetime by finding alternatives to allow for null tests, or driven by the theoretical need to solve these conceptual problems, the construction of concrete alternatives to BHs is a very active field. Such exotic compact objects (ECOs)  mimic both the exterior spacetime as well as their dark nature, but do not possess event horizons, enabling external observers to probe the interior of these objects.
Many ECOs have been constructed~\cite{Colpi:1986ye,Seidel:1991zh,Mazur:2001fv,Giudice:2016zpa}, and their dynamics can lead to interesting observational consequences~\cite{Cardoso:2016rao,Cardoso:2016oxy,Cunha:2017qtt,Barausse:2018vdb,Maggio:2018ivz,Abedi:2020ujo,Maggio:2020jml}. 

One signature of ECOs is their response to nonradial perturbations~\cite{Regge:1957td,Zerilli:1971wd} and the associated quasinormal mode (QNM) spectra~\cite{Berti:2009kk,Cardoso:2019rvt}. The spectra are sensitive to the whole spacetime and are thus able to probe the interior of the ECO, potentially enabling a clean distinction between BHs and various ECO models. Moreover, QNMs are fundamentally tied to the stability of the ECO, a necessary criterion for any ECO that could describe BH candidates observed in our Universe. Hence, the QNM spectra of ECOs are a powerful tool in the study of BH alternatives.

Motivated by the BH information problem, a new kind of ``firewall'' ECO was proposed in Ref.~\cite{Kaplan:2018dqx}. We briefly review the central elements of its construction for completeness; the interested reader is referred to Ref.~\cite{Kaplan:2018dqx} for a fuller description.  First, if BHs are to release their information in a manner consistent with quantum mechanics, this information needs to propagate from the singularity to the horizon. Without propagation, one cannot transmit information. Such propagation violates GR. The most natural way to violate GR is if this propagation happens in a region of Planck density where GR is expected to break down. In order for information to propagate all the way from the singularity to the horizon, the entire spacetime between the singularity and the horizon must be at Planck density, leading to an object that is effectively a macroscopic singularity. All of this needs to happen without changing the observed parameters of the BH, such as the Arnowitt-Deser-Misner (ADM) mass. 

The second element informing the construction of Ref.~\cite{Kaplan:2018dqx} is the fact that macroscopic singularities are expected to exist in the inner horizons of Reissner-Nordstr\"om and Kerr BHs, without changing any of the ADM parameters of the BH. These structures need to exist, since the inner horizon is a Cauchy horizon. If GR were to hold at this horizon, the theory would no longer be predictive even classically, let alone quantum mechanically. Macroscopic singularities at the inner horizon can exist without changing the ADM parameters, since divergent blueshifts at the inner horizon can lead to enormous local energy densities at the inner horizon without changing the external parameters of the BH. 

\begin{figure}[t]
  \includegraphics[width=0.95\columnwidth]{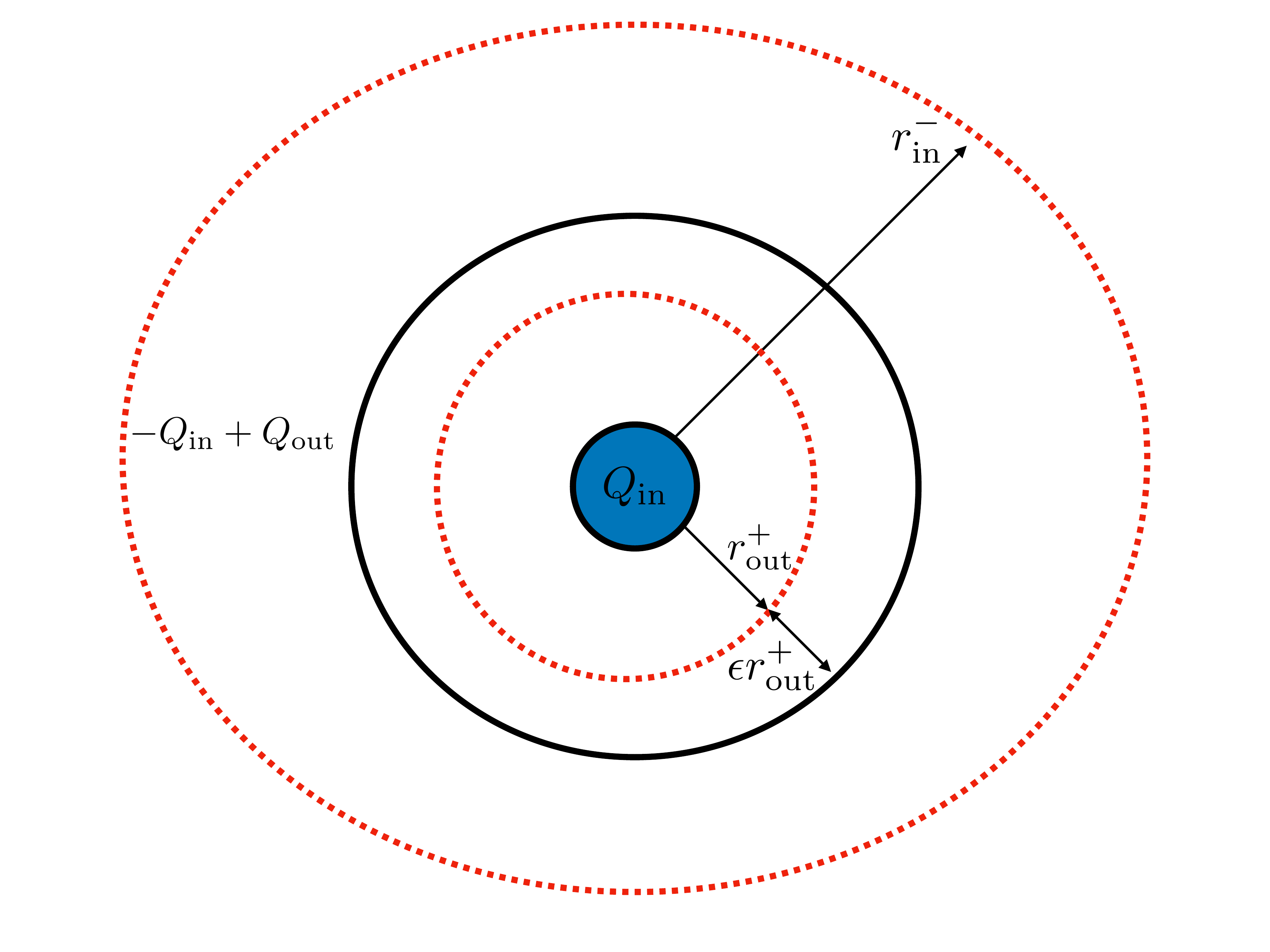}\\
  \caption{Schematic illustration of the firewall ECO of Ref.~\cite{Kaplan:2018dqx}. The inner horizon $r_{\rm in}^-$ of the interior metric is outside of the shell, while the outer horizon $r_{\rm out}^+$ of the exterior metric is within the shell, preventing horizons in the whole spacetime.}
\label{fig:SchBH}
\end{figure}

Using this property of Reissner-Nordstr\"om spacetimes,  Ref.~\cite{Kaplan:2018dqx} demonstrated the existence of a firewall ECO schematically illustrated in Fig.~\ref{fig:SchBH}. The ECO has the same exterior metric as the Schwarzschild solution all the way to a distance that is a Planck length from the horizon $r^+_{\rm out}$ of the Schwarzschild BH. A shell at Planck density is located a Planck length $\epsilon r^+_{\rm out}$ away from the horizon, and the interior of this shell is also an object that is at Planck density. Since the entire interior is at Planck density, this object can conceivably form due to the fact that its evolution is no longer controlled by GR.  

The construction of Ref.~\cite{Kaplan:2018dqx} relies on the fact that Reissner-Nordstr\"om BHs have both an inner and outer horizon, and so near the singularity the time direction remains timelike. As depicted in Fig.~\ref{fig:SchBH}, one can consider a spacetime where there is a core Reissner-Nordstr\"om singularity with charge $Q_{\rm in}$. Within its inner horizon $r_{\rm in}^-$, one may place a charged shell that carries the charge necessary so that the net charge of the full spacetime equals the ADM charge $Q_{\rm out}$ of the BH. With suitable choice of parameters, one can obtain the ECO described above, using sources of matter that obey the dominant energy condition. 

We examine the stability of this construction under radial and nonradial polar perturbations. We consider a more general case than that presented in Ref.~\cite{Kaplan:2018dqx}:
firstly, we allow for the exterior metric to be Reissner-Nordstr\"om, rather than just Schwarzschild.
Secondly, we allow the shell to be placed at any radius, rather than just a Planck length away from the corresponding Schwarzschild horizon of the exterior metric.

We work under the assumption that the theory of gravity across spacetime is GR.
We focus on polar perturbations because (unlike axial perturbations~\cite{1991RSPSA.434..449C,Benhar:1998au}) they couple to the matter content of the shell, and so we expect the polar QNM spectra to be more sensitive to the details of the model. 

Our key results are shown in Section~\ref{sec:NumericalResults}, which we recommend for the reader uninterested in the rather heavy derivations, with Figs.~\ref{fig:AllInternalBounds}, \ref{fig:SmallestpRegionPlot} and \ref{fig:AltAllInternalBounds} containing most of our key results. We find that the requirements of perturbative stability and subluminal speed of sound for matter on the shell constrain the internal mass $M_{\rm in}$ and charge $Q_{\rm in}$ of the firewall to a compact region which shrinks as $\epsilon\to 0$. This is of major interest to the firewall construction, as the distance between the shell and the horizon should be of the order of a Planck length. The stability of nonradial polar perturbations also depends upon boundary conditions chosen in the interior of the construction. Thus the requirement of stability offers a window into UV physics, both in terms of specifying the kinds of objects (internal mass and charge) that could be used to create the firewall and the possible boundary conditions that may be natural in a full UV-complete theory of gravity.

The paper is organized as follows.
In Section~\ref{sec:FirewallConstruction} we recap and generalize the firewall solution of Ref.~\cite{Kaplan:2018dqx}.
In Section~\ref{sec:StaticJunctionConditions} we present the Israel junction conditions \cite{Israel:1966rt} and their application to the firewall solution.
In Section~\ref{sec:RadialStability} we examine the radial stability of the firewall, mostly following Ref.~\cite{Visser:2003ge}, and we constrain some of the model parameters by imposing physical conditions on the speed of sound on the shell.
In Section~\ref{sec:PolarPertEqs} we summarize classic results on the perturbations of a Reissner–Nordström background, as presented in Refs.~\cite{Chandrasekhar:1985kt,Burko:1995gf}.
In Section~\ref{sec:Firstjunction} we find perturbative corrections to the junction conditions for both the metric and the Faraday tensor.
In Section~\ref{sec:BoundaryConditions} we discuss the physically appropriate boundary conditions needed to solve the perturbation equations.
In Section~\ref{sec:NumericalResults} we constrain the viable values of the internal metric parameters, and we examine the QNM spectra as functions of the free parameters of the firewall model.
In Section~\ref{sec:Conclusion} we summarize some of our results and outline possible directions for future work.

Throughout the paper we use geometrical units ($G=c=1$), a ``mostly plus'' metric signature $(-,\,+, \,+, \,+)$, and (unlike Chandrasekhar's book~\cite{Chandrasekhar:1985kt}) we use the traditional ordering $(t,\,r,\,\theta,\,\phi)$ for spherical polar coordinates.

\section{Firewall Construction}
\label{sec:FirewallConstruction}

The original firewall construction of Ref.~\cite{Kaplan:2018dqx} required an external Schwarzschild metric, but we will generalize it to allow for an external Reissner-Nordstr\"om metric.
The interior and exterior metrics are given by 
\begin{align}
\label{eq:InteriorMetric}
    ds^2_{\rm in}&=-\frac{\Delta_{\rm in} (r)}{r^2} dt'^2 + \frac{r^2}{\Delta_{\rm in} (r)}dr^2 + r^2d\Omega^2\,,\\
\label{eq:ExteriorMetric}
    ds^2_{\rm out}&=-\frac{\Delta_{\rm out} (r)}{r^2} dt^2 + \frac{r^2}{\Delta_{\rm out} (r)}dr^2 + r^2d\Omega^2\,,
\end{align}
where  the functions
\begin{align}
    \Delta_{\rm out} &= \left(r-r^+_{\rm out}\right)\left(r-r^-_{\rm out}\right) \,,\\
    r_{\rm out}^{\pm} &= M_{\rm out} \pm \sqrt{M_{\rm out}^2 - Q_{\rm out}^2}\,,
\end{align}
are related to the mass $M_{\rm out}$ and charge $Q_{\rm out}$ of the outer spacetime. Here and below we use a subscript ``out'' (``in'') to refer to the outer (inner) spacetime, and relations analogous to the ones above apply also to the ``in'' quantities.  The time coordinate $t'$ in the interior of the shell differs from the time coordinate $t$ in the exterior.  The external metric reduces to Schwarzschild, as in the original construction of Ref.~\cite{Kaplan:2018dqx}, in the limit $Q_{\rm out}\to 0$.

The shell is placed at a position $a=r^+_{\rm out}(1+\epsilon)$ beyond the outer horizon of the exterior metric, but within the inner horizon of the interior metric:
\begin{equation}
\label{eq:horizonOrder}
    r^+_{\rm out} < a < r^-_{\rm in}\,.
\end{equation}
This construction (cf. Fig.~\ref{fig:SchBH}) avoids the presence of an event horizon.

The relation between the interior and exterior time variables follows by imposing continuity of the metric across the shell, which is located at $r=a$.
This requires
\begin{align} \label{eq:tRedef}
    dt' &= C_{\rm in} dt\,,\\
    C_{\rm in} &= \sqrt{\frac{\Delta_{\rm out} (a)}{\Delta_{\rm in} (a)}}\,.
\end{align}
Then the metric in both the interior and the exterior of the shell can be written (in terms of the same time variable) in the form
\begin{align}
  ds_{\rm in}^2 &= - f_{\rm in}(r) dt^2 + \frac{dr^2}{h_{\rm in}(r)} + r^2 d\Omega^2\,,\nn\\
  ds_{\rm out}^2&= - f_{\rm out}(r) dt^2 + \frac{dr^2}{h_{\rm out}(r)} + r^2 d\Omega^2\,.
\label{eq:sphericalMetric}
\end{align}
We will drop the ``in'' and ``out'' subscripts when the equations under consideration are symmetric under exchange.

Besides the metric, we will also be interested in the Faraday tensor $F_{\mu\nu}$.  In the interior, the only nonzero component of $F_{\mu\nu}$ not dictated by symmetry is given by
\begin{align}
    F_{t r} = C_{\rm in} \frac{Q_{\rm in}}{r^2}\,,
\end{align}
while in the exterior we have
\begin{align}
    F_{t r} = \frac{Q_{\rm out}}{r^2}\,.
\end{align}

We will use units such that $r^+_{\rm out}=1$ and we will assume that $Q_{\rm in}>0$.
With these choices, the external metric is extremal when $M_{\rm out}=|Q_{\rm out}|=1$.

\section{Israel Junction Conditions}
\label{sec:StaticJunctionConditions}

We assume the shell to be infinitesimally thin and model it as a spacelike hypersurface with spacelike, outward-pointing normal vector $n_\mu = (0,1,0,0)/\sqrt{g^{rr}}$.
The normal vector allows us to define the induced metric on the surface 
\begin{equation}
\label{eq:inducedMetric}
    \gamma_{\mu\nu}=g_{\mu\nu}-n_\mu n_\nu\,.
\end{equation}
Since the vector $n_\mu$ is pointing in the radial direction, we may simplify the analysis by using the same coordinates used for the bulk (at fixed $r$) to describe intrinsic quantities on the shell, which we will denote with Latin indices running through $0,2,3$. The embedding of the shell into the bulk spacetime is described by the extrinsic curvature $K_{\mu\nu}$, given by 
\begin{equation}
\label{eq:ExtrinsicCurvitureGeneral}
    K_{\mu\nu} =  \gamma_\mu^\alpha \gamma_\nu^\beta \nabla_\alpha n_\beta\,.
\end{equation}
The surface stress-energy tensor is restricted to the shell. For a perfect fluid it can be written as 
\begin{equation}
\label{eq:StressEnergy}
    S_{ab} = (\rho - \Theta)u_a u_b - \Theta \gamma_{ab}\,,
\end{equation}
where $\rho$ is the surface mass density, $\Theta$ is the surface tension, and $u^a = (1,0,0)/\sqrt{-\gamma_{tt}}$ (or equivalently, $u^\mu = (1,0,0,0)/\sqrt{-g_{tt}}$) is the four-velocity of a mass element on the shell.
Note that the four-velocity $u^\mu$ is perpendicular to the normal vector $n^\mu$. 

The presence of the thin shell of matter causes spacetime curvature.  Einstein's equations imply that the shell's embedding in spacetime is related to the surface stress-energy, and they dictate how the derivative of the metric jumps at $r=a$.  The relation is given by the Israel junction conditions~\cite{Israel:1966rt}:
\begin{align}
    [[K_{ij}]] &= 8 \pi [[S_{ij} - \gamma_{ij} S/2]]=8\pi[[\overline{S}_{ij}]]\,,\\
\label{eq:JunctionConditions}
    [[S_{ij}]] &= \frac{1}{8\pi} [[K_{ij} - \gamma_{ij} K]]=\frac{1}{8\pi}[[\overline{K}_{ij}]]\,,
\end{align}
where  $[[\cdot]]$ denotes the jump in the corresponding quantity at $r=a$,
\begin{equation}
    [[A]] = A(a_+) - A(a_-).
\end{equation}

Inserting Eqs.~\eqref{eq:sphericalMetric} into the junction conditions \eqref{eq:JunctionConditions}, one finds 
\begin{align}
\label{eq:rho}
    \rho &= -\frac{1}{4 \pi a}[[\sqrt{h}]]\,,\\
\label{eq:p}
    \Theta &= -\frac{1}{16 \pi }\left[\left[ \sqrt{h}\left( \frac{2}{a} + \frac{f'}{f} \right) \right]\right]\,.
\end{align}
Note that the surface tension of the shell diverges as it approaches any horizon of the interior or exterior metrics.

Moreover, the Faraday tensor also changes across the shell. The junction conditions for the Faraday tensor in special relativity hold also in GR: namely, the tangential components are smooth across the surface, while the mixed components are discontinuous, the difference being proportional to the surface four-current
\begin{equation}
    s_a = \eta u_a\,.
\end{equation}
Here $\eta$ is the surface comoving charge density, and $u_a$ the surface three-velocity.
The jump in the Faraday tensor is given by~\cite{1968CzJPh..18..435K} 
\begin{align}
    [[F_{ab}]] &= 0\,,\\
    [[F_{ar}]] &= -4\pi s_a\,.
\end{align}
Hence we have
\begin{align}
\label{eq:eta}
    [[F_{tr}]]&=4\pi\eta\sqrt{f(a)}\,,\\
    \eta &= \frac{Q_{\rm out} - C_{\rm in} Q_{\rm in}}{4 \pi a^2 \sqrt{f(a)}}\,.
\end{align}
The difference in charge is the integral of the surface charge density across the shell.

We will assume the matter component of the shell to have a constant mass-to-charge ratio $\sigma$, so that we have the following relation between the charge and mass densities:
\begin{align}
\label{eq:MassToCharge}
    \rho &= \sigma \eta\,.
\end{align}

\section{Radial Stability}
\label{sec:RadialStability}

The radial stability of the shell can be studied by following the method outlined in Ref.~\cite{Visser:2003ge}.
We will initially work with a generic metric of the form \eqref{eq:sphericalMetric}, where the radius of the shell is a function of the time coordinate $a(t)$.
We will only outline the calculation for brevity. More details can be found in Ref.~\cite{Visser:2003ge}.

The general idea is to study the dynamics of $a(t)$.  The Israel junction conditions allow us to relate the spacetime geometry to the thermodynamic properties of the shell, and hence to determine the radial acceleration of the shell.  The equation for the surface mass density can be cast into an ``energy balance'' equation with some effective potential.  This balance equation allows us to ``invert'' the meaning of the junction conditions: rather than asking for the acceleration in terms of the thermodynamic properties of the shell, we can express the thermodynamic properties of the shell in terms of the potential, which can be chosen freely.  In particular, we can make the shell static at radius $a_0$ by through a choice of potential, and hence determine the thermodynamic parameters that correspond to a stable shell.  Of particular interest is the speed of sound corresponding to this stability condition.

If we express the radial position of the shell as a function of proper time $a(\tau)$, the four-velocity can be written as
\begin{align}
    u^\mu &= \left( \frac{d t}{d \tau},\frac{d r}{d \tau},0,0 \right)\bigg\vert_{r=a(\tau)}\nonumber\\
     &= \left(\frac{\sqrt{h+(\partial_\tau a)^2}}{\sqrt{hf}},\partial_\tau a,0,0 \right)\,,
\end{align}
where we have used the proper time of a shell element at fixed $\theta$ and $\phi$ to find $dt$~\cite{Visser:2003ge}.
The normal vector of a mass element on the shell, which is perpendicular to the four-velocity, is
\begin{align}
    n^\mu &= \left(\frac{\partial_\tau a}{\sqrt{fh}}, \sqrt{h+(\partial_\tau a)^2} , 0, 0 \right)\,.
\end{align}
It is now simple to compute the induced metric \eqref{eq:inducedMetric} and the extrinsic curvature \eqref{eq:ExtrinsicCurvitureGeneral} in terms of the vectors $u^\mu$ and $n^\mu$.

Using the junction conditions~\eqref{eq:JunctionConditions}, the surface mass density and the surface tension are (cf.~\cite{Visser:2003ge} for details):
\begin{equation}
\label{eq:radialSigma}
    \rho = -\frac{1}{4 \pi a} \bigg[\bigg[ 
    \sqrt{h+(\partial_\tau a)^2}
    \bigg]\bigg]\,,
\end{equation}
\begin{equation}
\label{eq:radialPresure}
    \Theta = -\frac{1}{8 \pi} \bigg[\bigg[ 
    A + \frac{1}{a}\sqrt{h+(\partial_\tau a)^2}
    \bigg]\bigg]\,,
\end{equation}
where $A$ is the magnitude of the four-acceleration:
\begin{equation}
    A = \frac{h^2 f' - f (\partial_\tau a)^2 h' + h \left[  (\partial_\tau a)^2 f' + 2 f \partial_\tau^2 a\right] }
    {  2  f h \sqrt{h + (\partial_\tau a)^2}}\,.
\end{equation}

The junction condition for the surface mass density~\eqref{eq:radialSigma} can be cast as an ``energy balance'' equation for $a(\tau)$ of the form 
\begin{equation}
\label{eq:master}
\frac{1}{2}(\partial_\tau a)^2+U(a)=0\,.
\end{equation}
This can be studied using standard one-dimensional effective potential methods for the motion of point particles. The ``potential'' $U$ can be considered as a function of $\rho$, and is given by
\begin{align}
    U &= -\frac{M_s^2}{8a^2} - \frac{a^2}{8M_s^2}(h_{\rm in}-h_{\rm out})^2+\frac{1}{4}(h_{\rm in}+h_{\rm out})\,,
\end{align}
where $M_s = 4 \pi \rho a^2$ is the mass of the shell.

We can now invert the potential to express $\rho$ in terms of $U$.
This procedure gives two roots:
\begin{align}
    \rho^2 &= \frac{1}{16 \pi^2 a^2 } \left[\sqrt{h_{\rm out}(h_{\rm out}+U)}\pm\sqrt{h_{\rm in}(h_{\rm in}+U)}\right]^2\,.
\end{align}
To find the physical root, consider the limit where there is no shell. Then we must have $\rho=0$ and $h_{\rm in}=h_{\rm out}$.
Inserting this into the solution for $\rho^2$ above, we conclude that the negative root corresponds to the physical branch.
The negative root can be simplified to the form
\begin{equation}
\label{eq:sigmaOfU}
    \rho = -\frac{1}{4 \pi a } \bigg[\bigg[
    \sqrt{h-2U}
    \bigg]\bigg]\,.
\end{equation}
Equation~\eqref{eq:master} can then be used to express the surface tension $\Theta$ in terms of $U$ and $\partial_a U$.

\begin{figure}[t]
  \includegraphics[width=0.9\columnwidth]{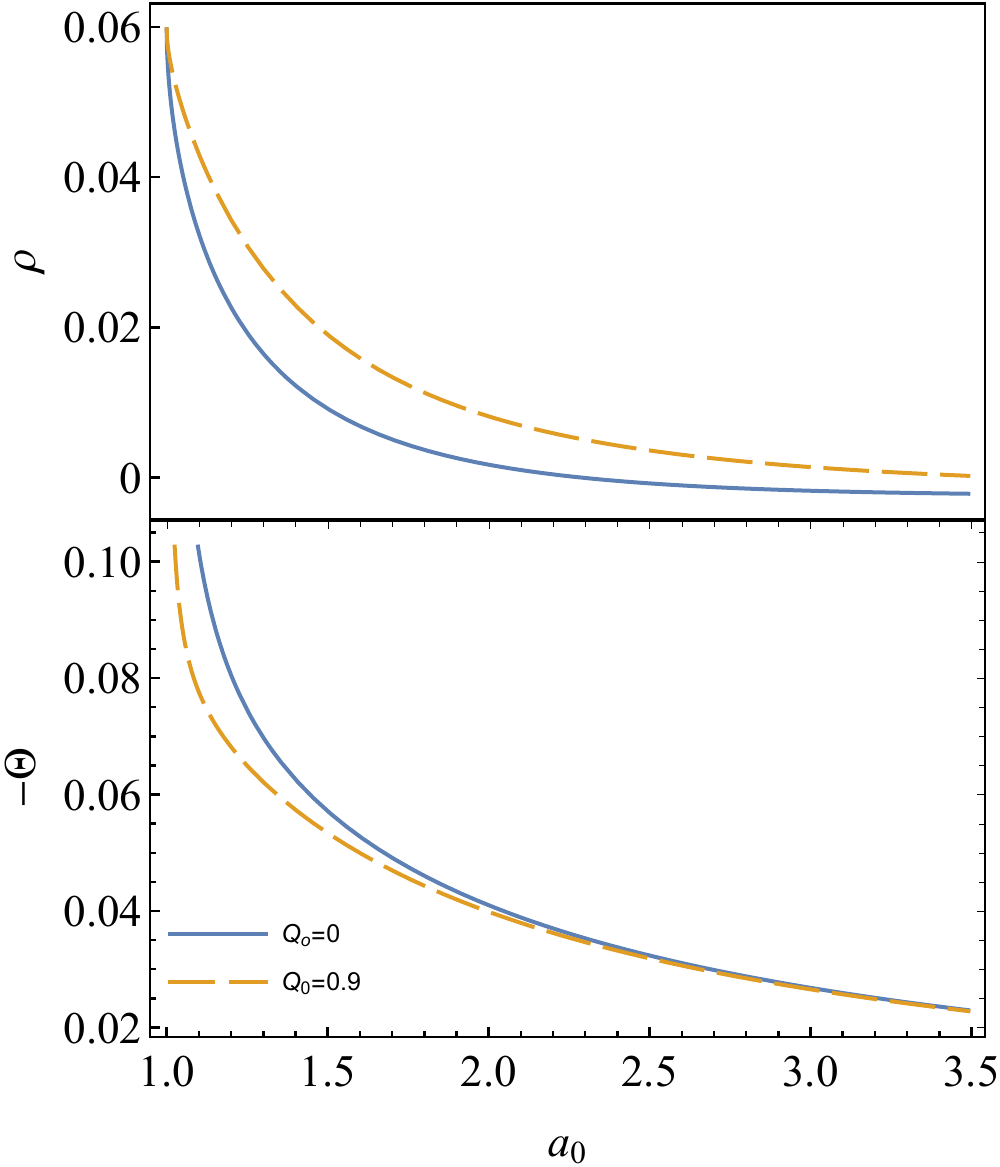}\\
  \caption{Surface energy density $\rho$ (Eq.~\eqref{eq:staticSurfaceEnergyDensity}: top panel) and surface tension $\Theta$ (Eq.~\eqref{eq:staticSurfaceTension}: bottom panel) for radially stable firewalls as functions of the shell's stable location $a_0$, assuming that the shell is at the minimum of the effective potential.  The blue solid line refers to a Schwarzschild exterior spacetime ($Q_{\rm out}=0$), the orange dashed line to a near-extremal Reissner-Nordstr\"om spacetime ($Q_{\rm out}=0.9$), both in units such that $r_{\rm out}^+=1$.  In this specific example we set $M_{\rm in}=Q_{\rm in}=7a_0/4$.  }
\label{fig:prhoVSa}
\end{figure}

\begin{figure*}
        \includegraphics[width=.65\columnwidth]{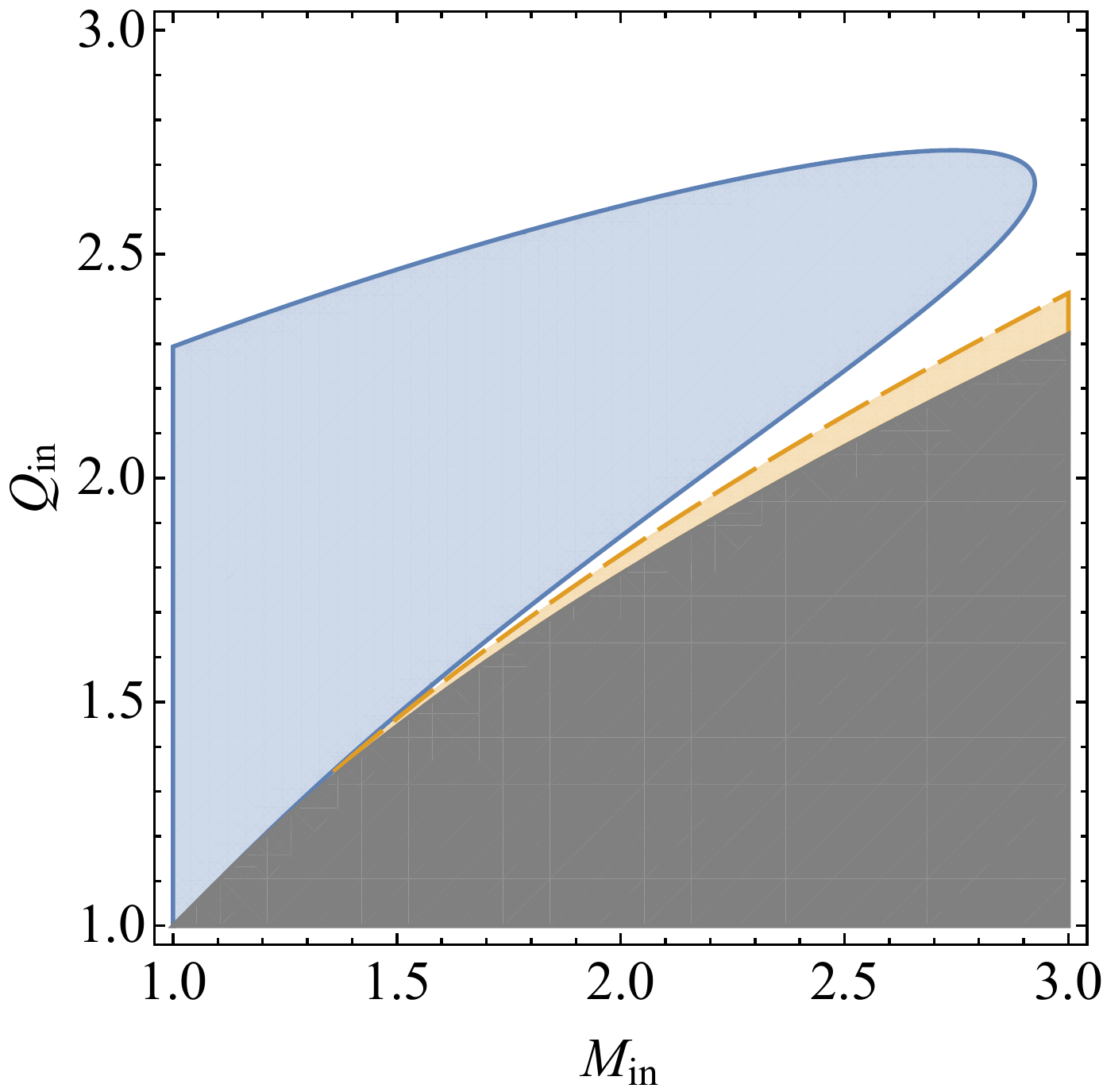}
        \includegraphics[width=.65\columnwidth]{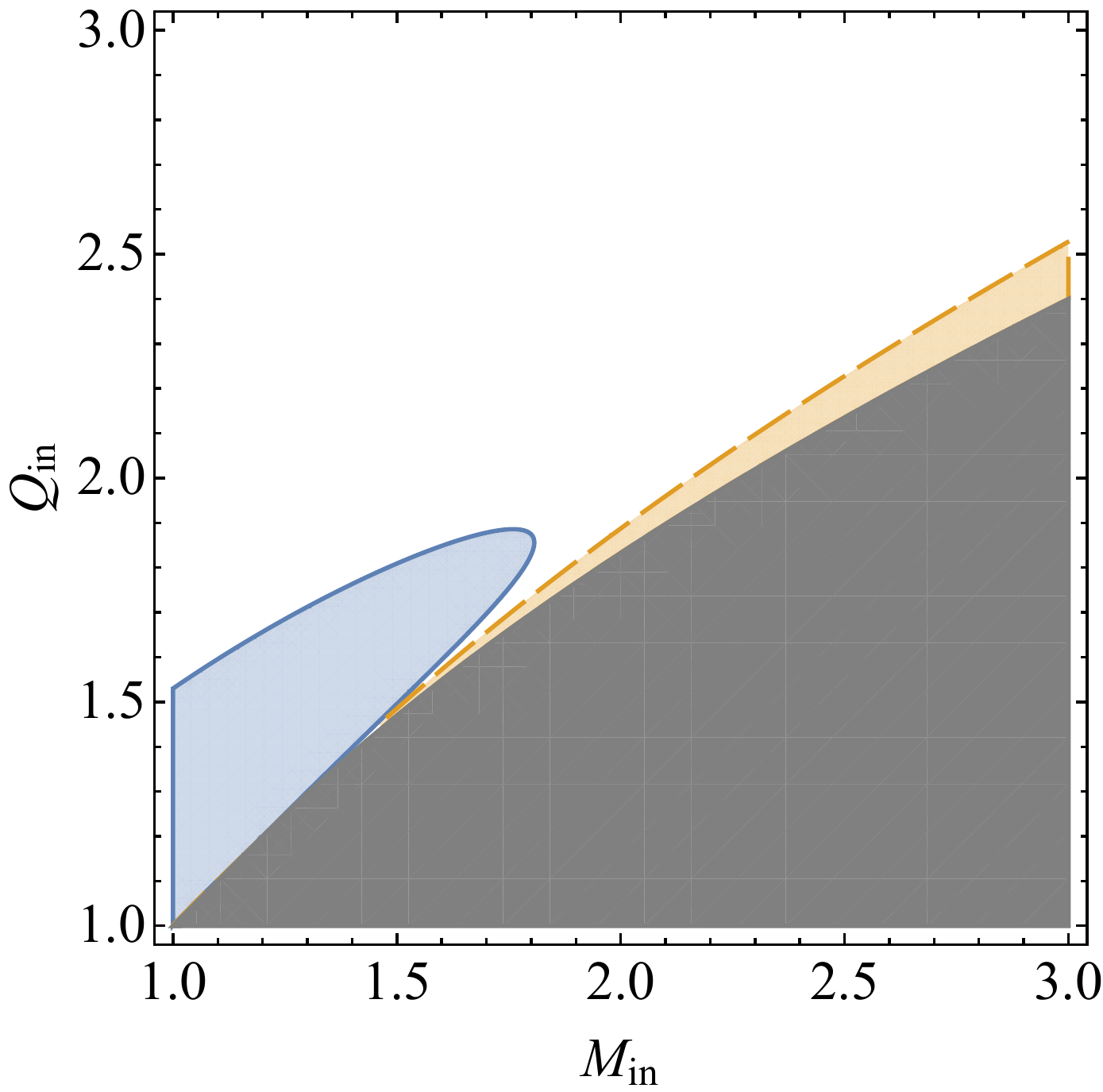}
        \includegraphics[width=.65\columnwidth]{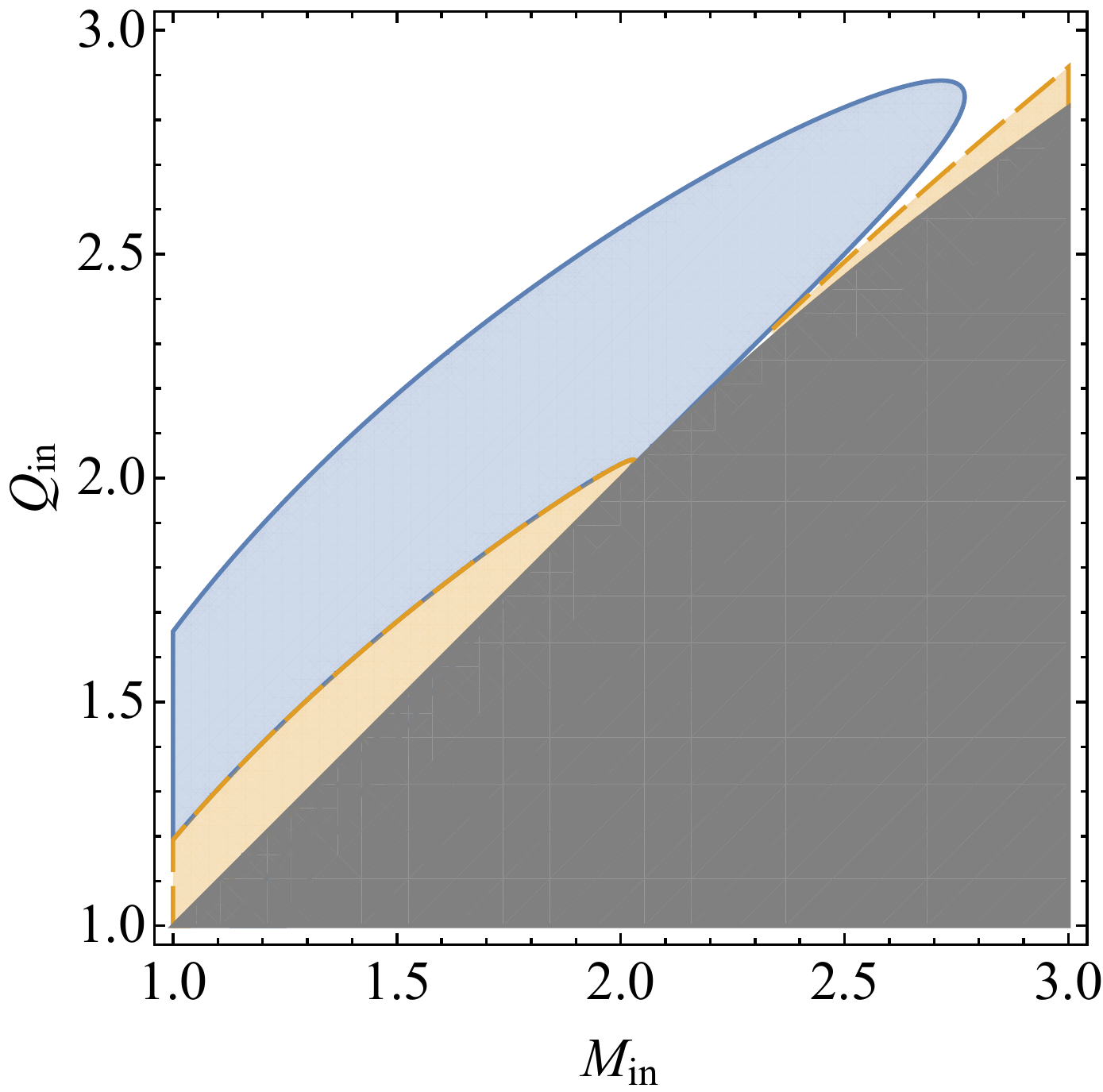}
        \includegraphics[width=.65\columnwidth]{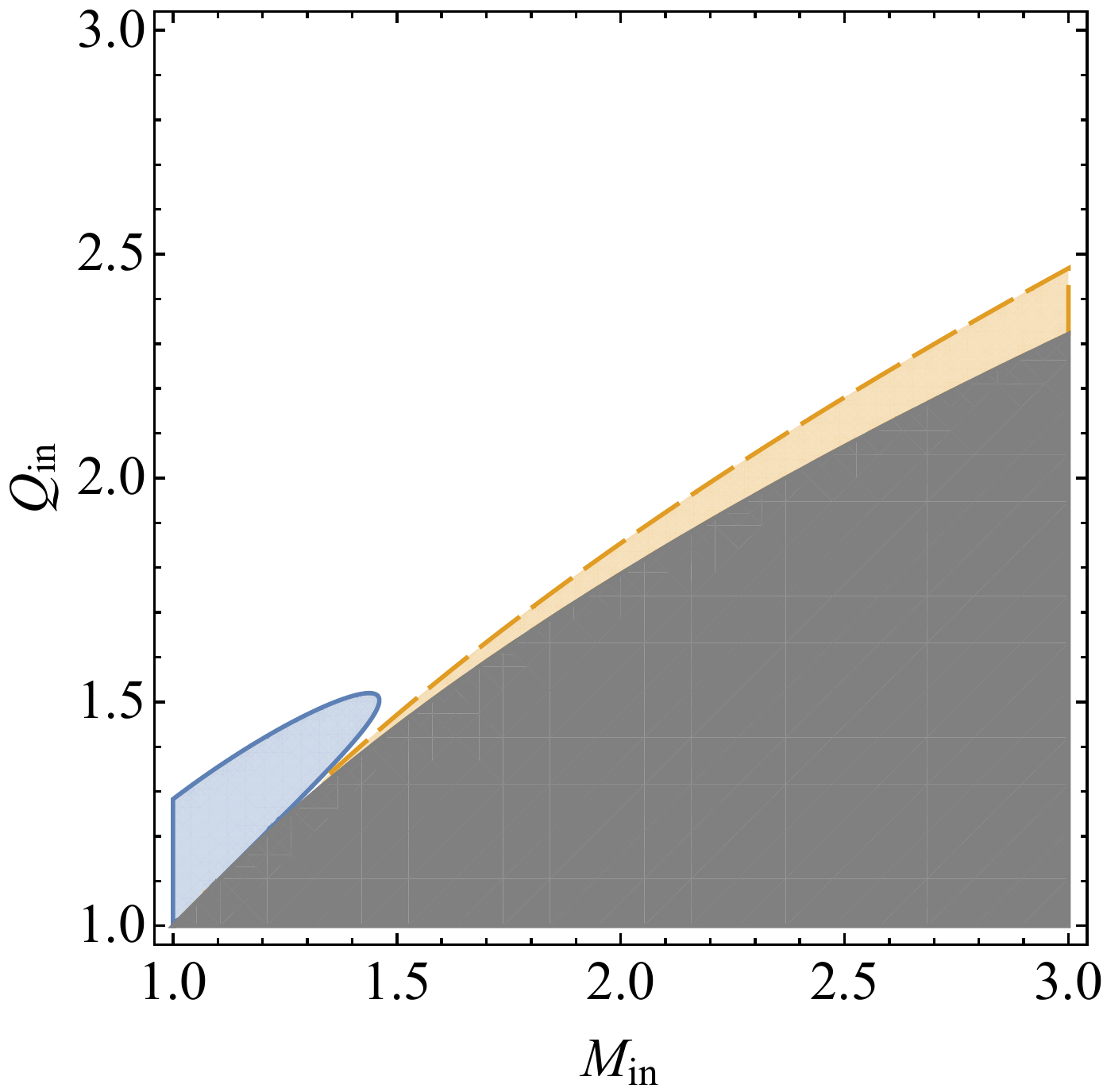}
        \includegraphics[width=.65\columnwidth]{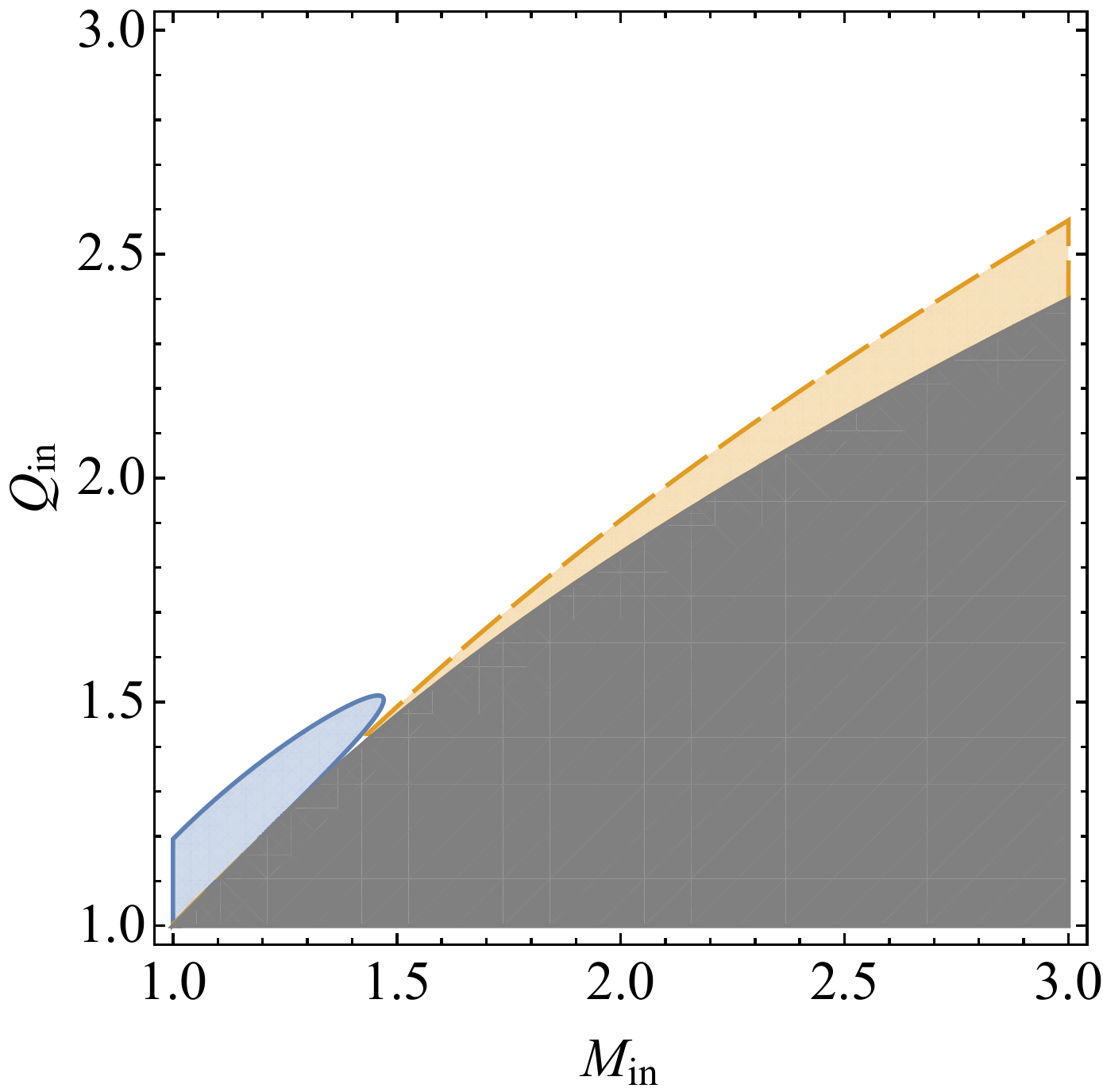}
        \includegraphics[width=.65\columnwidth]{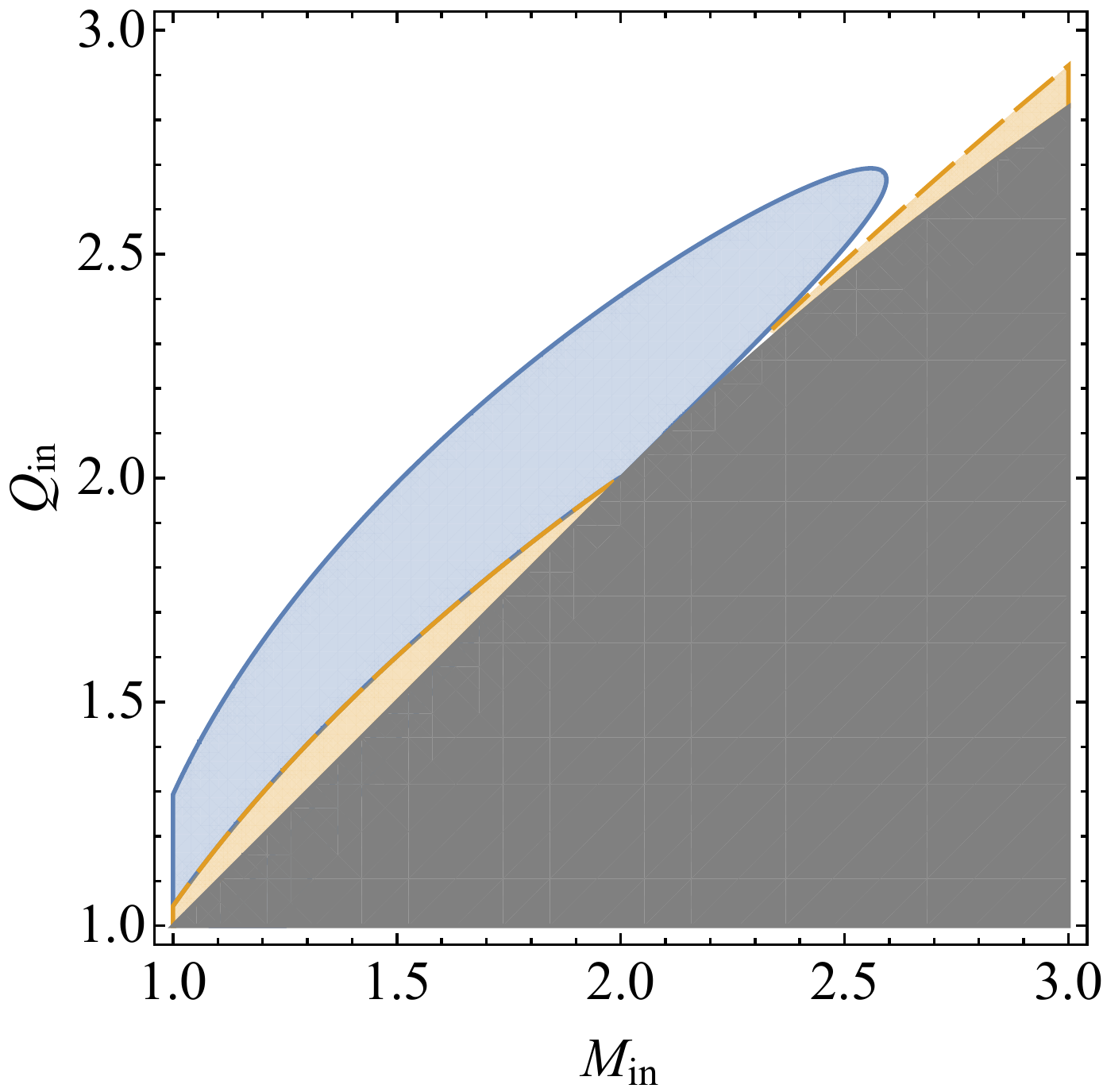}
        \caption{Constraints in the $(M_{\rm in},Q_{\rm in})$ plane associated with physical requirements on the speed of sound $v_s$ [Eq.~\eqref{eq:speedOfSound}] for three shell locations: $a_0=1.1$ (left), $a_0=1.2$ (center) and $a_0=2$ (right). The top row corresponds to a Schwarzschild exterior; the bottom row, to a Reissner-Nordstr\"om exterior with $Q_{\rm out}=0.9$, in units such that $r_{\rm out}^+=1$.  The blue region bounded by a solid line is unphysical because $v_s>1$, while the orange region bounded by a dashed line corresponds to $v_s<0$.  The orange region in the bottom-left corner of the right panels is due to a divergence in Eq.~\eqref{eq:speedOfSound}. The gray region is excluded because $r_{\rm in}^-<a_0$.}
        \label{fig:SpeedBounds}
\end{figure*}

Despite Eqs.~\eqref{eq:sigmaOfU} and \eqref{eq:radialSigma} looking similar, their physical meaning is different.  Equations~\eqref{eq:radialSigma} and~\eqref{eq:radialPresure} give the density $\rho$ and tension $\Theta$ of the shell in terms of the velocity and position, which are in principle prescribed by some model for the shell. 
However, we may consider Eq.~\eqref{eq:master} as an independent equation prescribing the dynamics of the shell, and hence \eqref{eq:sigmaOfU} prescribes the thermodynamic properties of the shell which are compatible with the specified dynamics. 

If we specialize these general results to the firewall metric of Eqs.~\eqref{eq:sphericalMetric}, we find
\begin{align}
    \label{eq:staticSurfaceEnergyDensity}
    \rho &= -\frac{1}{4 \pi a}\bigg[\bigg[
    \sqrt{1-2 M /a + Q^2 / a^2 - 2 U(a)}
    \bigg] \bigg] \,,\\
    \label{eq:staticSurfaceTension}
    \Theta &= -\frac{1}{8 \pi a}\bigg[\bigg[
    \frac{a-M-2aU(a)-a^2\partial_a U(a)}
    {\sqrt{a^2-2 a M+ Q^2 - 2 a^2 U(a)}}
    \bigg] \bigg] \,.
\end{align}
Note that these depend upon the potential and its first derivative, in contrast to  Eqs.~\eqref{eq:radialSigma} and~\eqref{eq:radialPresure}, which depend on the shell's position and velocity.

By analogy with classical mechanics, the shell will be stable under radial perturbations if there is some shell position $a_0$ such that
\begin{align}
    U(a_0) &= 0\,,\\
    \partial_a U(a_0) &= 0\,,\\
    \partial_a^2 U(a_0) &> 0\,.
\end{align}
Each value for $a_0$ would correspond to a different firewall model. Note that $\partial_a^2 U(a_0)$ does not appear in Eqs.~\eqref{eq:staticSurfaceEnergyDensity} and~\eqref{eq:staticSurfaceTension}.
We now consider the case of the shell resting at the minimum of $U(a)$ and examine its properties.

In Fig.~\ref{fig:prhoVSa}, for illustration, we show how the surface energy density and pressure vary with the shell's static position $a_0$ for $Q_{\rm out}=0$ (top) and $Q_{\rm out}=0.9$ (bottom). For illustration we set $M_{\rm in}=Q_{\rm in}=7a_0/4$, so that the inner radius $r_{\rm in}^-$ of the internal metric is located outside the shell.
In the limit $a_0\to1$, the energy density curves tend to a finite limit $\rho=3/(16\pi)\approx0.06$. Conversely, the surface tension diverges as the shell approaches the horizon. These results are consistent with the following analytic expansion of the surface energy density [Eq.~\eqref{eq:staticSurfaceEnergyDensity}] and surface tension [Eq.~\eqref{eq:staticSurfaceTension}] around $a_0=1$ for a radially stable firewall:
\begin{align}
\label{eq:rhoLimit}
    \rho &= \frac{\sqrt{1-2 M_{\rm in} + Q_{\rm in}^2}}{4 \pi} + \mathcal{O}((a_0-1)^{1/2})\,,\\
    \Theta &= - \frac{\sqrt{1-Q_{\rm out}^2}}{16 \pi \sqrt{a_0-1}} + \mathcal{O}((a_0-1)^0) \,.
\end{align}

The speed of sound on the shell is
\begin{align}
\label{eq:speedOfSound}
    v_s^2 &= -\frac{\partial \Theta}{\partial \rho} = -\frac{\partial \Theta}{\partial a} \frac{\partial a}{\partial \rho}\bigg\vert_{a=a_0}\,.
\end{align}
and for a Schwarzschild exterior we can write it as a series expansion around the horizon $a_0=1$ as follows:
\begin{align}
    v_s^2 &= \frac{1}{4(a_0-1)}+\frac{F_{-1/2}(M_{\rm in},Q_{\rm in})}{\sqrt{a_0-1}} \nonumber\\
    &+[F_{0}(M_{\rm in},Q_{\rm in})+\partial_a^2 U(a_0)]+\mathcal{O}((a_0-1)^{1/2})\,,
      \label{eq:vs}
\end{align}
for some functions $F_{-1/2}$ and $F_0$, whose explicit expression is not important for the present purposes. The first two terms clearly diverge as $a_0\to1$, while $\partial_a^2 U(a_0)$ enters at zeroth-order in the expansion
and with a positive sign. We will see momentarily that the requirement of a subluminal speed of sound ($0\leq v_s^2\leq 1$) places tight bounds on the model, and allowing $\partial_a^2 U(a_0)\neq 0$ can only increase the value of $v_s$.
Therefore we will set $\partial_a^2 U(a_0)=0$, which allows us to be as liberal as possible with the model.

The expansion \eqref{eq:vs} shows that the speed of sound becomes superluminal for a shell at the horizon for all model parameters. However, by fine-tuning the model we can impose a subluminal speed of sound for all finite values of $(a_0-1)$. In Fig.~\ref{fig:SpeedBounds} we show constraints in the $(M_{\rm in},Q_{\rm in})$ plane associated with physical requirements on the speed of sound $v_s$. We consider three shell locations ($a_0=1.1$, $a_0=1.2$ and $a_0=2$) and two exterior spacetimes: a Schwarzschild exterior (top row) and a Reissner-Nordstr\"om exterior with $Q_{\rm in}=0.9$ (bottom row). Recall that we use units such that $r_{\rm out}^+=1$.

The blue region, excluded by the requirement that the speed of sound should be subluminal, grows unbounded as $a_0\to0$.  As the shell position $a_0$ increases, this blue region initially shrinks, and then it grows again.  The orange region bounded by a dashed line corresponds to $v_s<0$.  The gray region is excluded because $r_{\rm in}^-<a_0$.  As $a_0$ gets larger and the blue region expands again, the allowed parameter space gets squeezed between the blue and orange ``forbidden regions.''  The emergence of the orange region in the bottom-left corner of the right panels is due to a divergence in Eq.~\eqref{eq:speedOfSound}: around this divergence, the speed of sound tends to $\pm\infty$ from either side.

In the following sections, when examining a particular shell position $a$ we will assume that the shell is static and located at a minimum of $U(a)$. For brevity we will drop the zero subscript from the static shell position and set $a=a_0$.

\section{Polar perturbations of the metric and Faraday tensor}
\label{sec:PolarPertEqs}

We now consider the nonradial stability of this firewall model, focusing on even-parity (or ``polar'') perturbations.  As both the interior and the exterior are described by the Reissner-Nordstr\"om metric, we can derive the perturbation equations in both cases following Refs.~\cite{Chandrasekhar:1985kt,Burko:1995gf}, but using two different time coordinates: $t'$ in the interior spacetime of Eq.~\eqref{eq:InteriorMetric}, and $t$ in the exterior spacetime of Eq.~\eqref{eq:ExteriorMetric}.  For brevity, we will only write down the perturbation equations in terms of the variable $t$.

We make use of Chandrasekhar's ``metric of sufficient generality'' for polar perturbations~\cite{Chandrasekhar:1985kt} 
\begin{align}
    ds^2 &= -e^{2\nu}dt^2
        +e^{2\mu_r}dr^2
        + e^{2\mu_\theta} d\theta^2
        + e^{2\psi}d\phi^2 \,,
\end{align}
where $\nu,\psi,\mu_r$ and $\mu_\theta$ are functions of $(t,\,r,\,\theta)$.
At leading order, these functions are chosen such that the metric is either of Eqs.~\eqref{eq:sphericalMetric}.
First-order perturbations to the metric are given by
\begin{align}
\label{eq:polarMetricPert}
    \delta g_{\mu\nu} = 2
    \begin{bmatrix}
    - \delta \nu \Delta / r^2 & 0 & 0  & 0   \\
    0 &  \delta \mu_2\, r^2 / \Delta & 0 & 0  \\
    0 & 0 &  \delta \mu_3\, r^2 & 0 \\
    0 & 0 & 0 &  \delta \psi\, r^2 \sin^2\theta  
\end{bmatrix}\,,
\end{align}
where 
\begin{subequations}
\label{eq:metricPerturbations}
\begin{align}
    \delta\nu &= N(r,t) P_\ell(\theta) \,,\\
    \delta\mu_2 &= L(r,t) P_\ell(\theta) \,,\\
    \delta\mu_3 &= T(r,t) P_\ell(\theta) + V(r,t)\partial^2_\theta P_\ell(\theta)\,,\\
    \delta\psi &= T(r,t) P_\ell(\theta) + V(r,t)\cot(\theta)\partial_\theta P_\ell(\theta) \,,
\end{align}
\end{subequations}
and $P_\ell(\theta)$ denotes Legendre polynomials. The functions $N,L,T$ and $V$ differ in the interior and in the exterior. They must be related by imposing continuity of the metric, the junction conditions and the $t\to t'$ time redefinition.

The polar components of the Faraday tensor are perturbed in a similar fashion:
\begin{align}
    \delta F_{(0)(1)} &= -\frac{r^2 h}{2 Q}B_{01}(t,r) P_\ell(\theta)
    \,,\label{eq:deltaF01}\\
    \delta F_{(0)(2)} &= \frac{r \sqrt{h}}{2 Q}B_{02}(t,r) P_\ell'(\theta)
    \,,\label{eq:deltaF02}\\
    \delta F_{(1)(2)} &= i \omega \frac{r }{2\sqrt{h} Q}B_{13}(t,r) P_\ell'(\theta)
    \,,
\label{eq:deltaF12}
\end{align}
where we use brackets to indicate tetrad components of the tensor, which are related to the perturbations in the coordinate frame via the relation $\delta F_{\mu\nu}=\delta F_{(\alpha)(\beta)}e^{(\alpha)}_\mu e^{(\beta)}_\nu$, where the tetrad (``vierbein'') vectors are defined in Appendix~\ref{app:tetrad} to improve readability. We introduced some notational changes with respect to Ref.~\cite{Chandrasekhar:1985kt}: Eqs.~\eqref{eq:deltaF01}, \eqref{eq:deltaF02} and \eqref{eq:deltaF12} differ by a minus sign with respect to the equations in Ref.~\cite{Chandrasekhar:1985kt} due to the change in metric signature, and the indices have a different meaning because of the different ordering of the coordinates. 

The time dependence can be separated as usual by a Fourier expansion, but we must use different frequencies in the interior and in the exterior: for example
\begin{align}
    N_{\rm out}(r,t) &= \hat{N}_{\rm out}(r)e^{-i \omega_{\rm out} t}\,,\\
    N_{\rm in}(r,t') &= \hat{N}_{\rm in}(r)e^{-i \omega_{\rm in} t'}\,.
\end{align}
Metric continuity demands $\omega_{\rm in}= C_{\rm in}^{-1} \omega_{\rm out}$ when we perform the change of variables $t'\to C_{\rm in} t$ [cf. Eq.~\eqref{eq:tRedef}].

While it may appear that there are several perturbation variables, there are only two polar degrees of freedom. Through a series of definitions and substitutions~\cite{Chandrasekhar:1985kt,Burko:1995gf}, the perturbation equations for the Reissner-Nordstr\"om metric can be found to be
\begin{equation}
\label{eq:RNPerturbationEquation}
    \left( \frac{d^2}{dr_*^2} + \omega^2 \right)Z^+_{1,2} = V_{1,2} Z^+_{1,2}\,,
\end{equation}
where
\begin{align}
    V_{1,2}(r) &= \frac{\Delta}{r^5}\left[U(r) \pm \frac{1}{2}(q_1 - q_2) W(r)\right]\,,\\
    U(r)&= (\Lambda r+3M)W(r)-\Lambda \frac{\Delta(r)}{\chi(r)}\nonumber \\ 
    &+[\chi(r)-\Lambda r/2-M]\,,\\
    W(r) &= \frac{\Delta(r)}{r\chi(r)^2}(\Lambda r + 3M) + \frac{\Lambda r+2M}{2\chi(r)}\,,\\
    \label{eq:PertDefRefStart}
    \chi(r) &= \Lambda r/2 + 3M - 2Q^2/r\,,\\
    q_1&=3M+\sqrt{9M^2+4Q^2\Lambda}\,,\\
    q_2&=3M-\sqrt{9M^2+4Q^2\Lambda}\,,\\
    \label{eq:PertDefRefEnd}
    \Lambda &= (\ell-1)(\ell+2)\,.
\end{align}
The potential $V_1$ ($V_2$) corresponds to choosing the plus (minus) sign.
The algebraic completion of the system, relating the wave functions $Z_{1,2}$ to the metric components, is given in Appendix~\ref{app:AlgebraicCompletion}.

It is easily seen that $V_2$ does not diverge away from the origin, while $V_1$ diverges as $\chi(r)\to0$. 
The divergence occurs at a radius $d_{\ell}$ that is a monotonically decreasing function of~$\ell$:
\begin{align}
\label{eq:v1div}
    d_{\ell} &= \frac{-3M+\sqrt{9M^2+4\Lambda Q^2}}{\Lambda}\,,\\
    d_{1} &=\frac{2Q^2}{3M} \,.
\end{align}
for $\ell>1$ and $\ell=1$, respectively.  These divergences all appear within the inner horizon, and so are not normally of physical interest (but see~\cite{Dotti:2010}).  All divergences cannot be moved outside of the shell, as for any set of metric perturbations there exists a finite $\ell$ such that $d_{\ell}<a$.
This will be important when discussing boundary conditions for the perturbation equation for $Z^+_1$ in Sec.~\ref{sec:BoundaryConditions} below.

The perturbation equations can be solved separately in the interior and in the exterior. The relation of the two sets of perturbation variables across the shell is prescribed by the junction conditions, as discussed in the next section.

\section{Junction conditions for the perturbations}
\label{sec:Firstjunction}

The perturbation equations~\eqref{eq:RNPerturbationEquation}, given initial conditions, allow us to find the form of the wave functions $Z_{1,2}^+$. Through the metric completions of Appendix~\ref{app:AlgebraicCompletion}, this determines the metric perturbations from the origin to the location of the shell. To integrate beyond the shell, we must impose continuity of the metric and find the jump in the derivatives of the metric perturbations across the shell.
Metric perturbations complicate the procedure because they modify the radius and the four-velocity of a given shell element.

In this section we find the conditions needed to describe the change in the perturbed metric and in the perturbed Faraday tensor across the charged shell.  To simplify the calculation, it is convenient to perform a coordinate transformation such that the shell remains static at a fixed radius (Sec.~\ref{sec:CoordChange}).  Then imposing metric continuity becomes trivial, as shown in Sec.~\ref{sec:MetricCont}.  In Section~\ref{sec:firstIsrael} we apply the Israel junction conditions, and finally in Sec.~\ref{sec:faradayFirst} we find the change in the Faraday tensor across the shell.

\subsection{Making the shell static}
\label{sec:CoordChange}

The shell lies at the transition between the internal and external metric, but the metric perturbations cause the shell to oscillate with four-velocity
\begin{equation}
  u^\alpha = (1+\dot{\delta t}, \dot{\delta r}, \dot{\delta \theta}, 0) /\sqrt{-g_{tt}(a)}\,,
\end{equation}
where dots denote derivatives with respect to $t$.
It is easier for our purposes to impose junction conditions when the shell is static, as in Ref.~\cite{Pani:2009} (see Ref.~\cite{Cardoso:2019upw} for a similar calculation were the shell remains dynamical).
Therefore we look for a coordinate transformation such that the shell is static, i.e. the four-velocity of an element on the shell becomes 
\begin{equation}
    u^\alpha = (1,0,0,0)/\sqrt{-g_{tt}(a)}\,.
\end{equation}
We will perform this transformation on the internal and the external metric, both expressed in terms of the $t$ coordinate. 

For polar perturbations, infinitesimal coordinate changes of the form $x_{\alpha}\to \bar x_{\alpha}=x_{\alpha}+\sum_{i=0}^2\xi_{\alpha}^{(i)}$ involve three independent functions $y(t)$, $z(t)$ and $w(t)$, which can differ in the interior and in the exterior:
\begin{align}
    \xi_{\alpha}^{(0)} &= (y(t)P_\ell(\theta),0,0,0)\,,\\
    \xi_{\alpha}^{(1)} &= (0,z(t)P_\ell(\theta),0,0)\,,\\
    \xi_{\alpha}^{(2)} &= (0,0,w(t)P_\ell(\theta)_{,\theta},0)\,,
\end{align}
where as usual $P_\ell(\theta)$ denotes Legendre polynomials, and we omit an $\ell$ subscript on the functions of time for notational simplicity.
We can now Fourier transform the three independent functions, e.g.
\begin{align}
    y_{\rm out}(t) &= \hat y_{\rm out} e^{-i \omega_{\rm out} t}\,,\\   
    y_{\rm in}(t) &= \hat y_{\rm in} e^{-i \omega_{\rm out} t}\,,
\end{align}
and set their amplitude at the shell by imposing the junction conditions.

The perturbed metric in the new coordinate system is given by 
\begin{equation}
\label{eq:pertebedMetricNewCoords}
    \overline{g}_{\alpha\beta} = g_{\alpha\beta} + \delta g_{\alpha\beta} + \sum_{i=0}^2 \Xi_{\alpha\beta}^{(i)}\,,
\end{equation}
where
\begin{equation}
    \Xi_{\alpha\beta}^{(i)} = \xi_{\alpha;\beta}^{(i)}+\xi_{\beta;\alpha}^{(i)}
\label{eq:gaugeshell}
\end{equation}
and covariant derivatives are computed with respect to the unperturbed metric $g_{\mu\nu}$. 
The Faraday tensor also changes under this change of coordinates:
\begin{equation}
    \overline{F}_{\mu\nu} =  F_{\mu\nu} + \delta F_{\mu\nu} +\sum_{i=0}^2 \mathcal{L}_{\xi^{(i)}} F_{\mu\nu}
\label{eq:lieshell}
\end{equation}
where $\mathcal{L}_{\xi^{(i)}} F_{\mu\nu}$ is the Lie derivative of $F_{\mu\nu}$ along the vector $\xi^{(i)}$.  Explicit expressions for $\Xi_{\alpha\beta}^{(i)}$ and $\mathcal{L}_{\xi^{(i)}} F_{\mu\nu}$ are given in Appendix~\ref{app:CoordinateChange}.

\subsection{Metric continuity}
\label{sec:MetricCont}

For the spacetime to be defined at the shell, the metric must be continuous across the shell, while discontinuities in the metric derivatives are controlled by the junction conditions.  These conditions impose relations between the metric perturbations and the coordinates on either side of the shell.  The $tt$, $t\theta$, $\theta\theta$, and $\phi\phi$ components of the junction conditions give
\begin{align}
\label{eq:MetricConttt}
    [[2\dot y - h z f']] &= 2 f [[N]]\,,\\
\label{eq:MetricContttheta}
    [[y+\dot w]] &=0\,,\\
\label{eq:MetricContthetatheta}
    [[h z] &= -a[[T]]\,,\\
\label{eq:MetricContphiphi}
    [[w]] &= -a^2 [[V]]\,,
\end{align}
respectively .
Note that Eq.~\eqref{eq:MetricConttt} is the same condition found when imposing the normalization of the four-velocity with the perturbed metric in the $\overline{x}^\mu$ coordinate system.

\subsection{Israel junction conditions}
\label{sec:firstIsrael}

The surface stress-energy tensor of the shell is 
\begin{equation}
    S_{ij} = \left[ \rho - \Theta + (\delta\rho - \delta \Theta) P_\ell\right]u_i u_j - \left[ \Theta + \delta \Theta P_\ell \right] \overline{g}_{ij}\,,
\end{equation}
where $\Theta$ is the surface tension.
The perturbations $\delta \Theta$ and $\delta \rho$ are related via 
\begin{equation}
\label{eq:speedOfPErt}
    \delta \Theta = - u_s^2 \delta \rho.
\end{equation}
Note that the speed of sound for the polar perturbations need not be the same as the speed of sound for radial oscillations of the shell in Eq.~\eqref{eq:speedOfSound}, and so we treat it as an independent parameter. 

We can decompose the angular dependence of the surface stress-energy tensor, of the metric and of the extrinsic curvature as in Appendix A of Ref.~\cite{Pani:2009}.
The polar components of a three-tensor with signature $(-,\,+,\,+)$ can be written as
\begin{align}
\label{eq:decomposition}
    T &=
    \left[
    \begin{array}{cccc}
         T_1 & 0 & 0 \\
         0 & T_2 & 0 \\
         0 & 0 & T_2 \sin^2\theta 
    \end{array}\right] P_\ell 
    +
    T_3 \left[
    \begin{array}{cccc}
         0 & \partial_\theta & 0 \\
         \partial_\theta & 0 & 0 \\
         0 & 0 & 0 
    \end{array}\right] P_\ell  \nonumber\\
    &+
    T_5 \left[
    \begin{array}{cccc}
         0 & 0 & 0 \\
         0 & \partial_\theta^2 & 0 \\
         0 & 0 & \cos\theta\, \sin\theta\, \partial_\theta
    \end{array}\right] P_\ell \,.
\end{align}
Given a tensor $T_{ij}$, there are four irreducible components $T_{(k)}$ $(k=1,\,2,\,3,\,5)$ that transform as polar quantities (we follow the notation of Ref.~\cite{Pani:2009}, where $T_{(4)}$ and $T_{(6)}$ denote axial quantities).
Through this decomposition we can relate the irreducible components under rotations of Eq.~\eqref{eq:JunctionConditions}.

For the stress-energy tensor of Eq.~\eqref{eq:StressEnergy}, one finds
\begin{align}
    S_{(1)} &= f(\delta \rho - 2(\Theta - 2 \rho) N(a,t)\nonumber\\ &-(\Theta-2\rho)(hzf'-2\dot{y})\,,\\
    S_{(2)} &= -a[a\delta \Theta + 2a \Theta T(a,t) + 2 \Theta h z]  \,,\\
    S_{(3)} &= -\rho(y+\dot{w})  \,,\\
    S_{(5)} &= -2 \Theta [a^2V(a,t)+w] \,.
\end{align}
The components of the stress-energy tensor should not be affected by coordinate changes: they are intrinsic to the shell, and they should only depend upon intrinsic quantities. The perturbations and changes to the metric inside and outside the shell should play no role because the shell is static.

From Eqs.~\eqref{eq:MetricConttt}, \eqref{eq:MetricContttheta} and \eqref{eq:MetricContphiphi} we find 
\begin{align}
    8 \pi f \delta \rho &= [[\bar K_1]]\,,\\
    -8 \pi a^2\delta \Theta &= [[\bar K_2]]\,,\\
    0 &= [[\bar K_3]]\,,\\
    0 &= [[\bar K_5]]\,.
\end{align}
This makes physical sense: the shell is static in the new coordinate system, and so we should not expect any shear or momentum flux.
The forms of $\bar K_i$, which are long and not enlightening, are presented in Appendix~\ref{app:KiFaraday}.

\subsection{Faraday tensor}
\label{sec:faradayFirst}

We now introduce perturbations $\delta\eta$ to the surface comoving charge density:
\begin{align}
    s_\alpha + \delta s_\alpha = \left[\eta +\delta\eta(t) P_\ell\right] u_\alpha\,.
\end{align}
The perturbation $\delta\eta$ can be related to the perturbation in the matter density $\delta \rho$ through the constant mass-to-charge ratio $\sigma$ introduced in Eq.~\eqref{eq:MassToCharge}:
\begin{align}
    \delta \rho &= \sigma \delta \eta\,.
\end{align}
By imposing metric continuity we find 
\begin{align}
\label{eq:JunctionFaraday}
    \big[\big[\delta F_{tr} +\sum_{i=0}^2 (\mathcal{L}_{\xi^{(i)}} F)_{tr}\big]\big] &= -4\pi \delta\eta\sqrt{f}P_\ell \,,\\
    \big[\big[\delta F_{t\theta} +\sum_{i=0}^2 (\mathcal{L}_{\xi^{(i)}} F)_{t\theta}\big]\big] &= 0 \,,\\
    \big[\big[\delta F_{r\theta} +\sum_{i=0}^2 (\mathcal{L}_{\xi^{(i)}} F)_{r\theta}\big]\big] &= 0 \,.
\end{align} 
The remaining conditions are trivial.
\section{Boundary conditions}
\label{sec:BoundaryConditions}

Now that we have the perturbation equations and the junction conditions, we need to solve the perturbation equations~\eqref{eq:RNPerturbationEquation}. To find the wave functions $Z^+_i$, we integrate outwards from $r=0$ to the shell, which is located at $r=a$.
The QNM frequencies of the spacetime are found by imposing outgoing-wave conditions in the limit $r\to\infty$, i.e.
\begin{equation}
    Z^+_{1,2}(r\to\infty) = e^{ i \omega r_*}\,.
\end{equation}

The boundary conditions at $r=0$ are more involved. To start the integrations we need an expansion of the wave functions close to the singularity ($r=0$), which can be found as follows.
The tortoise coordinate $r_*$, defined by $f = \frac{dr}{dr_*}$, is
\begin{equation}
    r_* = r + \frac{r_-^2}{r_--r_+} \log\bigg( \frac{r_--r}{r_-}\bigg) + \frac{r_+^2}{r_+-r_-} \log\bigg( \frac{r_+-r}{r_+}\bigg)\,.
\end{equation}
An expansion around $r=0$ yields
\begin{equation}
    r_* \approx \frac{r^3}{3 r_- r_+}\,,
\end{equation}
and hence one can easily find the Frobenius series solution~\cite{Dotti:2010}:
\begin{align}
    Z^+_{1,2} &= A_{1,2} r_*^{1/3}\bigg[1 + r_*^{1/3} \sum_{i=1}^{\infty} a^{(1,2)}_i r_*^{i/3} \bigg] \nonumber \\ 
    &+ B_{1,2} r_*^{2/3}\bigg[\sum_{i=0}^{\infty} b^{(1,2)}_i r_*^{i/3} \bigg]\,.
    \label{eq:FrobeniusSeries}
\end{align}
Note that we do not use the $r_*^{2/3}$ term within the Frobenius series for the $r_*^{1/3}$ solution on the first line.
This is as $r_*^{1/3}$ and $r_*^{2/3}$ are linearly independent solution, and so the coefficient of $r_*^{2/3}$ cannot depend upon the coefficient of $r_*^{1/3}$.
The system has two degrees of freedom, so we need two boundary conditions for the solutions to be unique.

Both of the independent solutions for $Z^+_{1,2}$ are regular at the singularity, seemingly leaving no room to impose physically motivated boundary conditions.  Note however that the perturbations must be small, and so their contribution to any physical quantity should also be small.  Following Ref.~\cite{Dotti:2006}, we will consider in particular the Kretschmann invariant $\mathcal{K}=R_{\alpha\beta\gamma\delta}R^{\alpha\beta\gamma\delta}$.  As $r\to0$, the background's contribution to $\mathcal{K}$ diverges as $56Q_{\rm in}^4r^{-8}$, while the Kretschmann invariant's perturbation diverges as $r^{-9}$.

The fact that perturbations to the Kretschmann invariant diverge faster than the background implies that the perturbations do not remain small near the singularity.  Insisting that the perturbations do not diverge faster than the background yields the following boundary conditions:
\begin{align} \label{eq:l2BoundaryCon}
    A_1 &= -\frac{q_{2, \rm in} A_2}{2Q_{\rm in}}, \quad (\ell\neq 1)\,,\\
\label{eq:l1BoundaryCon}
    A_1 &= A_2 =0, \quad (\ell=1)\,.
\end{align}
We could also apply the same argument to $F_{(a)(b)}F^{(a)(b)}$, but it would yield the same conditions. Note that we have the correct number of boundary conditions for $\ell=1$, but we are missing one boundary condition for $\ell\geq2$.

Importantly, the  two degrees of freedom $A_i$ ($i=1,\,2$) must mix in order to have physically meaningful perturbations of the Kretschmann invariant at the singularity.
This means that, in general, these boundary conditions imply that these perturbations cannot be excited independently.

The missing boundary condition for $\ell\geq2$ can be found by using again the divergence properties of the Kretschmann invariant, but this time at the finite value of $r=d_\ell$ were the potential $V_1$ diverges: cf. Eq.~\eqref{eq:v1div}.

The series solution for $Z_1^+$ around $r=d_\ell$ is
\begin{align}
    Z_1^+ &= A (r-d_\ell)^{-1}\big[ 1 + \sum_{i=1} a_i (r-d_\ell)^i \nonumber \\
    &+ (r-d_\ell)^3 \log(r-d_\ell)\sum_{i=1} c_i (r-d_\ell)^i \big] \nonumber\\
    &+  B (r-d_\ell)^2 \big[ 1 + \sum_{i=1} b_i (r-d_\ell)^i \big]\,.
    \label{eq:dlSeriesSol}
\end{align}
As $r\to d_\ell$ the perturbations (and hence the Kretschmann invariant) will diverge unless we impose a second boundary condition, i.e.
\begin{equation}
\label{eq:ABoundaryCondition}
    A=0\,.
\end{equation}
For this boundary condition to be valid, we need the divergence for all $\ell\geq 2$ to be within the shell. 
Moreover, to avoid the divergence for $\ell=1$, $d_1$ must lie outside the shell.
This restricts the space of allowed values of $(M_{\rm in},\,Q_{\rm in})$, as we must have
\begin{equation}
\label{eq:QiBoundsGeneral}
    d_2 < a < d_1\,.
\end{equation}
Thus we have four constraints on the interior metric: the two above, that the shell is locate within $r_{\rm in}^-$, and that $Q_{\rm in}<M_{\rm in}$.
A viable region exists for all $a>1$, but the space of allowed internal parameters becomes highly constrained as the shell position approaches the horizon.

\begin{figure}[t]
  \includegraphics[width=\columnwidth]{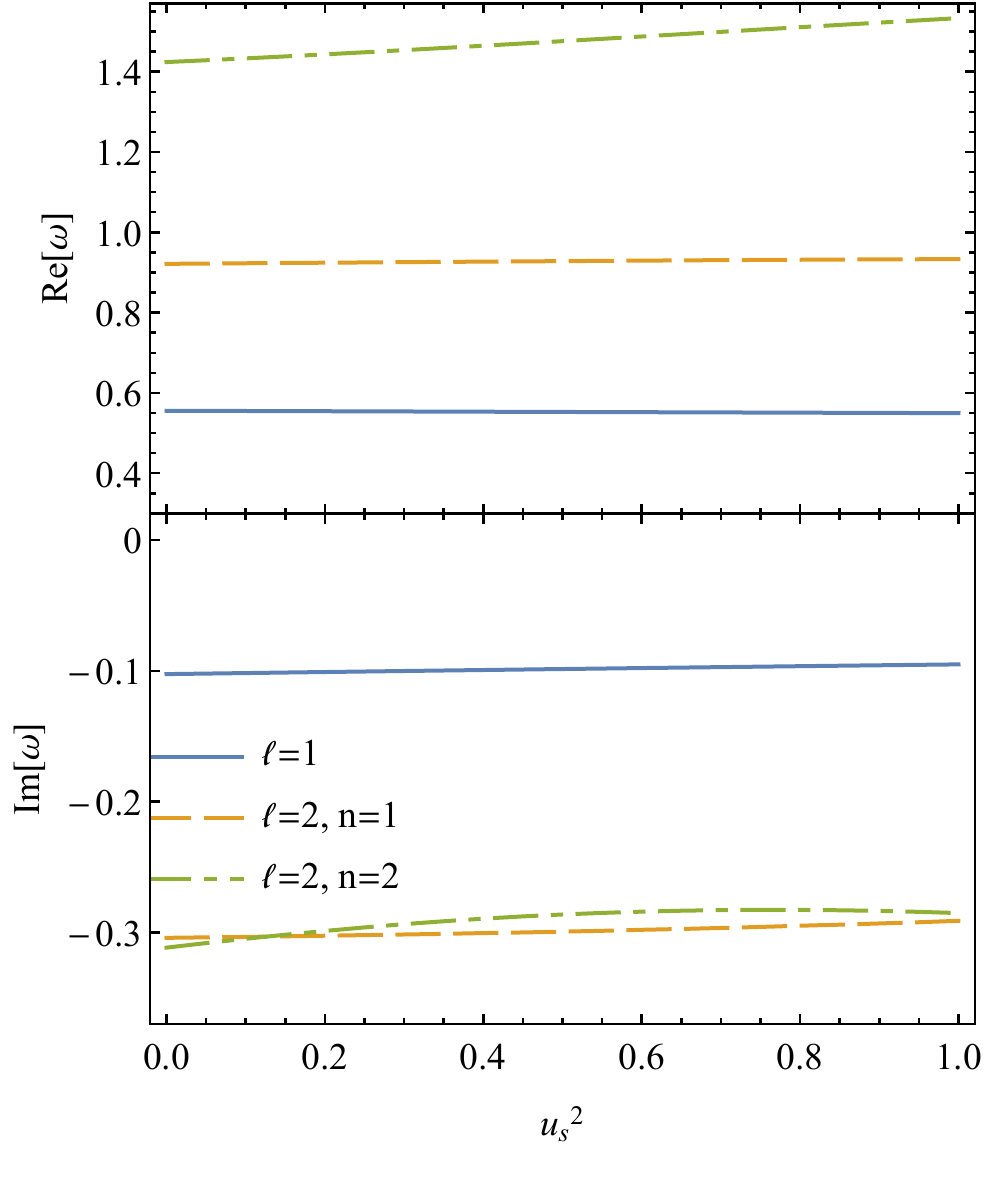}\\
  \caption{An example of the variation of the QNM frequencies with the speed of sound $u_s$ on the shell for a model with parameters $Q_{\rm out}=0$, $M_{\rm in}=Q_{\rm in}=7a/4$.  The blue solid line is the fundamental mode with $\ell=1$. The orange dashed and green dot-dashed lines correspond to the two dominant modes with $\ell=2$, whose imaginary parts cross at a finite speed of sound $u_s=0.12$.}
\label{fig:QNMvsVs2}
\end{figure}

\begin{figure*}
        \includegraphics[width=.65\columnwidth]{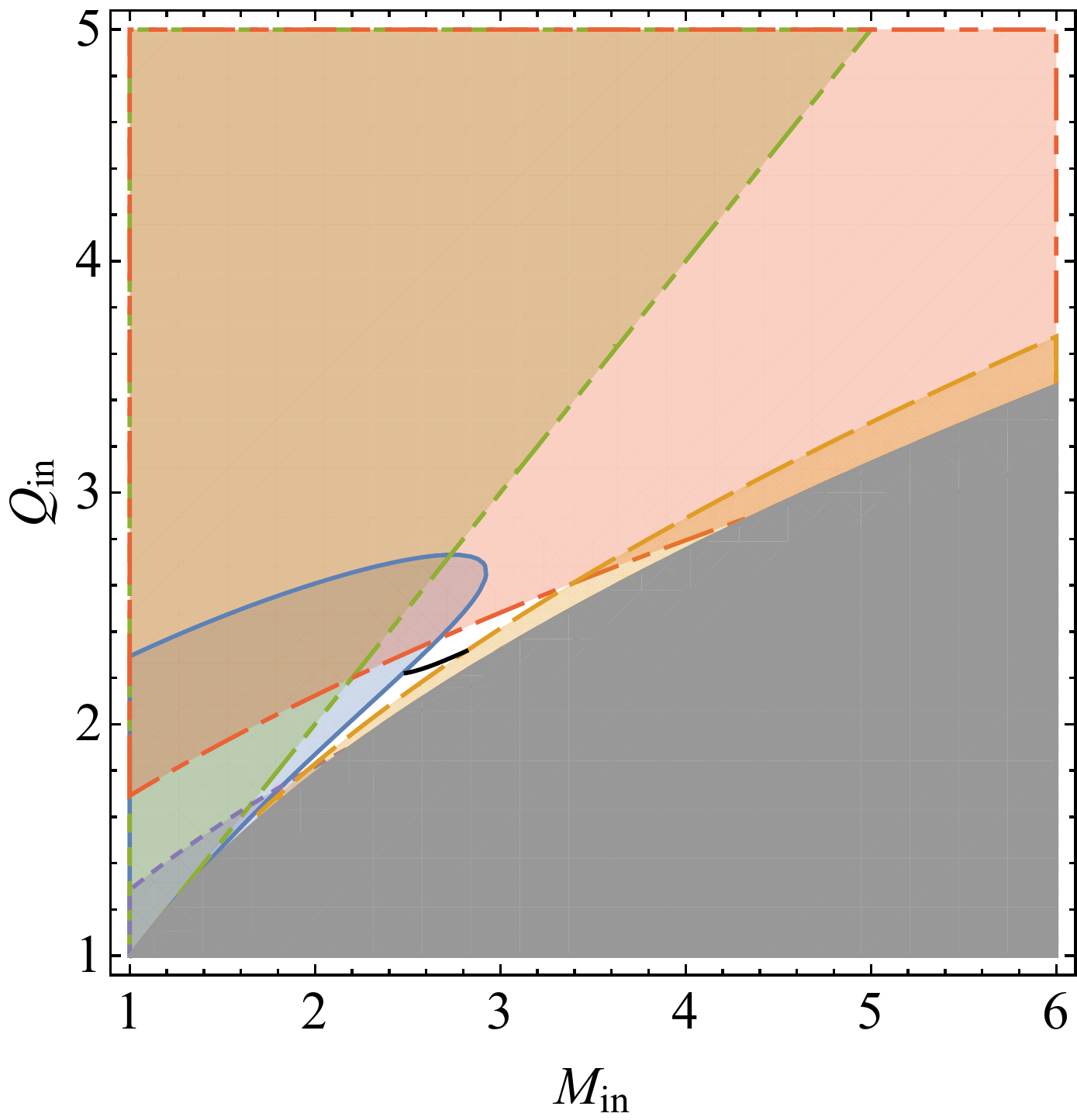}
        \includegraphics[width=.65\columnwidth]{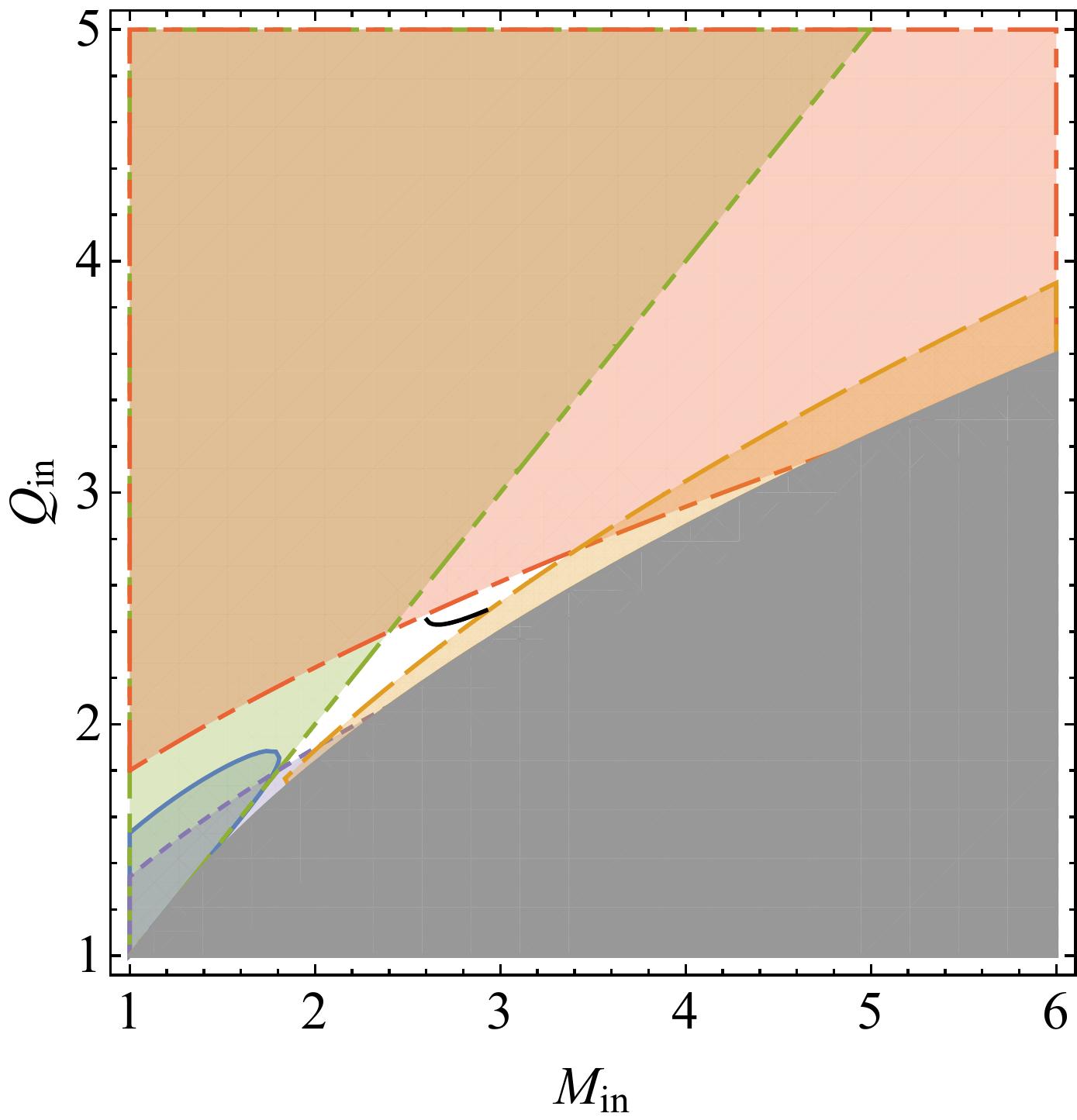}
        \includegraphics[width=.65\columnwidth]{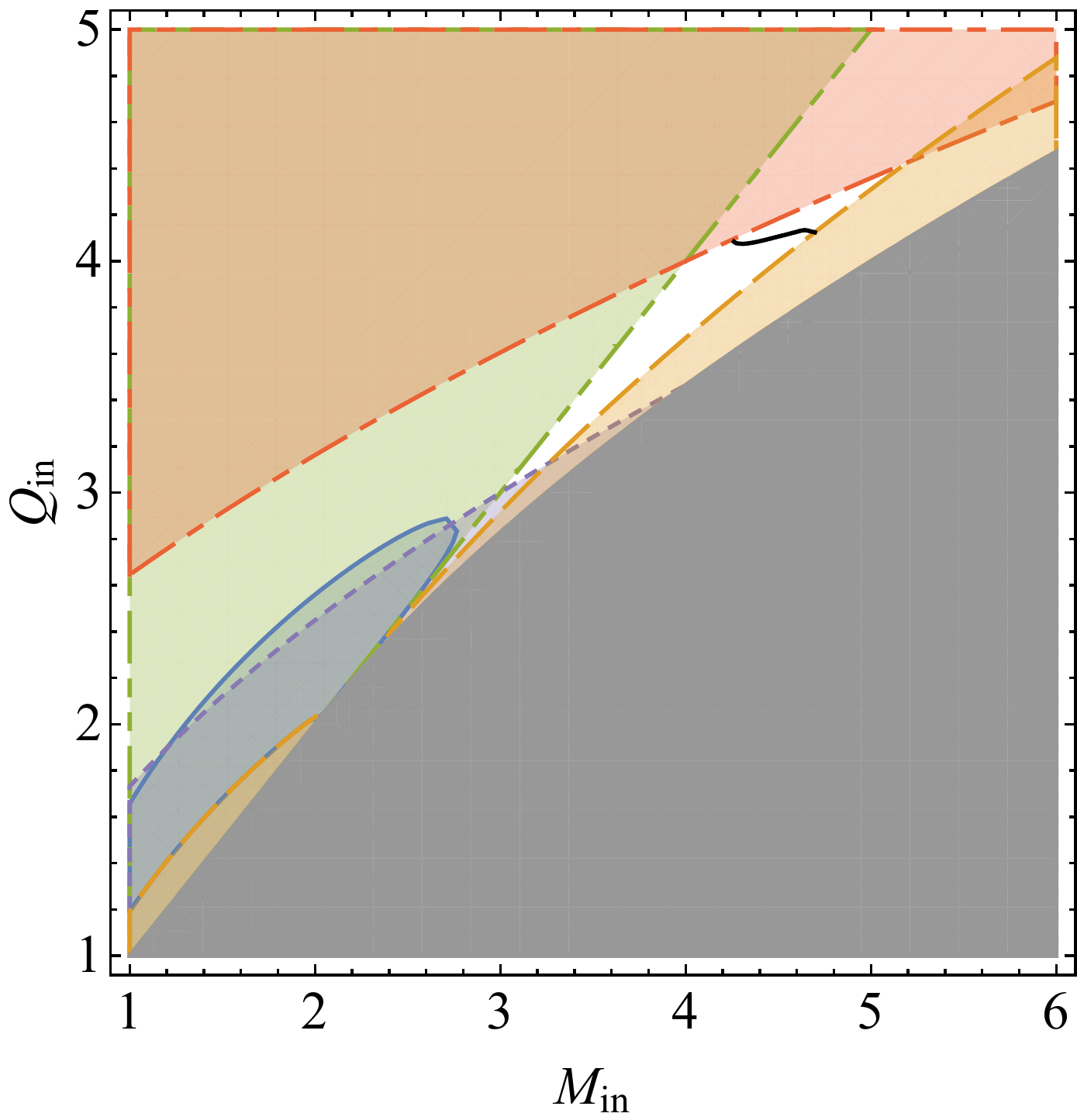}
        \caption{Bounds on the internal metric parameters $(M_{\rm in},Q_{\rm in})$ for a Schwarzschild exterior. The left, center and right plots correspond to $a=1.1$, $a=1.2$ and $a=2$, respectively. The white region represents allowed models. The black solid line within the white region corresponds to ${\rm Im}(\omega)=0$; points to the upper right of the black line correspond to ${\rm Im}(\omega)>0$, and therefore to unstable models. The white region is mostly determined by the requirement that $v_s<1$ (blue solid line, which excludes the bottom-left region in each plane) and $v_s>0$ (dashed orange line, which excludes the bottom-right region in each plane). In the grayed-out region the firewall model is not defined because $r_{\rm in}^-<a$. Points below the dotted line (corresponding to $a=d_{\ell=1}$) and above the dot-dashed line (corresponding to $a=d_{\ell=2}$) do not have well-defined boundary conditions for the perturbation equations. Points above the dot-dot-dashed line (top-left region) are excluded because $Q_{\rm in}>M_{\rm in}$.  The region of allowed parameters is therefore the bottom-left ``slice'' of the white region. It gradually shrinks and approaches a line segment as $a\to1$.}
        \label{fig:AllInternalBounds}
\end{figure*}
\section{Numerical results}
\label{sec:NumericalResults}

The integration from the origin to infinity is performed in two steps. We first integrate from the origin to the shell, where we apply the junction conditions; then we integrate from the shell to infinity.

The case $\ell=1$ is special: as discussed in Appendix~\ref{app:AlgebraicCompletion} there is only one physical degree of freedom, so we can work with the perturbation function $H_1^+$ defined in Eq.~\eqref{eq:H1p}.
We compute the QNM frequencies using both a shooting method and an adapted version of Leaver's method, as described in Refs.~\cite{Leaver:1985ax,Leins:1993zz,Benhar:1998au}.

Because of the junction conditions and of the more complex boundary conditions, gravitational and electromagnetic perturbations are coupled when $\ell \geq 2$.
In this case we find the spectra of coupled ordinary differential equations following the methods described in Ref.~\cite{Pani:2013pma}, and again we cross-check results by comparing a shooting method against an adapted version of Leaver's method.

The additional boundary condition in Eq.~\eqref{eq:ABoundaryCondition} causes a further complication when we integrate $Z^+_{1,2}$ from the origin to infinity, because of the divergence in the potential $V_1$ at $r=d_\ell$ (where $\ell>1$), as defined in Eq.~\eqref{eq:v1div}. Our integration procedure can be summarized as follows:
\begin{itemize}
    \item We use the Frobenius series \eqref{eq:FrobeniusSeries} to approximate both wave functions $Z^+_{1,2}$ from $r=0$ to $r=\epsilon$.
    \item We use the shooting method to find the value of $A_1$ which satisfies the boundary condition $A=0$ at $r=d_\ell$.
    \item We integrate the wave function $Z_1^+$ to $r=d_\ell-\delta$, and use the series solution \eqref{eq:dlSeriesSol} to extend the solution across the singularity to $r=d_\ell+\delta$.
    \item We integrate both wave functions $Z^+_{1,2}$ out to the location of the shell (note that there are no singularities in the integration of $Z^+_{2}$), where we apply the junction conditions.
    \item We integrate from the shell out to infinity, imposing outgoing-wave boundary conditions at infinity.
    \item We use the methods described in Ref.~\cite{Pani:2013pma} to find the QNM frequencies $\omega$.
\end{itemize}
After some numerical experimentation we decided to set $\epsilon=0.005$ and $\delta=0.01$ for numerical stability. Therefore the stability of our numerical scheme imposes a (nonphysical) lower bound on the shell position: $a\geq1.005$ for $\ell=1$, and $a\geq1.025$ for $\ell>1$.

Below we discuss the dependence of the QNM frequencies on the model parameters.

\subsection{Speed of sound}
\label{sec:SpeedOfSounds}

In Section.~\ref{sec:RadialStability} we derived the speed of sound $v_s$ for radial oscillations of the shell, Eq.~(\ref{eq:speedOfSound}), which is different from the speed of sound $u_s$ for polar perturbations of the shell, as defined in Eq.~\eqref{eq:speedOfPErt}. Here we examine the dependence of the QNM spectra upon $u_s$. 

In Fig.~\ref{fig:QNMvsVs2} we show how the fundamental $\ell=1$ mode and the two dominant $\ell=2$ modes vary with $u_s^2$.  For concreteness we consider a shell located at $a=1.2$, and we consider internal parameters $M_{\rm in}=Q_{\rm in}=7a/4$ which lie in region of allowed values (cf. Fig.~\ref{fig:AllInternalBounds}) and are halfway between the bounds \eqref{eq:QiBoundsGeneral}.

This example shows that the QNM frequencies typically have a weak dependence on the speed of sound. However, the imaginary parts of the two dominant modes with $\ell=2$ switch order at $u_s=0.12$ for $a=1.2$.  This weak dependence on the speed of sound makes this firewall model different from gravastars, where the structure of the spectra is more strongly dependent on the speed of sound~\cite{Pani:2009}.  Since the spectra show little dependence on $u_s$, and to explore the most extreme cases allowed by this model, from now on we will set $u_s=1$.

\begin{figure*}
        \includegraphics[width=.9\columnwidth]{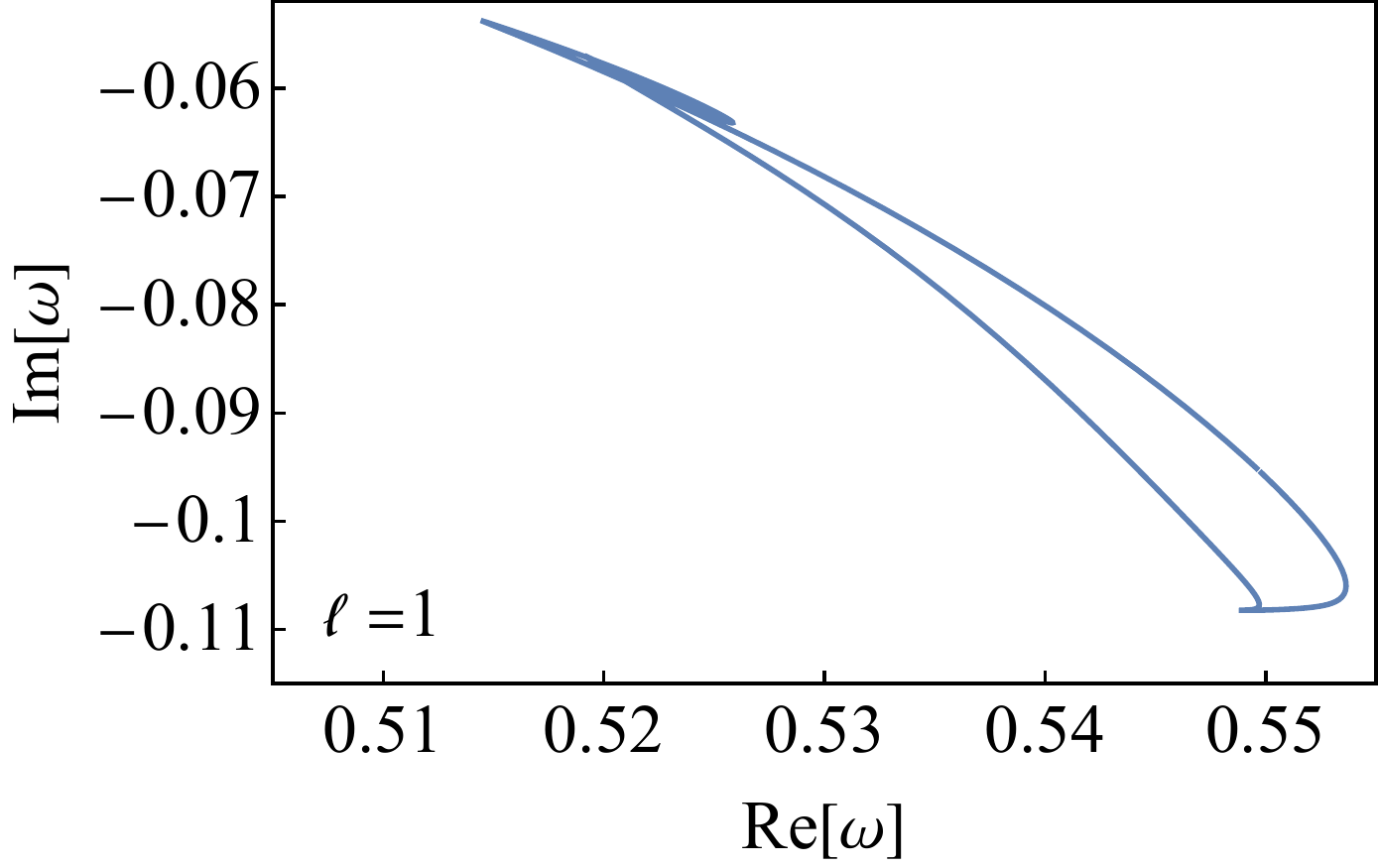}
        \includegraphics[width=.9\columnwidth]{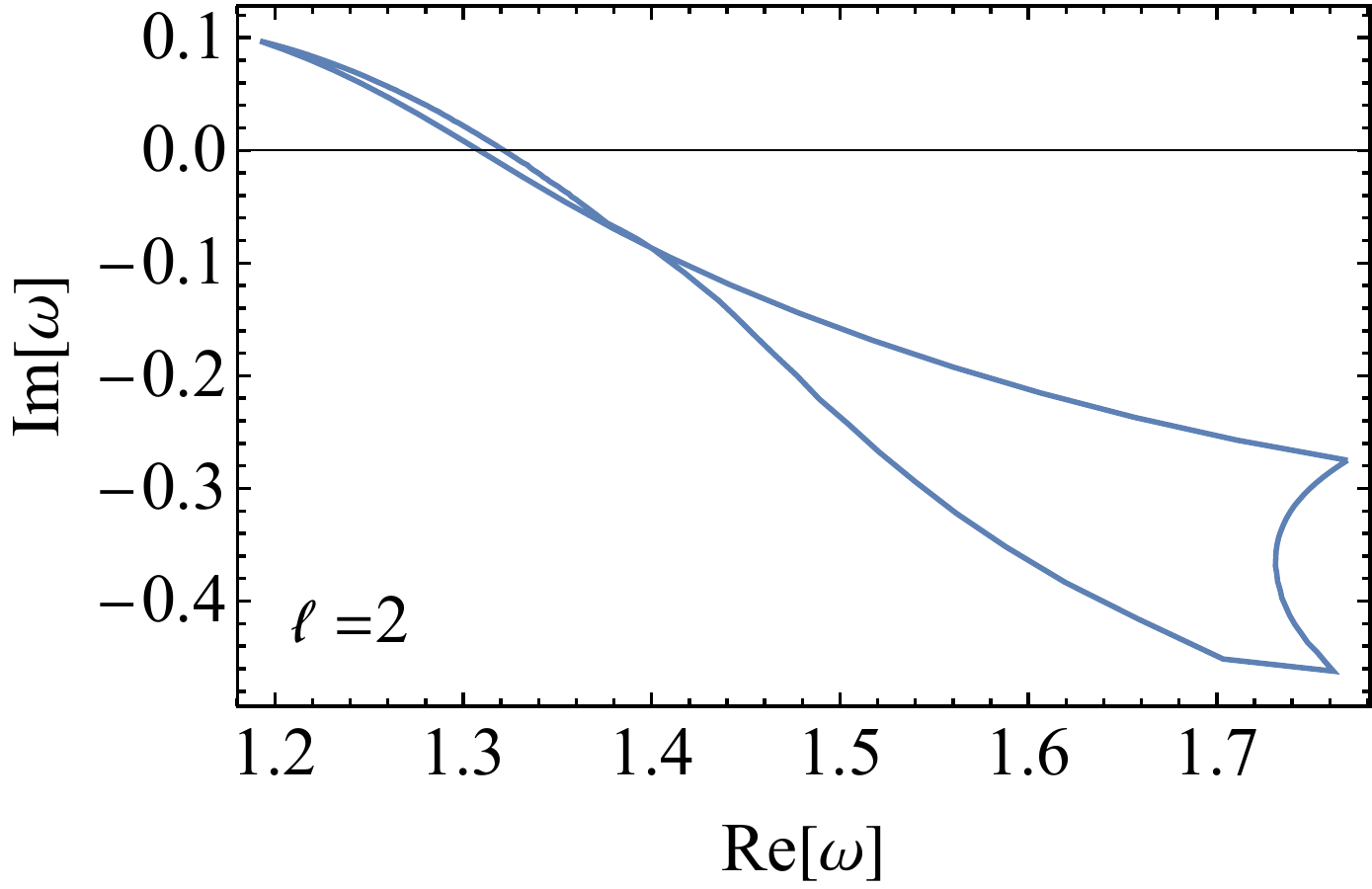}
        \caption{Change in the complex QNM frequencies (in units $r^+_{\rm out}=1$) as we track $(M_{\rm in},\, Q_{\rm in})$ around the white allowed region shown in Fig.~\ref{fig:AllInternalBounds} for a ``Schwarzschild-like'' model with $Q_{\rm out}=0$, $u_s=1$ and $a=1.2$. The left (right) plot refers to the fundamental mode with $\ell=1$ ($\ell=2$). The $\ell=2$ modes become unstable (the imaginary part crosses zero) for two specific values of the internal parameters, corresponding to the edges of the black solid line in the middle panel of Fig.~\ref{fig:AllInternalBounds}.}
        \label{fig:MiQiQNM}
\end{figure*}

\begin{figure*}
        \includegraphics[width=.65\columnwidth]{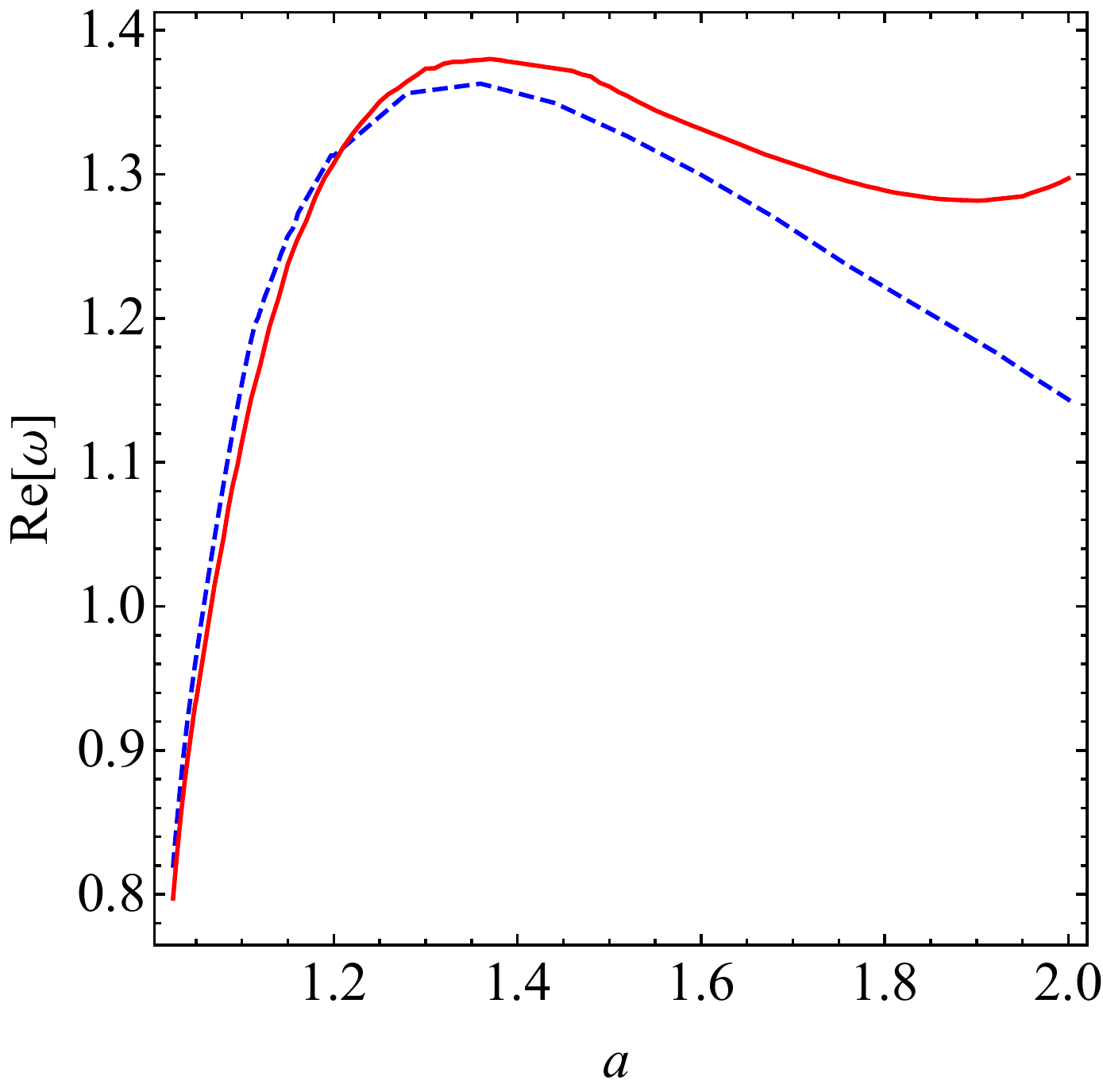}
        \includegraphics[width=.65\columnwidth]{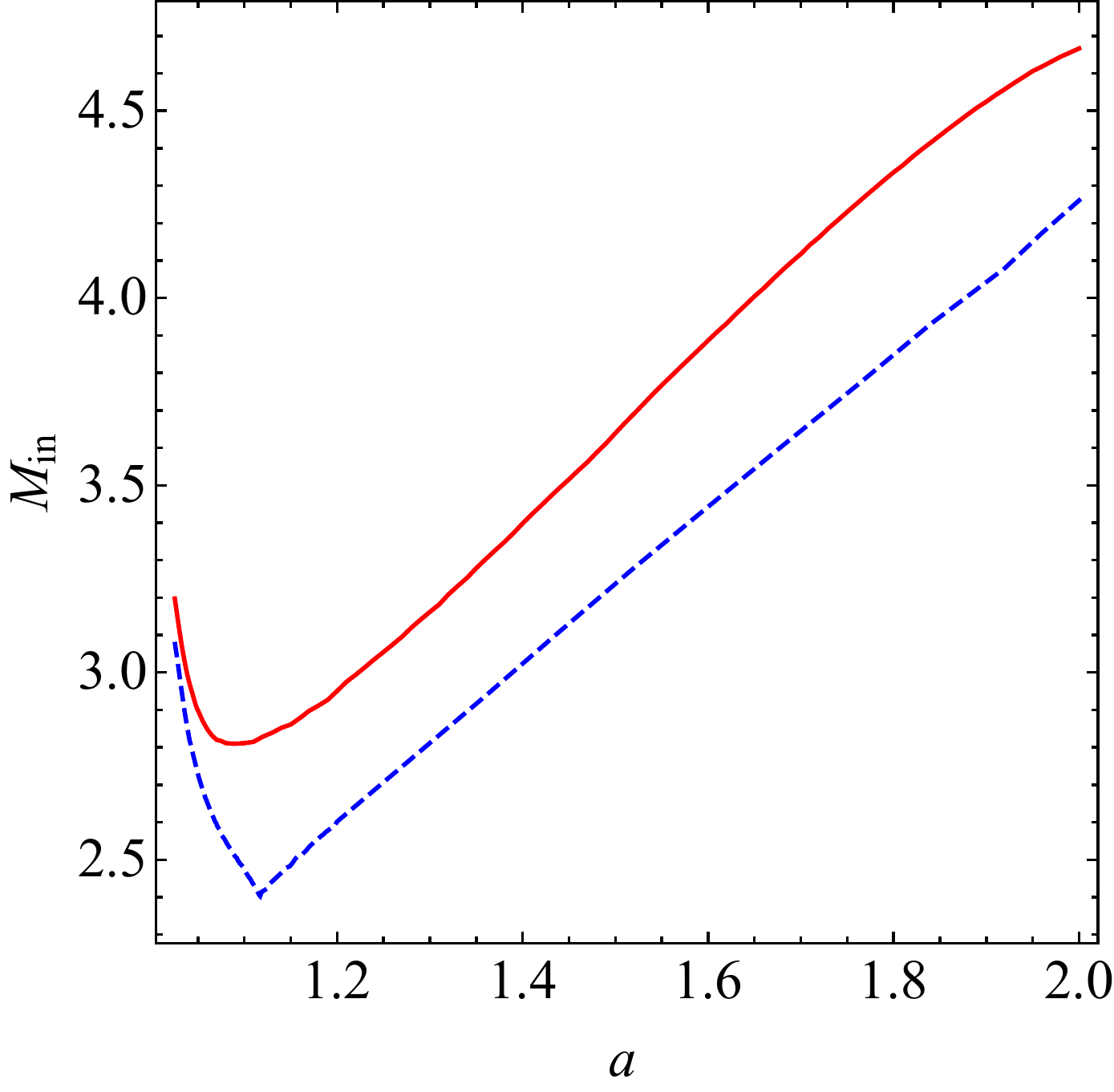}
        \includegraphics[width=.65\columnwidth]{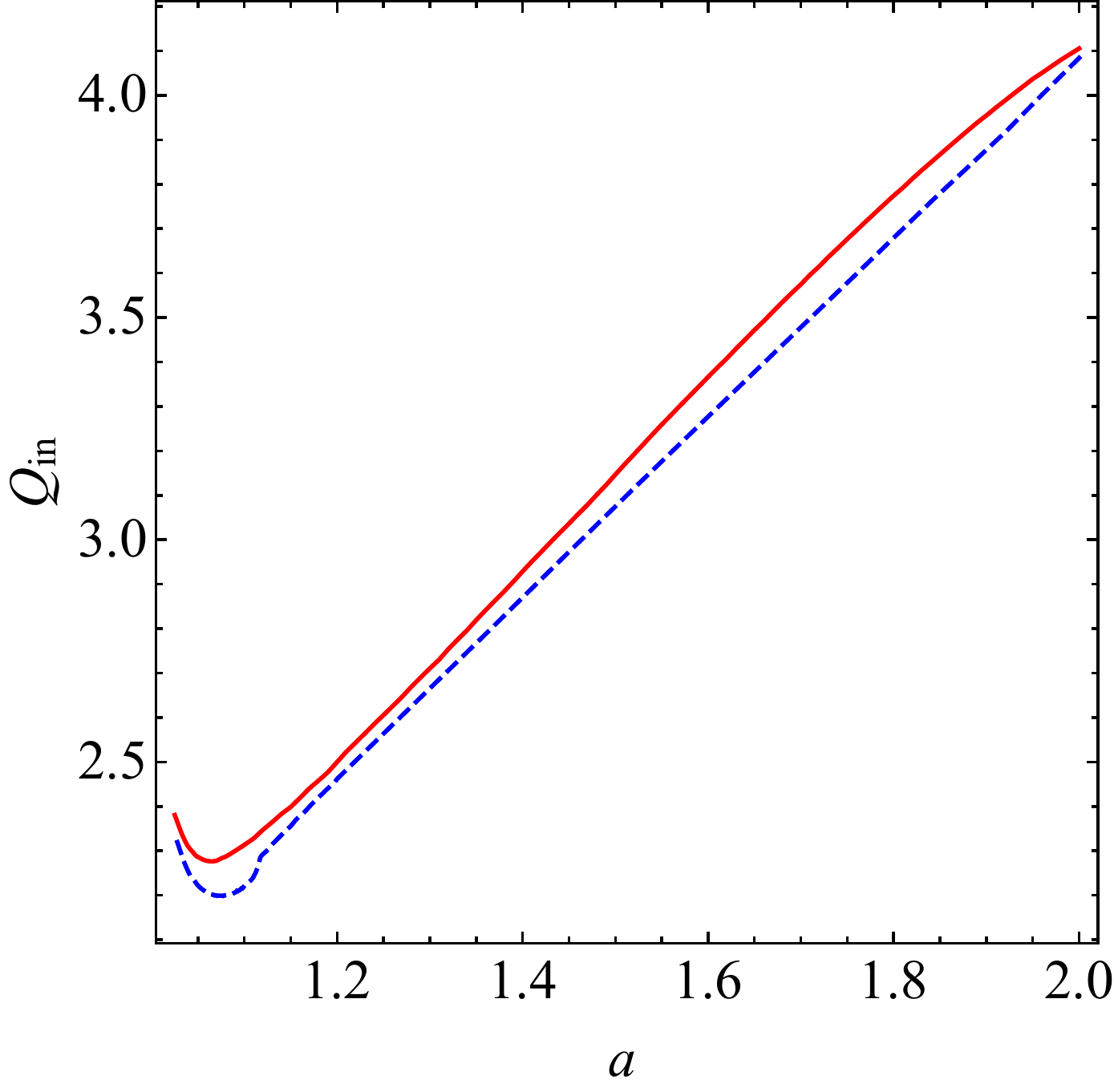}
        \caption{Crossover frequency (i.e., the value of ${\rm Re}(\omega)$ at which ${\rm Im}(\omega)=0$) and corresponding internal parameters ($M_{\rm in}$ and $Q_{\rm in}$) as a function of the shell position $a$. The solid red line corresponds to the rightmost allowed point in the $(M_{\rm in},\,Q_{\rm in})$ plane; the dashed blue line corresponds to the leftmost point (cf. Fig.~\ref{fig:AllInternalBounds}).}
        \label{fig:pStabilityTransition}
\end{figure*}

\subsection{Interior parameters}
\label{sec:ExtremalInterior}

The firewall model has two seemingly arbitrary parameters, the internal charge $M_{\rm in}$ and mass $Q_{\rm in}$.  As there is no {\em a priori} relation between the values for the interior mass and charge and the other model parameters, each choice for these internal parameters corresponds to a different realization of the firewall model, and therefore to a different QNM spectrum. In principle the internal parameters $(M_{\rm in},\,Q_{\rm in})$ could depend on the specifics of the firewall's formation channel and take a wide variety of values. If so, the QNM spectra of these objects would be nearly unpredictable. However there are a number of constraints on the internal parameters: the inner horizon of the internal metric must be located beyond the shell [Eq.~\eqref{eq:horizonOrder}]; the radial speed of sound must satisfy the conditions $0\leq v_s\leq 1$; and further constraints come from the boundary conditions for the perturbation equations [Eq.~\eqref{eq:QiBoundsGeneral}].  These bounds, and therefore the shape and extension of the allowed region, depend upon the shell position $a$, as shown in Fig.~\ref{fig:AllInternalBounds}. Quite remarkably, the allowed region gradually shrinks (approaching a line segment) as $a\to 1$.

The space of valid internal parameters is compact, and so we may examine the change of the spectra as we traverse a loop in the $(M_{\rm in},Q_{\rm in})$ parameter space, keeping fixed the other model parameters.
This represents the extreme values that different realizations of a firewall (which looks otherwise identical to an external observer) may take.
In Fig.~\ref{fig:MiQiQNM} we set $Q_{\rm out}=0$, $u_s=1$ and $a=1.2$, corresponding to the middle panel of Fig.~\ref{fig:AllInternalBounds}, and we show how the fundamental ($n=1$) complex QNM frequencies for $\ell=1$ and $\ell=2$ change as we ``loop around'' the allowed internal parameters in the $(M_{\rm in},\,Q_{\rm in})$ plane.
Importantly, the spectrum for these ``edge models'' is also compact: an arbitrary point in the allowed ``white region'' for $(M_{\rm in},Q_{\rm in})$ would have its QNM frequency within the loops shown in Fig.~\ref{fig:MiQiQNM}.  For the range of values of $a$ that we explored, the $\ell=1$ mode remains stable for all allowed values of the internal parameters, while the $\ell=2$ modes can be unstable. This somewhat counterintuitive result is most likely due to the different boundary conditions for $\ell=1$ and $\ell>1$ perturbations.  The potential instability of the fundamental $\ell=2$ mode further restricts the allowed region of the internal parameters for the firewall model: the marginal stability line with ${\rm Im}(\omega)=0$ is shown as a solid black line in Fig.~\ref{fig:AllInternalBounds} for the three selected values of the shell position $a$, and only the bottom-left ``slice'' of each white region corresponds to nonradially stable solutions.

\begin{figure}
        \includegraphics[width=\columnwidth]{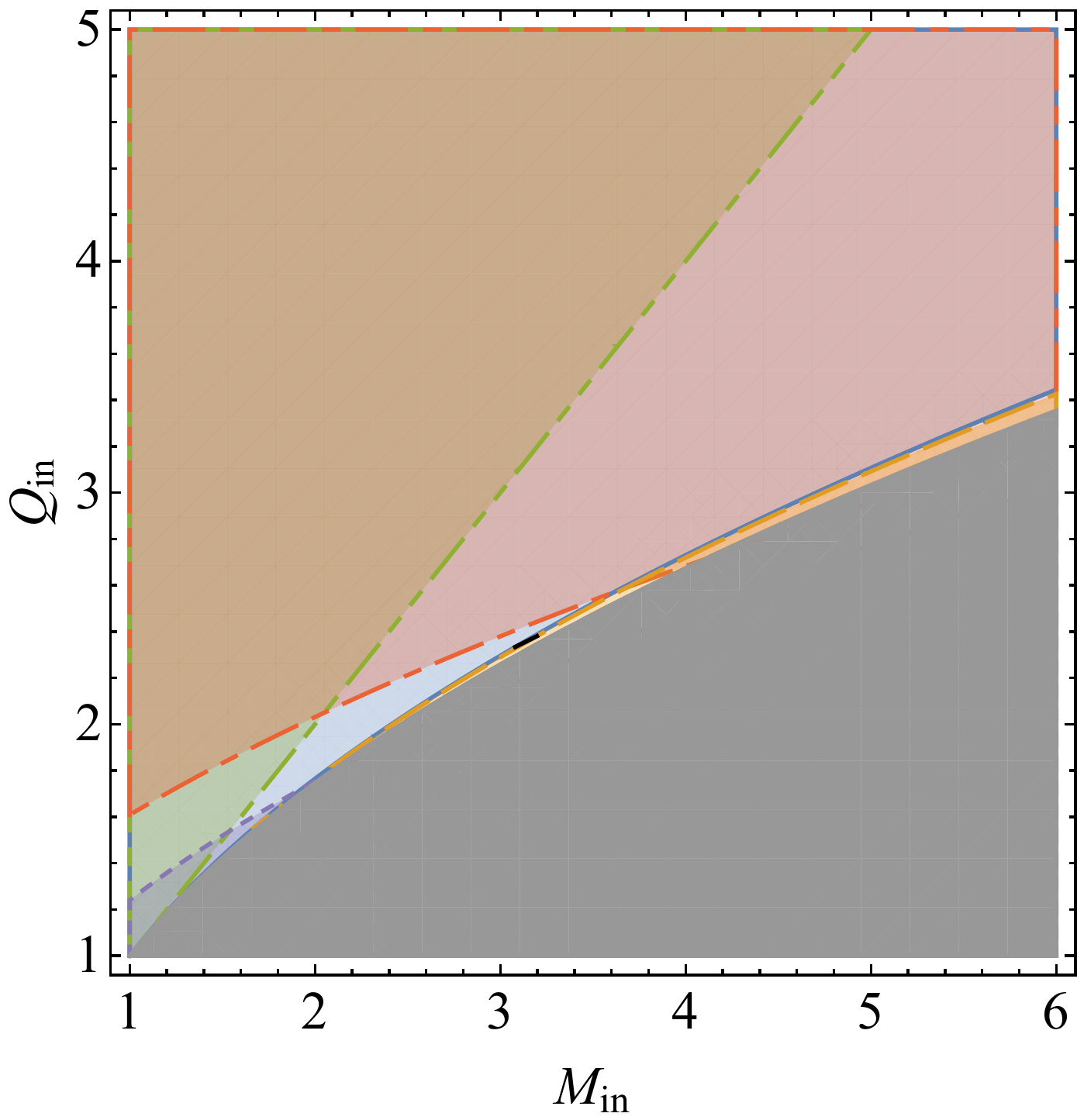}
        \caption{Same as Fig.~\ref{fig:AllInternalBounds}, but for $a=1.025$, the smallest value for the shell location that we could consider within our numerical scheme. The white region is barely visible and it spans the range $M_{\rm in}\simeq [3.08,\,3.20]$, $Q_{\rm in}\simeq [2.33,\,2.38]$.} 
        \label{fig:SmallestpRegionPlot}
\end{figure}

\begin{figure}[t]
  \includegraphics[width=\columnwidth]{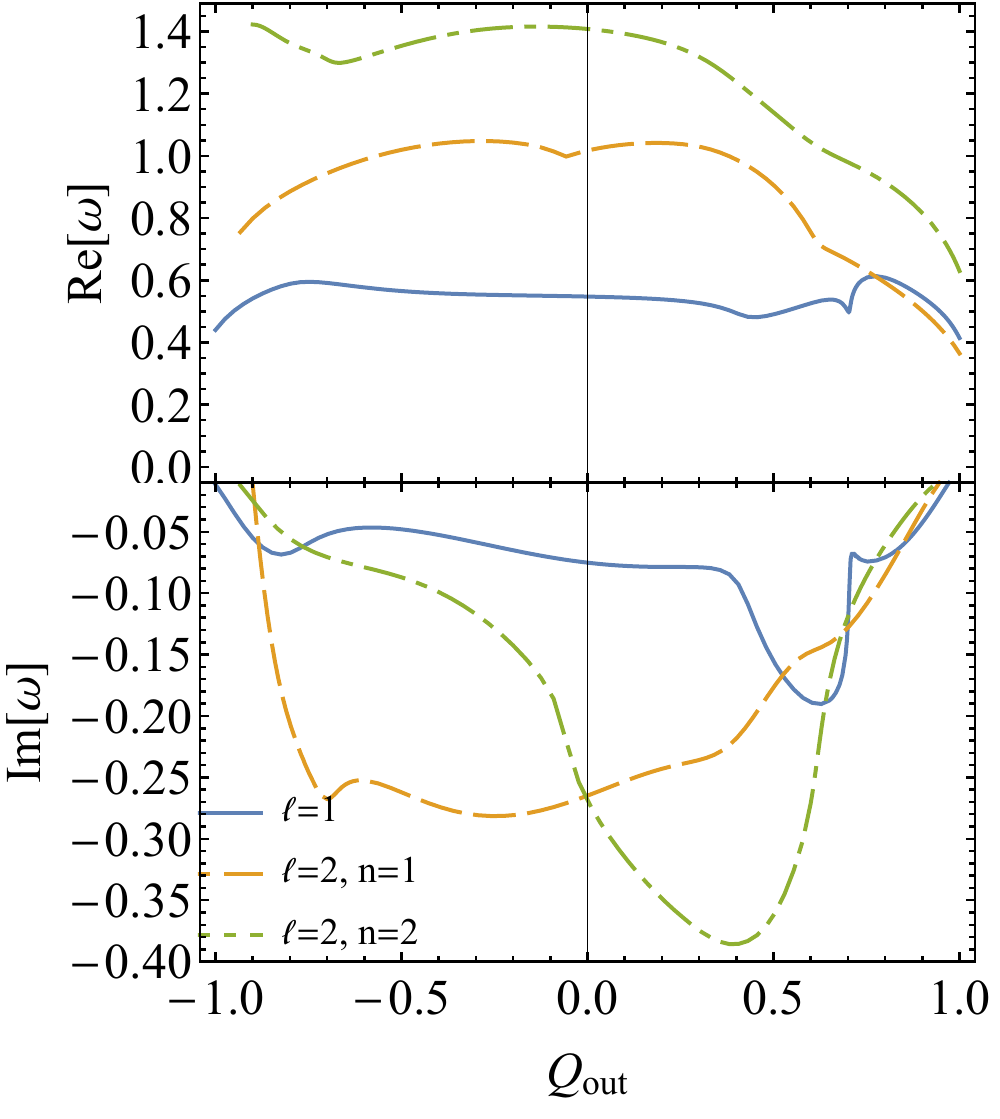}\\
  \caption{Real and imaginary parts of the QNM frequencies as a function of the external charge $Q_{\rm out}$ for $a=1.15$, $M_{\rm in}=Q_{\rm in}=7a/4$. Different lines refer to the fundamental mode with $\ell=1$ and to the two lowest lying modes with $\ell=2$.}
\label{fig:QNMvsQo}
\end{figure}

\subsection{Allowed interior parameters for near-Planckian shell locations}
\label{sec:ShellPosition}

In the firewall model of Ref.~\cite{Kaplan:2018dqx}, the shell is placed a Planck length away from the horizon. This is an important feature of the model: one can justify the firewall as being due to (as yet unknown) high-energy physics in the limit where the surface mass density also becomes Planckian.  Recall that we work in units such that $r_{\rm out}^+=1$, and so the physical position of the shell has been scaled out.  For an astrophysically relevant exterior mass of (say) $10 M_\odot$ and in our units, a shell located a Planck length away from the horizon corresponds to $a=1+10^{-40}$.  However, as discussed in the introduction to this section, our numerical procedure limits us to a value of $a=1.025$, which would be equivalent to the shell being placed at a physical distance of approximately 600\,m from the event horizon -- a far cry from the physical value required in the original model, but still indicative of what may happen at small separations between the horizon and the shell.

To illustrate the dependence of the instability crossing on the shell position, in Fig.~\ref{fig:pStabilityTransition} we show the crossover frequency (i.e., the value of ${\rm Re}(\omega)$ at which ${\rm Im}(\omega)=0$) and the corresponding internal parameters ($M_{\rm in}$ and $Q_{\rm in}$) as a function of the shell position in the range $a\in(1.025,2)$ where we can trust our numerical procedure. As seen in Fig.~\ref{fig:AllInternalBounds}, the instability crossing is approximately a straight line in the $M_{\rm in}(a)$  and $Q_{\rm in}(a)$ plane in most of the range, with a turnover below $a\simeq 1.2$.
This turnover occurs between $a=1.1$ and $a=1.2$, when one of the extrema of the marginal instability black line in Fig.~\ref{fig:AllInternalBounds} crosses the ``apex'' of the triangular allowed white region.

In Fig.~\ref{fig:SmallestpRegionPlot} we show a ``stability diagram'' similar to Fig.~\ref{fig:AllInternalBounds} for the smallest shell location we could consider ($a=1.025$). This places stringent bounds on firewall models that are stable under nonradial perturbations. Even at this relatively large (non Planckian) shell location, allowed models are restricted to a very limited range of internal parameters: $M_{\rm in}\simeq [3.08,\,3.20]$, $Q_{\rm in}\simeq [2.33,\,2.38]$. By extrapolating our numerics to $a \rightarrow 1$, we estimate that the allowed parameter space shrinks to the line segment between 
$M_{\rm in}\simeq [2,\,4]$, $Q_{\rm in}\simeq [1.73,\,2.65]$ in the $\left(M_{\rm in},\,Q_{\rm in}\right)$ plane. 

\begin{figure}[t]
  \includegraphics[width=\columnwidth]{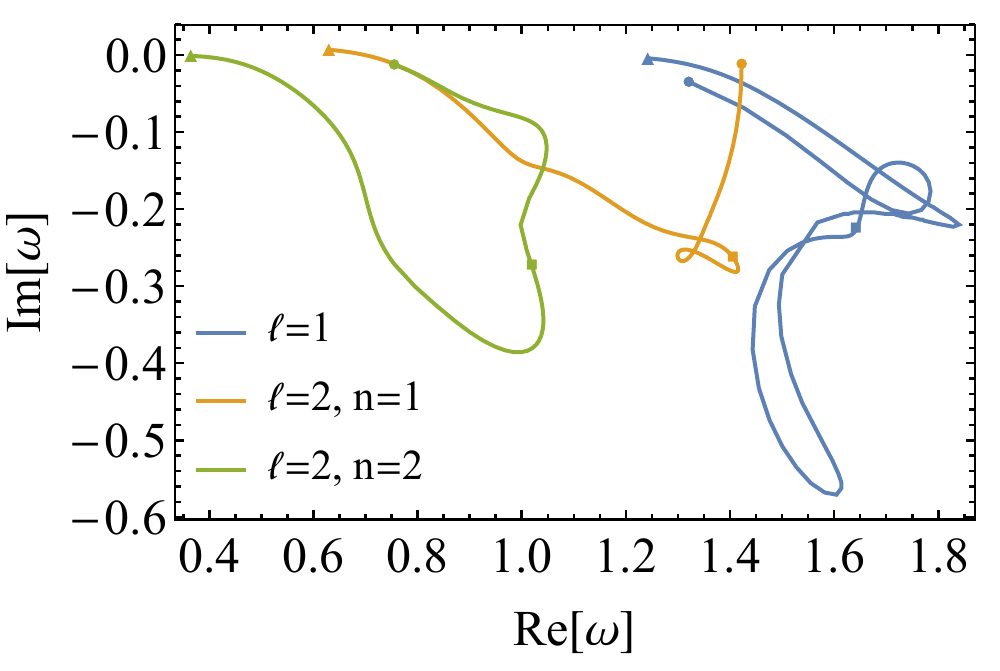}\\
  \caption{Tracks in the complex plane of the same QNM frequencies shown in Fig.~\ref{fig:QNMvsQo}.  For clarity, we have multiplied the $\ell=1$ QNM frequencies by 3.
  Filled circles (squares) denote the QNM frequencies at which the system becomes unstable for negative (positive) $Q_{\rm out}$.}
\label{fig:QoQNMwrvswi}
\end{figure}

\begin{figure*}[t]
        \includegraphics[width=.665\columnwidth]{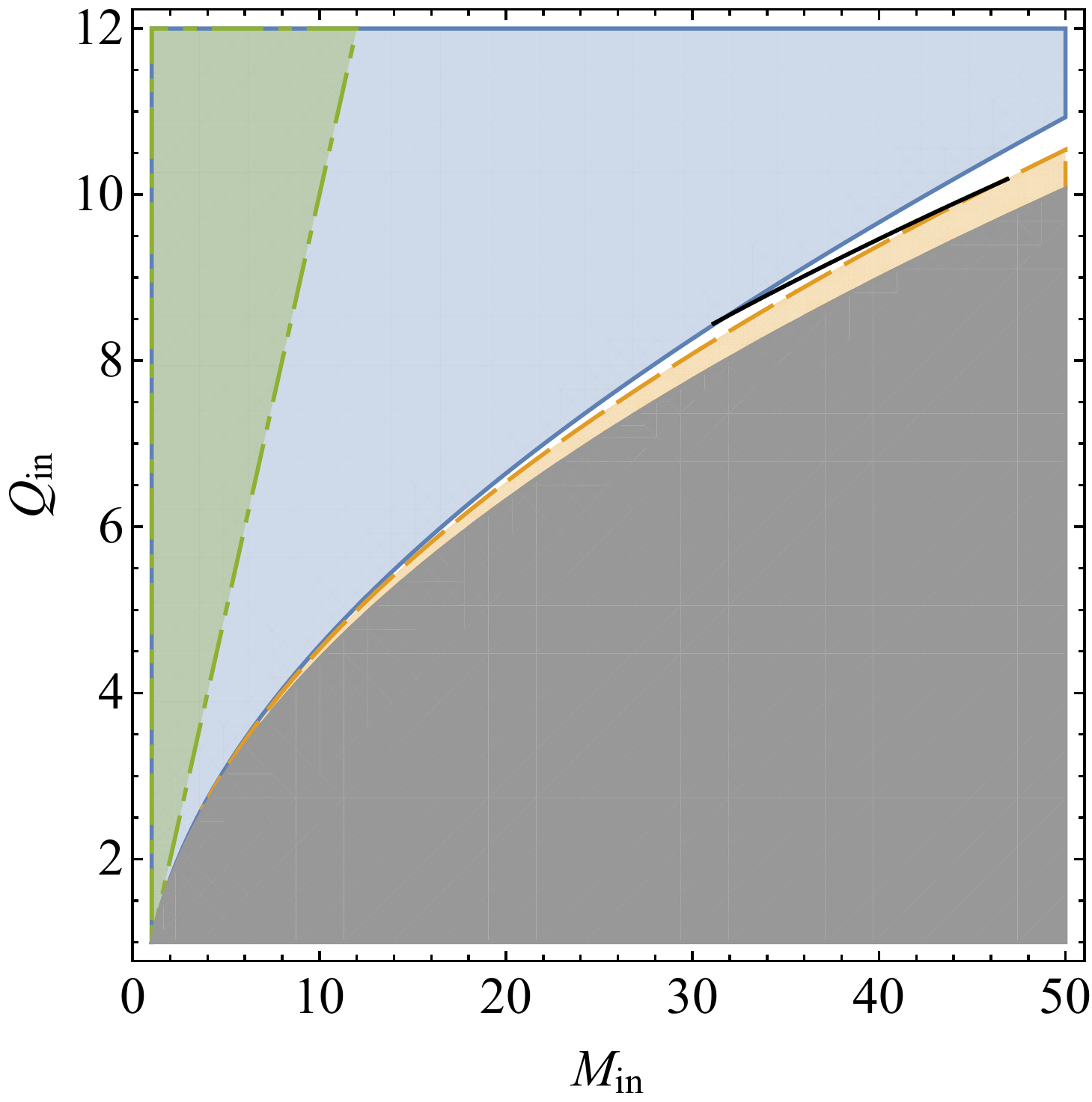}
        \includegraphics[width=.65\columnwidth]{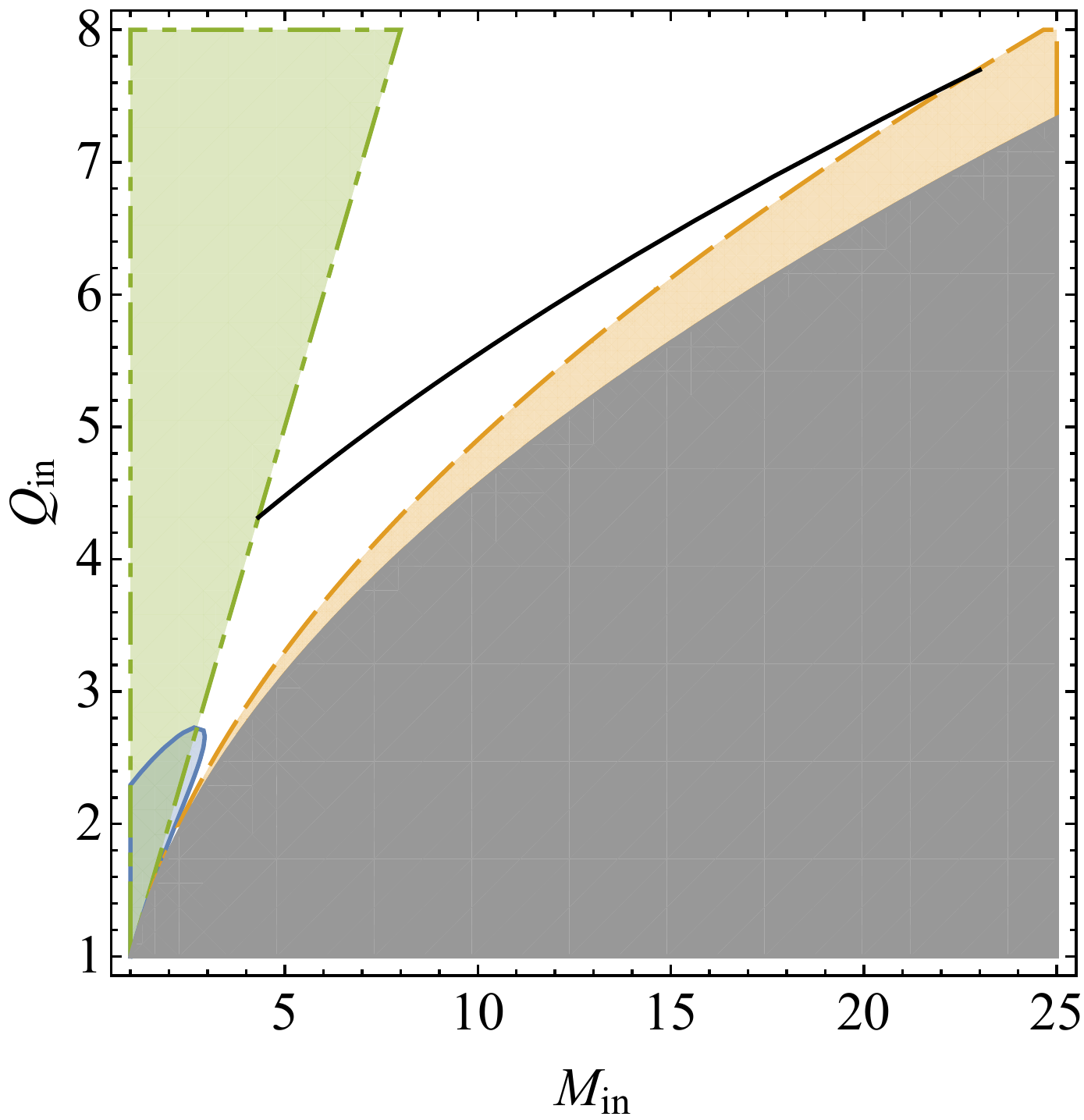}
        \includegraphics[width=.65\columnwidth]{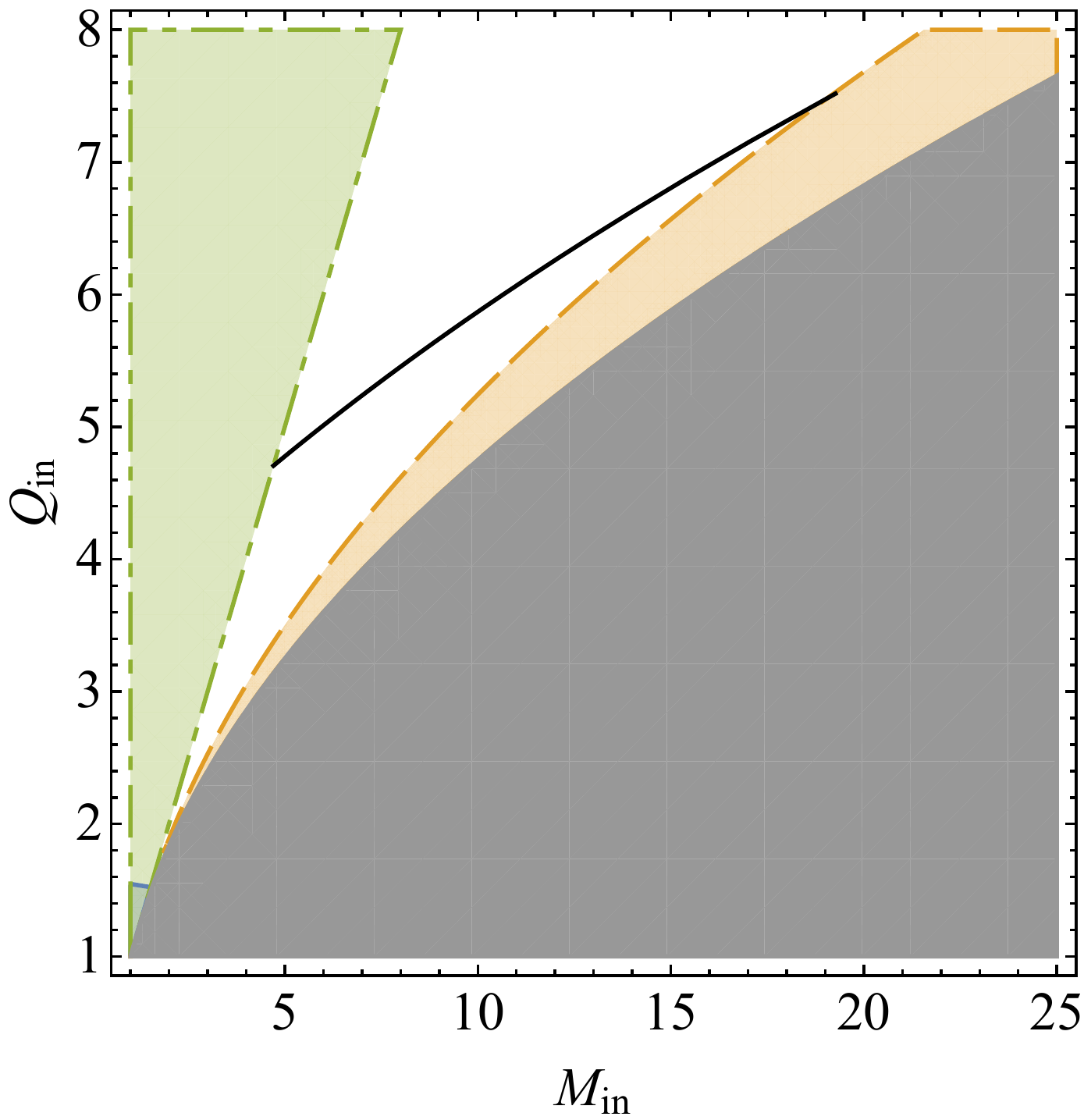}
        \caption{Bounds on the internal metric parameters $(M_{\rm in},Q_{\rm in})$ for a Schwarzschild exterior. Different panels correspond to $a=1.025$ (left), $a=1.1$ (center) and $a=1.2$ (right). The speed of sound imposes two bounds: $v_s<1$ (blue solid line) and $v_s>0$ (dashed orange line). In the grayed-out region the firewall model is not defined because $r_{\rm in}^-<a$. The black line corresponds to marginally instability of the $\ell=1$ perturbations, i.e. ${\rm Im}(\omega)=0$, when we use with the alternative boundary conditions of Sec.~\ref{sec:DifferentBoundaryConditions}. Note the different scales of the axes relative to Fig.~\ref{fig:AllInternalBounds}.}
        \label{fig:AltAllInternalBounds}
\end{figure*}

\subsection{Spacetimes with nonzero external charge}
\label{sec:exteriorCharge}

While the most astrophysically relevant case is that of a neutrally charged exterior spacetime, it is interesting (at least academically) to investigate the stability of charged exterior metrics. For concreteness, in this analysis we set $a=1.15$, $u_s=1$, $M_{\rm in}=Q_{\rm in}=7a/4$.
While these results are not generic, the dimensionality of the parameter space of solutions is too large, and this exploratory study is sufficient for a qualitative understanding of the stability properties of charged spacetimes. 

In Fig.~\ref{fig:QNMvsQo} we plot the real and imaginary part of the QNM frequencies as functions of the external charge $Q_{\rm out}$.
We focus on the dominant mode with $\ell=1$ and the two dominant modes with $\ell=2$.
Note their equal imaginary part when $Q_{\rm out}=0$.
We find an interesting structure with kinks in the spectra.
The imaginary part of the frequency crosses zero when the charge is extremal ($|Q_{\rm out}|\approx1$), indicating that extremal systems would be marginally unstable even in the absence of an event horizon.

To further illuminate the structure of the spectrum, in Fig.~\ref{fig:QoQNMwrvswi} we show the QNM tracks in the complex plane.
The $\ell=1$ QNM frequency tracks show multiple loops: similar features were also found in the case of gravastars when the speed of sound on the gravastar ``shell'' varies~\cite{Pani:2009}, and also for ``ordinary'' Reissner-Nordstr\"om BHs: see e.g. Fig.~3 in Ref.~\cite{Berti:2003zu}.

\subsection{Alternative boundary conditions}
\label{sec:DifferentBoundaryConditions}

The bound upon the internal parameters which restricts the parameter space from reaching large values of $M_{\rm in}$ and $Q_{\rm in}$ (dot-dashed line in Figs.~\ref{fig:AllInternalBounds} and \ref{fig:SmallestpRegionPlot}) is primarily due to Eq.~\eqref{eq:QiBoundsGeneral}, which follows from the boundary condition~\eqref{eq:l2BoundaryCon}. 
This condition is set for modes with $\ell>1$, as otherwise the system is not well defined.
We also placed requirements upon the perturbations of the Kretschmann invariant at the origin, but one may still raise valid concerns. 
Specifically, at the origin the curvature diverges, and so presumably GR will break down at some finite radius. If this is the case, the perturbation equations and their series solution at the origin are no longer meaningful, and presumably the boundary conditions which place tight constraints upon $M_{\rm in}$ and $Q_{\rm in}$ would no longer apply. 

Effective field theory arguments suggest that higher-order curvature terms would become relevant in the field equations as we approach the singularity and the curvature grows.
A conservative requirement would be that the perturbations of all curvature scalars do not diverge quicker than the background. 
As an ansatz for boundary conditions that would satisfy such a constraint, we assume that $A_1=A_2=0$ for all $\ell$.
We must also assume that any modification to the perturbation equations remains negligible, so that the problem is tractable and the RN perturbation equations~\eqref{eq:RNPerturbationEquation} are still valid, at least in some perturbative sense. This alternative formulation of the problem is equivalent to the $\ell=1$ system derived earlier, with the main difference that the $(M_{\rm in},\,Q_{\rm in})$ plane is no longer bounded by Eq.~\eqref{eq:l2BoundaryCon}, and so we are free to examine larger values of these internal parameters.

To study the stability of the system under these alternative boundary conditions we will focus on perturbations with $\ell=1$. These perturbations can be expected to yield the strongest instabilities, because now all modes (including those with $\ell>1$) have the same boundary conditions. 
In Fig.~\ref{fig:AltAllInternalBounds} we show again the bounds  in the $(M_{\rm in},\,Q_{\rm in})$ plane that correspond to the radial speed of sound being subluminal, together with the marginal instability line identified by the condition ${\rm Im}(\omega)=0$ under these new boundary conditions. We could not check whether there is an unstable mode for all values of the internal parameters above this line, but we conjecture that the system is unstable in this region.
When we impose these alternative boundary conditions the allowed region is larger than in Figs.~\ref{fig:AllInternalBounds}, but it remains compact.

Even with these alternative boundary conditions, the internal parameters $M_{\rm in}$ and  $Q_{\rm in}$ are still restricted to small (sub-Planckian) values for the values of $a$ considered here. Note however that the allowed region does grow as $a \rightarrow 1$, potentially permitting larger values of $M_{\rm in}$ and  $Q_{\rm in}$. It is the asymptotic nature of this region that is of interest to the firewall model. Due to the limitations of our numerical approach, we are not able to definitively establish whether there are points of stability in this asymptotic region.

The firewall model we considered is a horizonless object where the interior is causally connected with the exterior, so one would expect the allowed range of parameters describing the firewall to depend on the physics in the interior. 
By necessity we assumed GR to be valid, but UV physics should be relevant near $r=0$.
In the absence of a UV-complete theory of gravity we cannot derive the perturbative formalism within the firewall, and exact predictions cannot be made.
However, one may attempt the inverse problem and fit the model to future observations of QNMs where: (i) we impose some other set of boundary conditions at a finite radius where GR is still applicable, or (ii) we assume (hypothetical, or EFT-inspired) field equations different from GR within the shell. This may allow us to impose requirements on the UV theory in the interior for Planckian firewalls.

\section{Conclusions}
\label{sec:Conclusion}
Given a fixed set of ADM parameters, the firewall ECO matches onto the exterior geometry of a Schwarzschild BH with those ADM parameters. In the construction described in Ref.~\cite{Kaplan:2018dqx}, the interior geometry is a function of two unconstrained parameters $\left(M_{\rm in}, Q_{\rm in}\right)$. In this work, we have examined the radial stability and polar QNM spectra of this model. The condition of having a radially stable firewall with subluminal sound speed significantly constrains the internal parameters  $\left(M_{\rm in}, Q_{\rm in}\right)$. In the limit that the firewall is placed just a Planck length away from the Schwarzschild horizon, the allowed range of parameters appears to shrink to a line in the $\left(M_{\rm in}, Q_{\rm in}\right)$ plane. The formation of the firewall ECO from collapsing matter relies on unknown UV physics. Presumably, in any UV-complete model the evolution of the collapsing matter into the firewall would proceed through specific objects that exist in the UV, which would correspond to a unique point in  the $\left(M_{\rm in}, Q_{\rm in}\right)$ plane. The requirement of stability and subluminality narrows down the range of possible objects in the UV that could conceivably allow the firewall to exist. Moreover, the nonradial stability of this object is dependent on the boundary conditions chosen in the interior of the firewall. There is no natural way to choose these conditions, since they arise from unknown UV physics. Our results suggest that stability may provide an insight into this physics. It would be interesting to see if these stability arguments would eliminate the range of parameters in Ref.~\cite{Kaplan:2018dqx} that could be used to construct negative-mass (and therefore presumably unstable~\cite{Gleiser:2006yz,Cardoso:2006bv}) Schwarzschild solutions.

In this paper, we have only examined the polar nonradial QNM spectrum of the firewall. Polar perturbations couple to the stress-energy tensor of the shell, and so they are sensitive to the properties of the assumed exotic matter on the shell. In principle, the axial spectrum can also easily be found following the methods described in this work to further clarify the stability properties of the firewall.
However, expressions for the algebraic completion  of the axial modes analogous to the known results for the Reissner-Nordstr\"om background~\cite{1981RSPSA.378...73X} are not readily available. Once such expressions are derived, the junction conditions can be applied to find how the perturbations change across the shell and the stability of these modes can be studied. 

There are two notable limitations of our analysis.
First, we have modelled the shell as a perfect fluid. 
It might be interesting to perform these analyses using relativistic objects (such as branes) that could exist in the UV and support a richer class of fluids. 
Second, our analysis relies on using GR to describe the spacetime, but the formation of the firewall ECO explicitly requires the use of unspecified physics beyond GR. 
In the absence of a full theory, we cannot perform a more complete analysis, but the spirit of our present inquiry is similar in vein to the approach taken in Ref.~\cite{Kaplan:2018dqx}. 
We could construct such objects in the limit where the densities are sub-Planckian and demonstrate stability, then take the limit where the objects approach Planck density and see if there are instabilities. 
It is encouraging that radial stability is possible for such objects even in GR. We cannot conclusively establish the stability of nonradial polar perturbations of the Planckian firewall in Ref.~\cite{Kaplan:2018dqx}, but we find hints that the object is unstable within GR when $M_{\rm in}$ and $Q_{\rm in}$ are large. It is possible that the UV completion of gravity may provide structures that could stabilize this system.

Our calculations are also of broader phenomenological relevance. As an object without a horizon, a firewall ECO can support hair. One may use our perturbative solutions at a finite radius and integrate them back to the location of the firewall to find the deformations of this firewall away from spherical symmetry and estimate the ``size'' of such hair. It would be interesting to map out the observational signatures of hair, either through its effects on orbits of test bodies or through the gravitational-wave emission of hairy BHs. Moreover, the existence of deformations at the surface of the BH can lead to mixing between various exterior angular momentum modes around the BH. This could have significant impact on phenomena such as superradiance and tests of the reflectivity of BH surfaces, since such mixing may cause low angular momentum modes to evolve into high angular momentum modes, trapping them behind the light ring. In light of upcoming experimental probes of BH spacetimes, it is important to analyze these effects in order to robustly test the firewall framework. 

\noindent{\bf{\em Acknowledgments.}}
We thank Paolo Pani, Vitor Cardoso and Francisco Duque for useful discussions. E.B. and R.M. are supported by NSF Grants No. PHY-1912550 and AST-1841358, NASA ATP Grants No. 17-ATP17-0225 and 19-ATP19-0051, and NSF-XSEDE Grant No. PHY-090003. The authors would like to acknowledge networking support by the GWverse COST Action CA16104, ``Black holes, gravitational waves and fundamental physics.'' We acknowledge support from the Amaldi Research Center funded by the MIUR program ``Dipartimento di Eccellenza''~(CUP: B81I18001170001). D.E.K and S.R. are supported by NSF Grant No. PHY-1818899.

\appendix

\section{Tetrad for Reissner-Nordstr\"om perturbations}
\label{app:tetrad}

The vierbein or tetrad for the perturbed metric \eqref{eq:polarMetricPert} in the primed coordinate system is defined as 
\begin{subequations}
\begin{align}
    e_{(0)}^\mu &= (\,h^{-1/2}(1-N P_\ell)\,,0\,,0\,,0\,)\,,\\
    e_{(1)}^\mu &= (\,0\,,h^{1/2}(1-L P_\ell)\,,0\,,0\,)\,,\\
    e_{(2)}^\mu &= (\,0\,,0\,,\,r^{-1}(1-T P_\ell - V P_\ell'')\,,\,0\,)\,,\\
    e_{(3)}^\mu &= \left(\,0\,,\,0\,,\,0\,,\,\frac{1- T P_\ell - V \cot\theta\, P_\ell'}{\sin\theta\, r}\,\right)\,.
\end{align}
\end{subequations}

The inverse tetrad $e^{(a)}_\mu$, such that $e^{(a)}_\mu e_{(a)}^\nu=\delta^\mu _\nu$ and $e^{(a)}_\mu e_{(b)}^\mu=\delta^a _b$, is given by
\begin{subequations}
\begin{align}
    e^{(0)}_\mu &= (\,h^{1/2}(1+N P_\ell)\,,0\,,0\,,0\,)\,,\\
    e^{(1)}_\mu &= (\,0\,,h^{-1/2}(1+L P_\ell)\,,0\,,0\,)\,,\\
    e^{(2)}_\mu &= (\,0\,,0\,,\,r(1+T P_\ell + V P_\ell'')\,,\,0\,)\,,\\
    e^{(3)}_\mu &= \left(\,0\,,\,0\,,\,0\,,\,r \sin\theta\,(1 + T P_\ell + V \cot\theta\, P_\ell')\,\right)\,.
\end{align}
\end{subequations}

The relations between the coordinate basis components of the Faraday tensor and their tetrad components are thus given by
\begin{align}
\label{eq:deltaFtrCoordBasis}
    \delta F_{tr} &=  
    C \left[\delta F_{(0)(1)}
    + \frac{Q}{r^2}(L h^{-1} + N h )P_\ell \right]
    \,,\\
    \delta F_{t\theta} &= C r \delta F_{(0)(2)} \sqrt{h}
    \,,\\
    \delta F_{r\theta} &= \frac{r \delta F_{(1)(2)}}{\sqrt{h}}
    \,.
\end{align}

 \section{Algebraic completion}
 \label{app:AlgebraicCompletion}

The perturbations of electrovacuum spacetimes yield wave equations corresponding to only two degrees of freedom in both the even and odd sector, respectively, which corresponding to ``tensor'' and ``vector'' modes. 
There is no simple relation between the perturbation equations \eqref{eq:RNPerturbationEquation} and the metric perturbations \eqref{eq:metricPerturbations}.
The fact that the field equations give a larger series of equations than there are degrees of freedom implies the existence of special solutions, which in turn give the completions (see Eqs.~(191)-(196) of~\cite{Chandrasekhar:1985kt} and Eqs.~(71)-(83) of~\cite{Burko:1995gf}):
\begin{subequations}
\label{eq:specialSoltions}
\begin{align}
    N^0 &= \frac{\Delta^{1/2}}{r^3} \left[M-\frac{r \left(M^2-Q^2+r^4 \omega ^2\right)}{\Delta }-\frac{2 Q^2}{r}\right]\,,\\
    L^0 &= \frac{\Delta^{1/2}}{r^4} \left(3 M r-4 Q^2\right)\,,\\
    V^0 &= \frac{\Delta^{1/2}}{r^2}\,,\\
    B^0_{r\theta} &= -\frac{2 Q^2 \Delta^{1/2}}{r^4}\,,\\
    B^0_{t \theta} &= \frac{2 Q^2}{\Delta^{1/2} r^5} \left(-3 M r+2 Q^2+r^2\right)\,,\\
\end{align}
\end{subequations}
for the special solutions.

The completions are then given by 
\begin{subequations}
\label{eq:AlgComp}
\begin{align}
    N &= N^0(r) \Phi (r) 
    + \Lambda\frac{\Delta}{r^2 \chi} H_2^+ \nonumber\\
    &- \frac{\Delta}{r^2 \chi} \partial_r\left( \frac{\Lambda r}{2} H_2^+ + Q H_1^+\right) \nonumber \\
      &+ \frac{1}{r \chi^2}\left[ \frac{\Delta}{r^2} ( \chi-\Lambda r-3M) - \left(\frac{\Lambda}{2} + 1\right)\chi \right]\nonumber\\
      &\times \left( \frac{\Lambda r}{2} H_2^+ + Q H_1^+ \right)\,,\\
    L &= L^0 \Phi 
    -\frac{1}{r^2}\left(\frac{\Lambda r}{2} H_2^+ + Q H_1^+ \right)\,,\\
    V &= V^0 \Phi + \frac{1}{r}H_2^+\,,\\
    B_{r\theta} &=  B^0_{r\theta} \Phi - \frac{Q}{r^2}H_1^+\,,\\
  B_{t\theta} &=  B^0_{t\theta} \Phi - \frac{Q}{r^2}\partial_r H_1^+ \nonumber\\
  &- 2 \frac{Q^2}{r^4\chi}\left( \frac{\Lambda r}{2} H_2^+ + Q H_1^+ \right)\,,\\
    T &= B_{r\theta} + V - L\,,\\
    B_{tr} &= \frac{1}{\Delta r^2}\left[2Q^2(2 T - \ell(\ell+1)V) - \ell (\ell+1) r^2 B_{r\theta}\right]\,,
\end{align}
\end{subequations}
 where
\begin{equation}
    \Phi(r) = \int \left[\frac{(\ell-1)(\ell+1)}{2} r H_2^+ + Q H_1^+\right]\frac{1}{\Delta^{1/2}\chi }dr
\end{equation}
and 
\begin{align}
\label{eq:H1p}
    H_1^+ &= \frac{1}{q_1}Z_1^+ + \frac{q_2}{2 q_1 Q}Z_2^+\,,\\
    H_2^+ &= \frac{Q}{3Mq_1}Z_1^+ + \frac{1}{q_1}Z_2^+\,,
\end{align}
recalling the definitions in Eq.~\eqref{eq:PertDefRefStart}-\eqref{eq:PertDefRefEnd}.
Note that for $\ell=1$, $H_2^+$ plays no role in the algebraic completions. This is expected, as in this case there should be only one degree of freedom.

\section{Infinitesimal metric and Faraday tensor gauge transformations}
\label{app:CoordinateChange}

The quantities appearing in the gauge transformations of Eq.~\eqref{eq:gaugeshell} read
\begin{widetext}
\begin{equation}
    \Xi_{\alpha\beta}^{(0)} = 
    \left[
\begin{array}{cccc}
 2 \dot y(t)    &    -\frac{f'(r)}{f(r)}y(t)    &   y(t) \partial_\theta & 0 \\
-\frac{f'(r)}{f(r)}y(t) & 0 & 0 & 0 \\
y(t) \partial_\theta & 0 & 0 & 0 \\
 0 & 0 & 0 & 0 \\
\end{array}
\right] P_\ell(\theta )\,,
\end{equation}
\begin{equation}
    \Xi_{\alpha\beta}^{(1)} = \left[
\begin{array}{cccc}
 -f'(r)h(r)z(t)     &  \dot z(t) & 0 & 0 \\
\dot z(t)               &   \frac{h'(r)}{h(r)} z(t)   &   z(t) \partial_\theta & 0 \\
 0                  &    z(t) \partial_\theta   & 2 r h(r) z(t) & 0 \\
 0                  & 0 & 0 & 2 r h(r) z(t) \sin ^2\theta \\
\end{array}
\right] P_\ell(\theta )\,,
\end{equation}
\begin{equation}
    \Xi_{\alpha\beta}^{(2)} = 
    \left[
\begin{array}{cccc}
 0 & 0 & \dot w(t) \partial_\theta & 0 \\
 0 & 0 & -\frac{2}{r}w(t)\partial_\theta & 0 \\
 \dot w(t) \partial_\theta & -\frac{2}{r}w(t)\partial_\theta & 2 w(t) \partial^2_{\theta\theta} & 0 \\
 0 & 0 & 0 &  \sin (2 \theta ) w(t) \partial_\theta  \\
\end{array}
\right] P_\ell(\theta)\,,
\end{equation}
\end{widetext}

The Lie derivatives of $F_{\mu\nu}$ along the infinitesimal gauge transformation vectors in Eq.~\eqref{eq:lieshell} are:
\begin{align}
    (\mathcal{L}_{\xi^{(0)}}F)_{tr} &= -F_{tr} \frac{\dot y }{f} P_\ell\,,\\
    (\mathcal{L}_{\xi^{(0)}}F)_{r\theta} &= F_{tr} \frac{ y }{ f }P_\ell'\,,\\
    (\mathcal{L}_{\xi^{(1)}}F)_{tr} &= 
    F_{tr} \frac{z}{r}(rh'-2h)P_\ell\,,\\
    (\mathcal{L}_{\xi^{(1)}}F)_{t\theta} &= F_{tr} z h P_\ell'\,.
\end{align}
We do not list terms which are either dictated by symmetry or zero.
  
\section{Junction conditions for the intrinsic curvature and Faraday tensor}
\label{app:KiFaraday}

The trace-reversed extrinsic curvature at the shell on the right-hand side of Eq.~\eqref{eq:JunctionConditions} is given by 
\begin{widetext}
\begin{align}
    \overline{K}_1 &= \sqrt{h} \left(\frac{2 f L}{a}-\frac{4 f N}{a}-2 f T' + \ell(\ell+1) f V' \right) +\frac{4 \sqrt{h} \dot y}{a}
    +z \left[h^{3/2} \left(\frac{2 f}{a^2}-\frac{2 f'}{a}\right) - \sqrt{h} \left(\frac{\ell(\ell+1)  f}{a^2}+\frac{f h'}{a}\right)\right]
    \,,\\
    \overline{K}_2 &= 
    \sqrt{h} \left\{a \left[2 T - L + a \left(N' + T' - \ell(\ell+1)  V' \right) \right]-\frac{a^2 f' }{2 f}(L - 2 T)\right\}
                     +\frac{a^2 \sqrt{h} \ddot z}{f} \nonumber \\
                     &+ z \bigg\{\sqrt{h} \left[\frac{1}{4} a h' \left(\frac{a f'}{f} + 2 \right)  + \ell(\ell+1) \right] 
    + \frac{1}{2} h^{3/2} \left[-\frac{a^2 f'^2}{f^2}+\frac{a \left(a f''+2 f'\right)}{f}+2\right]\bigg\}
    \,,\\
    \overline{K}_3 &= 
    \sqrt{h} \left(\frac{f'}{2 f}+\frac{1}{a}\right) \dot w + \sqrt{h} \left(\frac{2 y}{a}+\dot z \right)
    \,,\\
    \overline{K}_5 &= 
    \sqrt{h} \left[-a^2 V' + a \left(\frac{a f'}{f}+2 \right) V + w \left(\frac{f'}{f}+\frac{2}{a}\right)+z\right]
    \,.
\end{align}
\end{widetext}
Equating the irreducible components above with those of the stress-energy give four junction conditions. 
The simplifications in Ref.~\cite{Pani:2009} do not apply because $h(r)$ is not continuous across the shell.

\bibliography{Firewall_refs}

\begin{thebibliography}{41}%
\makeatletter
\providecommand \@ifxundefined [1]{%
 \@ifx{#1\undefined}
}%
\providecommand \@ifnum [1]{%
 \ifnum #1\expandafter \@firstoftwo
 \else \expandafter \@secondoftwo
 \fi
}%
\providecommand \@ifx [1]{%
 \ifx #1\expandafter \@firstoftwo
 \else \expandafter \@secondoftwo
 \fi
}%
\providecommand \natexlab [1]{#1}%
\providecommand \enquote  [1]{``#1''}%
\providecommand \bibnamefont  [1]{#1}%
\providecommand \bibfnamefont [1]{#1}%
\providecommand \citenamefont [1]{#1}%
\providecommand \href@noop [0]{\@secondoftwo}%
\providecommand \href [0]{\begingroup \@sanitize@url \@href}%
\providecommand \@href[1]{\@@startlink{#1}\@@href}%
\providecommand \@@href[1]{\endgroup#1\@@endlink}%
\providecommand \@sanitize@url [0]{\catcode `\\12\catcode `\$12\catcode
  `\&12\catcode `\#12\catcode `\^12\catcode `\_12\catcode `\%12\relax}%
\providecommand \@@startlink[1]{}%
\providecommand \@@endlink[0]{}%
\providecommand \url  [0]{\begingroup\@sanitize@url \@url }%
\providecommand \@url [1]{\endgroup\@href {#1}{\urlprefix }}%
\providecommand \urlprefix  [0]{URL }%
\providecommand \Eprint [0]{\href }%
\providecommand \doibase [0]{http://dx.doi.org/}%
\providecommand \selectlanguage [0]{\@gobble}%
\providecommand \bibinfo  [0]{\@secondoftwo}%
\providecommand \bibfield  [0]{\@secondoftwo}%
\providecommand \translation [1]{[#1]}%
\providecommand \BibitemOpen [0]{}%
\providecommand \bibitemStop [0]{}%
\providecommand \bibitemNoStop [0]{.\EOS\space}%
\providecommand \EOS [0]{\spacefactor3000\relax}%
\providecommand \BibitemShut  [1]{\csname bibitem#1\endcsname}%
\let\auto@bib@innerbib\@empty
\bibitem [{\citenamefont {Berti}(2019)}]{Berti:2019tcy}%
  \BibitemOpen
  \bibfield  {author} {\bibinfo {author} {\bibfnamefont {E.}~\bibnamefont
  {Berti}},\ }\href {\doibase 10.1007/s10714-019-2622-2} {\bibfield  {journal}
  {\bibinfo  {journal} {Gen. Rel. Grav.}\ }\textbf {\bibinfo {volume} {51}},\
  \bibinfo {pages} {140} (\bibinfo {year} {2019})},\ \Eprint
  {http://arxiv.org/abs/1911.00541} {arXiv:1911.00541 [gr-qc]} \BibitemShut
  {NoStop}%
\bibitem [{\citenamefont {Berti}\ \emph {et~al.}(2015)\citenamefont {Berti}
  \emph {et~al.}}]{Berti:2015itd}%
  \BibitemOpen
  \bibfield  {author} {\bibinfo {author} {\bibfnamefont {E.}~\bibnamefont
  {Berti}} \emph {et~al.},\ }\href {\doibase 10.1088/0264-9381/32/24/243001}
  {\bibfield  {journal} {\bibinfo  {journal} {Class. Quant. Grav.}\ }\textbf
  {\bibinfo {volume} {32}},\ \bibinfo {pages} {243001} (\bibinfo {year}
  {2015})},\ \Eprint {http://arxiv.org/abs/1501.07274} {arXiv:1501.07274
  [gr-qc]} \BibitemShut {NoStop}%
\bibitem [{\citenamefont {Abbott}\ \emph {et~al.}(2016)\citenamefont {Abbott},
  \citenamefont {Abbott}, \citenamefont {Abbott}, \citenamefont {Abernathy},
  \citenamefont {Acernese}, \citenamefont {Ackley}, \citenamefont {Adams},
  \citenamefont {Adams}, \citenamefont {Addesso}, \citenamefont {Adhikari},\
  and\ \citenamefont {et~al.}}]{Abbott:2016}%
  \BibitemOpen
  \bibfield  {author} {\bibinfo {author} {\bibfnamefont {B.}~\bibnamefont
  {Abbott}}, \bibinfo {author} {\bibfnamefont {R.}~\bibnamefont {Abbott}},
  \bibinfo {author} {\bibfnamefont {T.}~\bibnamefont {Abbott}}, \bibinfo
  {author} {\bibfnamefont {M.}~\bibnamefont {Abernathy}}, \bibinfo {author}
  {\bibfnamefont {F.}~\bibnamefont {Acernese}}, \bibinfo {author}
  {\bibfnamefont {K.}~\bibnamefont {Ackley}}, \bibinfo {author} {\bibfnamefont
  {C.}~\bibnamefont {Adams}}, \bibinfo {author} {\bibfnamefont
  {T.}~\bibnamefont {Adams}}, \bibinfo {author} {\bibfnamefont
  {P.}~\bibnamefont {Addesso}}, \bibinfo {author} {\bibfnamefont
  {R.}~\bibnamefont {Adhikari}}, \ and\ \bibinfo {author} {\bibnamefont
  {et~al.}},\ }\href {\doibase 10.1103/physrevlett.116.221101} {\bibfield
  {journal} {\bibinfo  {journal} {Physical Review Letters}\ }\textbf {\bibinfo
  {volume} {116}} (\bibinfo {year} {2016}),\
  10.1103/physrevlett.116.221101}\BibitemShut {NoStop}%
\bibitem [{\citenamefont {Berti}\ \emph
  {et~al.}(2018{\natexlab{a}})\citenamefont {Berti}, \citenamefont {Yagi},\
  and\ \citenamefont {Yunes}}]{Berti:2018a}%
  \BibitemOpen
  \bibfield  {author} {\bibinfo {author} {\bibfnamefont {E.}~\bibnamefont
  {Berti}}, \bibinfo {author} {\bibfnamefont {K.}~\bibnamefont {Yagi}}, \ and\
  \bibinfo {author} {\bibfnamefont {N.}~\bibnamefont {Yunes}},\ }\href
  {\doibase 10.1007/s10714-018-2362-8} {\bibfield  {journal} {\bibinfo
  {journal} {General Relativity and Gravitation}\ }\textbf {\bibinfo {volume}
  {50}} (\bibinfo {year} {2018}{\natexlab{a}}),\
  10.1007/s10714-018-2362-8}\BibitemShut {NoStop}%
\bibitem [{\citenamefont {Berti}\ \emph
  {et~al.}(2018{\natexlab{b}})\citenamefont {Berti}, \citenamefont {Yagi},
  \citenamefont {Yang},\ and\ \citenamefont {Yunes}}]{Berti:2018b}%
  \BibitemOpen
  \bibfield  {author} {\bibinfo {author} {\bibfnamefont {E.}~\bibnamefont
  {Berti}}, \bibinfo {author} {\bibfnamefont {K.}~\bibnamefont {Yagi}},
  \bibinfo {author} {\bibfnamefont {H.}~\bibnamefont {Yang}}, \ and\ \bibinfo
  {author} {\bibfnamefont {N.}~\bibnamefont {Yunes}},\ }\href {\doibase
  10.1007/s10714-018-2372-6} {\bibfield  {journal} {\bibinfo  {journal}
  {General Relativity and Gravitation}\ }\textbf {\bibinfo {volume} {50}}
  (\bibinfo {year} {2018}{\natexlab{b}}),\
  10.1007/s10714-018-2372-6}\BibitemShut {NoStop}%
\bibitem [{\citenamefont {Cunha}\ and\ \citenamefont
  {Herdeiro}(2018)}]{Cunha:2018}%
  \BibitemOpen
  \bibfield  {author} {\bibinfo {author} {\bibfnamefont {P.~V.~P.}\
  \bibnamefont {Cunha}}\ and\ \bibinfo {author} {\bibfnamefont {C.~A.~R.}\
  \bibnamefont {Herdeiro}},\ }\href {\doibase 10.1007/s10714-018-2361-9}
  {\bibfield  {journal} {\bibinfo  {journal} {General Relativity and
  Gravitation}\ }\textbf {\bibinfo {volume} {50}} (\bibinfo {year} {2018}),\
  10.1007/s10714-018-2361-9}\BibitemShut {NoStop}%
\bibitem [{\citenamefont {Mathur}(2009)}]{Mathur:2009hf}%
  \BibitemOpen
  \bibfield  {author} {\bibinfo {author} {\bibfnamefont {S.~D.}\ \bibnamefont
  {Mathur}},\ }\href {\doibase 10.1088/0264-9381/26/22/224001} {\bibfield
  {journal} {\bibinfo  {journal} {Class. Quant. Grav.}\ }\textbf {\bibinfo
  {volume} {26}},\ \bibinfo {pages} {224001} (\bibinfo {year} {2009})},\
  \Eprint {http://arxiv.org/abs/0909.1038} {arXiv:0909.1038 [hep-th]}
  \BibitemShut {NoStop}%
\bibitem [{\citenamefont {Colpi}\ \emph {et~al.}(1986)\citenamefont {Colpi},
  \citenamefont {Shapiro},\ and\ \citenamefont {Wasserman}}]{Colpi:1986ye}%
  \BibitemOpen
  \bibfield  {author} {\bibinfo {author} {\bibfnamefont {M.}~\bibnamefont
  {Colpi}}, \bibinfo {author} {\bibfnamefont {S.~L.}\ \bibnamefont {Shapiro}},
  \ and\ \bibinfo {author} {\bibfnamefont {I.}~\bibnamefont {Wasserman}},\
  }\href {\doibase 10.1103/PhysRevLett.57.2485} {\bibfield  {journal} {\bibinfo
   {journal} {Phys. Rev. Lett.}\ }\textbf {\bibinfo {volume} {57}},\ \bibinfo
  {pages} {2485} (\bibinfo {year} {1986})}\BibitemShut {NoStop}%
\bibitem [{\citenamefont {Seidel}\ and\ \citenamefont
  {Suen}(1991)}]{Seidel:1991zh}%
  \BibitemOpen
  \bibfield  {author} {\bibinfo {author} {\bibfnamefont {E.}~\bibnamefont
  {Seidel}}\ and\ \bibinfo {author} {\bibfnamefont {W.~M.}\ \bibnamefont
  {Suen}},\ }\href {\doibase 10.1103/PhysRevLett.66.1659} {\bibfield  {journal}
  {\bibinfo  {journal} {Phys. Rev. Lett.}\ }\textbf {\bibinfo {volume} {66}},\
  \bibinfo {pages} {1659} (\bibinfo {year} {1991})}\BibitemShut {NoStop}%
\bibitem [{\citenamefont {Mazur}\ and\ \citenamefont
  {Mottola}(2001)}]{Mazur:2001fv}%
  \BibitemOpen
  \bibfield  {author} {\bibinfo {author} {\bibfnamefont {P.~O.}\ \bibnamefont
  {Mazur}}\ and\ \bibinfo {author} {\bibfnamefont {E.}~\bibnamefont
  {Mottola}},\ }\href@noop {} {\  (\bibinfo {year} {2001})},\ \Eprint
  {http://arxiv.org/abs/gr-qc/0109035} {arXiv:gr-qc/0109035 [gr-qc]}
  \BibitemShut {NoStop}%
\bibitem [{\citenamefont {Giudice}\ \emph {et~al.}(2016)\citenamefont
  {Giudice}, \citenamefont {McCullough},\ and\ \citenamefont
  {Urbano}}]{Giudice:2016zpa}%
  \BibitemOpen
  \bibfield  {author} {\bibinfo {author} {\bibfnamefont {G.~F.}\ \bibnamefont
  {Giudice}}, \bibinfo {author} {\bibfnamefont {M.}~\bibnamefont {McCullough}},
  \ and\ \bibinfo {author} {\bibfnamefont {A.}~\bibnamefont {Urbano}},\ }\href
  {\doibase 10.1088/1475-7516/2016/10/001} {\bibfield  {journal} {\bibinfo
  {journal} {JCAP}\ }\textbf {\bibinfo {volume} {1610}},\ \bibinfo {pages}
  {001} (\bibinfo {year} {2016})},\ \Eprint {http://arxiv.org/abs/1605.01209}
  {arXiv:1605.01209 [hep-ph]} \BibitemShut {NoStop}%
\bibitem [{\citenamefont {Cardoso}\ \emph
  {et~al.}(2016{\natexlab{a}})\citenamefont {Cardoso}, \citenamefont
  {Franzin},\ and\ \citenamefont {Pani}}]{Cardoso:2016rao}%
  \BibitemOpen
  \bibfield  {author} {\bibinfo {author} {\bibfnamefont {V.}~\bibnamefont
  {Cardoso}}, \bibinfo {author} {\bibfnamefont {E.}~\bibnamefont {Franzin}}, \
  and\ \bibinfo {author} {\bibfnamefont {P.}~\bibnamefont {Pani}},\ }\href
  {\doibase 10.1103/PhysRevLett.117.089902, 10.1103/PhysRevLett.116.171101}
  {\bibfield  {journal} {\bibinfo  {journal} {Phys. Rev. Lett.}\ }\textbf
  {\bibinfo {volume} {116}},\ \bibinfo {pages} {171101} (\bibinfo {year}
  {2016}{\natexlab{a}})},\ \bibinfo {note} {[Erratum: Phys. Rev.
  Lett.117,no.8,089902(2016)]},\ \Eprint {http://arxiv.org/abs/1602.07309}
  {arXiv:1602.07309 [gr-qc]} \BibitemShut {NoStop}%
\bibitem [{\citenamefont {Cardoso}\ \emph
  {et~al.}(2016{\natexlab{b}})\citenamefont {Cardoso}, \citenamefont {Hopper},
  \citenamefont {Macedo}, \citenamefont {Palenzuela},\ and\ \citenamefont
  {Pani}}]{Cardoso:2016oxy}%
  \BibitemOpen
  \bibfield  {author} {\bibinfo {author} {\bibfnamefont {V.}~\bibnamefont
  {Cardoso}}, \bibinfo {author} {\bibfnamefont {S.}~\bibnamefont {Hopper}},
  \bibinfo {author} {\bibfnamefont {C.~F.~B.}\ \bibnamefont {Macedo}}, \bibinfo
  {author} {\bibfnamefont {C.}~\bibnamefont {Palenzuela}}, \ and\ \bibinfo
  {author} {\bibfnamefont {P.}~\bibnamefont {Pani}},\ }\href {\doibase
  10.1103/PhysRevD.94.084031} {\bibfield  {journal} {\bibinfo  {journal} {Phys.
  Rev.}\ }\textbf {\bibinfo {volume} {D94}},\ \bibinfo {pages} {084031}
  (\bibinfo {year} {2016}{\natexlab{b}})},\ \Eprint
  {http://arxiv.org/abs/1608.08637} {arXiv:1608.08637 [gr-qc]} \BibitemShut
  {NoStop}%
\bibitem [{\citenamefont {Cunha}\ \emph {et~al.}(2017)\citenamefont {Cunha},
  \citenamefont {Berti},\ and\ \citenamefont {Herdeiro}}]{Cunha:2017qtt}%
  \BibitemOpen
  \bibfield  {author} {\bibinfo {author} {\bibfnamefont {P.~V.~P.}\
  \bibnamefont {Cunha}}, \bibinfo {author} {\bibfnamefont {E.}~\bibnamefont
  {Berti}}, \ and\ \bibinfo {author} {\bibfnamefont {C.~A.~R.}\ \bibnamefont
  {Herdeiro}},\ }\href {\doibase 10.1103/PhysRevLett.119.251102} {\bibfield
  {journal} {\bibinfo  {journal} {Phys. Rev. Lett.}\ }\textbf {\bibinfo
  {volume} {119}},\ \bibinfo {pages} {251102} (\bibinfo {year} {2017})},\
  \Eprint {http://arxiv.org/abs/1708.04211} {arXiv:1708.04211 [gr-qc]}
  \BibitemShut {NoStop}%
\bibitem [{\citenamefont {Barausse}\ \emph {et~al.}(2018)\citenamefont
  {Barausse}, \citenamefont {Brito}, \citenamefont {Cardoso}, \citenamefont
  {Dvorkin},\ and\ \citenamefont {Pani}}]{Barausse:2018vdb}%
  \BibitemOpen
  \bibfield  {author} {\bibinfo {author} {\bibfnamefont {E.}~\bibnamefont
  {Barausse}}, \bibinfo {author} {\bibfnamefont {R.}~\bibnamefont {Brito}},
  \bibinfo {author} {\bibfnamefont {V.}~\bibnamefont {Cardoso}}, \bibinfo
  {author} {\bibfnamefont {I.}~\bibnamefont {Dvorkin}}, \ and\ \bibinfo
  {author} {\bibfnamefont {P.}~\bibnamefont {Pani}},\ }\href {\doibase
  10.1088/1361-6382/aae1de} {\bibfield  {journal} {\bibinfo  {journal} {Class.
  Quant. Grav.}\ }\textbf {\bibinfo {volume} {35}},\ \bibinfo {pages} {20LT01}
  (\bibinfo {year} {2018})},\ \Eprint {http://arxiv.org/abs/1805.08229}
  {arXiv:1805.08229 [gr-qc]} \BibitemShut {NoStop}%
\bibitem [{\citenamefont {Maggio}\ \emph {et~al.}(2019)\citenamefont {Maggio},
  \citenamefont {Cardoso}, \citenamefont {Dolan},\ and\ \citenamefont
  {Pani}}]{Maggio:2018ivz}%
  \BibitemOpen
  \bibfield  {author} {\bibinfo {author} {\bibfnamefont {E.}~\bibnamefont
  {Maggio}}, \bibinfo {author} {\bibfnamefont {V.}~\bibnamefont {Cardoso}},
  \bibinfo {author} {\bibfnamefont {S.~R.}\ \bibnamefont {Dolan}}, \ and\
  \bibinfo {author} {\bibfnamefont {P.}~\bibnamefont {Pani}},\ }\href {\doibase
  10.1103/PhysRevD.99.064007} {\bibfield  {journal} {\bibinfo  {journal} {Phys.
  Rev.}\ }\textbf {\bibinfo {volume} {D99}},\ \bibinfo {pages} {064007}
  (\bibinfo {year} {2019})},\ \Eprint {http://arxiv.org/abs/1807.08840}
  {arXiv:1807.08840 [gr-qc]} \BibitemShut {NoStop}%
\bibitem [{\citenamefont {Abedi}\ \emph {et~al.}(2020)\citenamefont {Abedi},
  \citenamefont {Afshordi}, \citenamefont {Oshita},\ and\ \citenamefont
  {Wang}}]{Abedi:2020ujo}%
  \BibitemOpen
  \bibfield  {author} {\bibinfo {author} {\bibfnamefont {J.}~\bibnamefont
  {Abedi}}, \bibinfo {author} {\bibfnamefont {N.}~\bibnamefont {Afshordi}},
  \bibinfo {author} {\bibfnamefont {N.}~\bibnamefont {Oshita}}, \ and\ \bibinfo
  {author} {\bibfnamefont {Q.}~\bibnamefont {Wang}},\ }\href {\doibase
  10.3390/universe6030043} {\bibfield  {journal} {\bibinfo  {journal}
  {Universe}\ }\textbf {\bibinfo {volume} {6}},\ \bibinfo {pages} {43}
  (\bibinfo {year} {2020})},\ \Eprint {http://arxiv.org/abs/2001.09553}
  {arXiv:2001.09553 [gr-qc]} \BibitemShut {NoStop}%
\bibitem [{\citenamefont {Maggio}\ \emph {et~al.}(2020)\citenamefont {Maggio},
  \citenamefont {Buoninfante}, \citenamefont {Mazumdar},\ and\ \citenamefont
  {Pani}}]{Maggio:2020jml}%
  \BibitemOpen
  \bibfield  {author} {\bibinfo {author} {\bibfnamefont {E.}~\bibnamefont
  {Maggio}}, \bibinfo {author} {\bibfnamefont {L.}~\bibnamefont {Buoninfante}},
  \bibinfo {author} {\bibfnamefont {A.}~\bibnamefont {Mazumdar}}, \ and\
  \bibinfo {author} {\bibfnamefont {P.}~\bibnamefont {Pani}},\ }\href@noop {}
  {\  (\bibinfo {year} {2020})},\ \Eprint {http://arxiv.org/abs/2006.14628}
  {arXiv:2006.14628 [gr-qc]} \BibitemShut {NoStop}%
\bibitem [{\citenamefont {Regge}\ and\ \citenamefont
  {Wheeler}(1957)}]{Regge:1957td}%
  \BibitemOpen
  \bibfield  {author} {\bibinfo {author} {\bibfnamefont {T.}~\bibnamefont
  {Regge}}\ and\ \bibinfo {author} {\bibfnamefont {J.~A.}\ \bibnamefont
  {Wheeler}},\ }\href {\doibase 10.1103/PhysRev.108.1063} {\bibfield  {journal}
  {\bibinfo  {journal} {Phys. Rev.}\ }\textbf {\bibinfo {volume} {108}},\
  \bibinfo {pages} {1063} (\bibinfo {year} {1957})}\BibitemShut {NoStop}%
\bibitem [{\citenamefont {Zerilli}(1970)}]{Zerilli:1971wd}%
  \BibitemOpen
  \bibfield  {author} {\bibinfo {author} {\bibfnamefont {F.~J.}\ \bibnamefont
  {Zerilli}},\ }\href {\doibase 10.1103/PhysRevD.2.2141} {\bibfield  {journal}
  {\bibinfo  {journal} {Phys. Rev.}\ }\textbf {\bibinfo {volume} {D2}},\
  \bibinfo {pages} {2141} (\bibinfo {year} {1970})}\BibitemShut {NoStop}%
\bibitem [{\citenamefont {Berti}\ \emph {et~al.}(2009)\citenamefont {Berti},
  \citenamefont {Cardoso},\ and\ \citenamefont {Starinets}}]{Berti:2009kk}%
  \BibitemOpen
  \bibfield  {author} {\bibinfo {author} {\bibfnamefont {E.}~\bibnamefont
  {Berti}}, \bibinfo {author} {\bibfnamefont {V.}~\bibnamefont {Cardoso}}, \
  and\ \bibinfo {author} {\bibfnamefont {A.~O.}\ \bibnamefont {Starinets}},\
  }\href {\doibase 10.1088/0264-9381/26/16/163001} {\bibfield  {journal}
  {\bibinfo  {journal} {Class. Quant. Grav.}\ }\textbf {\bibinfo {volume}
  {26}},\ \bibinfo {pages} {163001} (\bibinfo {year} {2009})},\ \Eprint
  {http://arxiv.org/abs/0905.2975} {arXiv:0905.2975 [gr-qc]} \BibitemShut
  {NoStop}%
\bibitem [{\citenamefont {Cardoso}\ and\ \citenamefont
  {Pani}(2019)}]{Cardoso:2019rvt}%
  \BibitemOpen
  \bibfield  {author} {\bibinfo {author} {\bibfnamefont {V.}~\bibnamefont
  {Cardoso}}\ and\ \bibinfo {author} {\bibfnamefont {P.}~\bibnamefont {Pani}},\
  }\href {\doibase 10.1007/s41114-019-0020-4} {\bibfield  {journal} {\bibinfo
  {journal} {Living Rev. Rel.}\ }\textbf {\bibinfo {volume} {22}},\ \bibinfo
  {pages} {4} (\bibinfo {year} {2019})},\ \Eprint
  {http://arxiv.org/abs/1904.05363} {arXiv:1904.05363 [gr-qc]} \BibitemShut
  {NoStop}%
\bibitem [{\citenamefont {Kaplan}\ and\ \citenamefont
  {Rajendran}(2019)}]{Kaplan:2018dqx}%
  \BibitemOpen
  \bibfield  {author} {\bibinfo {author} {\bibfnamefont {D.~E.}\ \bibnamefont
  {Kaplan}}\ and\ \bibinfo {author} {\bibfnamefont {S.}~\bibnamefont
  {Rajendran}},\ }\href {\doibase 10.1103/PhysRevD.99.044033} {\bibfield
  {journal} {\bibinfo  {journal} {Phys. Rev.}\ }\textbf {\bibinfo {volume}
  {D99}},\ \bibinfo {pages} {044033} (\bibinfo {year} {2019})},\ \Eprint
  {http://arxiv.org/abs/1812.00536} {arXiv:1812.00536 [hep-th]} \BibitemShut
  {NoStop}%
\bibitem [{\citenamefont {{Chandrasekhar}}\ and\ \citenamefont
  {{Ferrari}}(1991)}]{1991RSPSA.434..449C}%
  \BibitemOpen
  \bibfield  {author} {\bibinfo {author} {\bibfnamefont {S.}~\bibnamefont
  {{Chandrasekhar}}}\ and\ \bibinfo {author} {\bibfnamefont {V.}~\bibnamefont
  {{Ferrari}}},\ }\href {\doibase 10.1098/rspa.1991.0104} {\bibfield  {journal}
  {\bibinfo  {journal} {Proceedings of the Royal Society of London Series A}\
  }\textbf {\bibinfo {volume} {434}},\ \bibinfo {pages} {449} (\bibinfo {year}
  {1991})}\BibitemShut {NoStop}%
\bibitem [{\citenamefont {Benhar}\ \emph {et~al.}(1999)\citenamefont {Benhar},
  \citenamefont {Berti},\ and\ \citenamefont {Ferrari}}]{Benhar:1998au}%
  \BibitemOpen
  \bibfield  {author} {\bibinfo {author} {\bibfnamefont {O.}~\bibnamefont
  {Benhar}}, \bibinfo {author} {\bibfnamefont {E.}~\bibnamefont {Berti}}, \
  and\ \bibinfo {author} {\bibfnamefont {V.}~\bibnamefont {Ferrari}},\
  }\bibfield  {booktitle} {\emph {\bibinfo {booktitle} {{Gravitational waves: A
  challenge to theoretical astrophysics. Proceedings, Trieste, Italy, June 6-9,
  2000}}},\ }\href {\doibase 10.1046/j.1365-8711.1999.02983.x} {\bibfield
  {journal} {\bibinfo  {journal} {Mon. Not. Roy. Astron. Soc.}\ }\textbf
  {\bibinfo {volume} {310}},\ \bibinfo {pages} {797} (\bibinfo {year}
  {1999})},\ \bibinfo {note} {[ICTP Lect. Notes Ser.3,35(2001)]},\ \Eprint
  {http://arxiv.org/abs/gr-qc/9901037} {arXiv:gr-qc/9901037 [gr-qc]}
  \BibitemShut {NoStop}%
\bibitem [{\citenamefont {Israel}(1966)}]{Israel:1966rt}%
  \BibitemOpen
  \bibfield  {author} {\bibinfo {author} {\bibfnamefont {W.}~\bibnamefont
  {Israel}},\ }\href {\doibase 10.1007/BF02710419, 10.1007/BF02712210}
  {\bibfield  {journal} {\bibinfo  {journal} {Nuovo Cim.}\ }\textbf {\bibinfo
  {volume} {B44S10}},\ \bibinfo {pages} {1} (\bibinfo {year} {1966})},\
  \bibinfo {note} {[Erratum: Nuovo Cim.B48,463(1967); Nuovo
  Cim.B44,1(1966)]}\BibitemShut {NoStop}%
\bibitem [{\citenamefont {Visser}\ and\ \citenamefont
  {Wiltshire}(2004)}]{Visser:2003ge}%
  \BibitemOpen
  \bibfield  {author} {\bibinfo {author} {\bibfnamefont {M.}~\bibnamefont
  {Visser}}\ and\ \bibinfo {author} {\bibfnamefont {D.~L.}\ \bibnamefont
  {Wiltshire}},\ }\href {\doibase 10.1088/0264-9381/21/4/027} {\bibfield
  {journal} {\bibinfo  {journal} {Class. Quant. Grav.}\ }\textbf {\bibinfo
  {volume} {21}},\ \bibinfo {pages} {1135} (\bibinfo {year} {2004})},\ \Eprint
  {http://arxiv.org/abs/gr-qc/0310107} {arXiv:gr-qc/0310107 [gr-qc]}
  \BibitemShut {NoStop}%
\bibitem [{\citenamefont {Chandrasekhar}(1985)}]{Chandrasekhar:1985kt}%
  \BibitemOpen
  \bibfield  {author} {\bibinfo {author} {\bibfnamefont {S.}~\bibnamefont
  {Chandrasekhar}},\ }in\ \href@noop {} {\emph {\bibinfo {booktitle} {{Oxford,
  UK: Clarendon (1992) 646 p., OXFORD, UK: CLARENDON (1985) 646 P.}}}}\
  (\bibinfo {year} {1985})\BibitemShut {NoStop}%
\bibitem [{\citenamefont {Burko}(1995)}]{Burko:1995gf}%
  \BibitemOpen
  \bibfield  {author} {\bibinfo {author} {\bibfnamefont {L.~M.}\ \bibnamefont
  {Burko}},\ }\href {\doibase 10.1103/PhysRevD.52.4518} {\bibfield  {journal}
  {\bibinfo  {journal} {Phys. Rev.}\ }\textbf {\bibinfo {volume} {D52}},\
  \bibinfo {pages} {4518} (\bibinfo {year} {1995})},\ \Eprint
  {http://arxiv.org/abs/gr-qc/9507012} {arXiv:gr-qc/9507012 [gr-qc]}
  \BibitemShut {NoStop}%
\bibitem [{\citenamefont {{Kucha{\v{r}}}}(1968)}]{1968CzJPh..18..435K}%
  \BibitemOpen
  \bibfield  {author} {\bibinfo {author} {\bibfnamefont {K.}~\bibnamefont
  {{Kucha{\v{r}}}}},\ }\href {\doibase 10.1007/BF01698208} {\bibfield
  {journal} {\bibinfo  {journal} {Czechoslovak Journal of Physics}\ }\textbf
  {\bibinfo {volume} {18}},\ \bibinfo {pages} {435} (\bibinfo {year}
  {1968})}\BibitemShut {NoStop}%
\bibitem [{\citenamefont {Dotti}\ and\ \citenamefont
  {Gleiser}(2010)}]{Dotti:2010}%
  \BibitemOpen
  \bibfield  {author} {\bibinfo {author} {\bibfnamefont {G.}~\bibnamefont
  {Dotti}}\ and\ \bibinfo {author} {\bibfnamefont {R.~J.}\ \bibnamefont
  {Gleiser}},\ }\href {\doibase 10.1088/0264-9381/27/18/185007} {\bibfield
  {journal} {\bibinfo  {journal} {Class. Quant. Grav.}\ }\textbf {\bibinfo
  {volume} {27}},\ \bibinfo {pages} {185007} (\bibinfo {year} {2010})},\
  \Eprint {http://arxiv.org/abs/1001.0152} {arXiv:1001.0152 [gr-qc]}
  \BibitemShut {NoStop}%
\bibitem [{\citenamefont {Pani}\ \emph {et~al.}(2009)\citenamefont {Pani},
  \citenamefont {Berti}, \citenamefont {Cardoso}, \citenamefont {Chen},\ and\
  \citenamefont {Norte}}]{Pani:2009}%
  \BibitemOpen
  \bibfield  {author} {\bibinfo {author} {\bibfnamefont {P.}~\bibnamefont
  {Pani}}, \bibinfo {author} {\bibfnamefont {E.}~\bibnamefont {Berti}},
  \bibinfo {author} {\bibfnamefont {V.}~\bibnamefont {Cardoso}}, \bibinfo
  {author} {\bibfnamefont {Y.}~\bibnamefont {Chen}}, \ and\ \bibinfo {author}
  {\bibfnamefont {R.}~\bibnamefont {Norte}},\ }\href {\doibase
  10.1103/PhysRevD.80.124047} {\bibfield  {journal} {\bibinfo  {journal} {Phys.
  Rev.}\ }\textbf {\bibinfo {volume} {D80}},\ \bibinfo {pages} {124047}
  (\bibinfo {year} {2009})},\ \Eprint {http://arxiv.org/abs/0909.0287}
  {arXiv:0909.0287 [gr-qc]} \BibitemShut {NoStop}%
\bibitem [{\citenamefont {Cardoso}\ and\ \citenamefont
  {Duque}(2020)}]{Cardoso:2019upw}%
  \BibitemOpen
  \bibfield  {author} {\bibinfo {author} {\bibfnamefont {V.}~\bibnamefont
  {Cardoso}}\ and\ \bibinfo {author} {\bibfnamefont {F.}~\bibnamefont
  {Duque}},\ }\href {\doibase 10.1103/PhysRevD.101.064028} {\bibfield
  {journal} {\bibinfo  {journal} {Phys. Rev.}\ }\textbf {\bibinfo {volume}
  {D101}},\ \bibinfo {pages} {064028} (\bibinfo {year} {2020})},\ \Eprint
  {http://arxiv.org/abs/1912.07616} {arXiv:1912.07616 [gr-qc]} \BibitemShut
  {NoStop}%
\bibitem [{\citenamefont {Dotti}\ \emph {et~al.}(2007)\citenamefont {Dotti},
  \citenamefont {Gleiser},\ and\ \citenamefont {Pullin}}]{Dotti:2006}%
  \BibitemOpen
  \bibfield  {author} {\bibinfo {author} {\bibfnamefont {G.}~\bibnamefont
  {Dotti}}, \bibinfo {author} {\bibfnamefont {R.}~\bibnamefont {Gleiser}}, \
  and\ \bibinfo {author} {\bibfnamefont {J.}~\bibnamefont {Pullin}},\ }\href
  {\doibase 10.1016/j.physletb.2006.12.004} {\bibfield  {journal} {\bibinfo
  {journal} {Phys. Lett.}\ }\textbf {\bibinfo {volume} {B644}},\ \bibinfo
  {pages} {289} (\bibinfo {year} {2007})},\ \Eprint
  {http://arxiv.org/abs/gr-qc/0607052} {arXiv:gr-qc/0607052 [gr-qc]}
  \BibitemShut {NoStop}%
\bibitem [{\citenamefont {Leaver}(1985)}]{Leaver:1985ax}%
  \BibitemOpen
  \bibfield  {author} {\bibinfo {author} {\bibfnamefont {E.~W.}\ \bibnamefont
  {Leaver}},\ }\href {\doibase 10.1098/rspa.1985.0119} {\bibfield  {journal}
  {\bibinfo  {journal} {Proc. Roy. Soc. Lond.}\ }\textbf {\bibinfo {volume}
  {A402}},\ \bibinfo {pages} {285} (\bibinfo {year} {1985})}\BibitemShut
  {NoStop}%
\bibitem [{\citenamefont {Leins}\ \emph {et~al.}(1993)\citenamefont {Leins},
  \citenamefont {Nollert},\ and\ \citenamefont {Soffel}}]{Leins:1993zz}%
  \BibitemOpen
  \bibfield  {author} {\bibinfo {author} {\bibfnamefont {M.}~\bibnamefont
  {Leins}}, \bibinfo {author} {\bibfnamefont {H.~P.}\ \bibnamefont {Nollert}},
  \ and\ \bibinfo {author} {\bibfnamefont {M.~H.}\ \bibnamefont {Soffel}},\
  }\href {\doibase 10.1103/PhysRevD.48.3467} {\bibfield  {journal} {\bibinfo
  {journal} {Phys. Rev.}\ }\textbf {\bibinfo {volume} {D48}},\ \bibinfo {pages}
  {3467} (\bibinfo {year} {1993})}\BibitemShut {NoStop}%
\bibitem [{\citenamefont {Pani}(2013)}]{Pani:2013pma}%
  \BibitemOpen
  \bibfield  {author} {\bibinfo {author} {\bibfnamefont {P.}~\bibnamefont
  {Pani}},\ }\bibfield  {booktitle} {\emph {\bibinfo {booktitle} {{Proceedings,
  Spring School on Numerical Relativity and High Energy Physics (NR/HEP2):
  Lisbon, Portugal, March 11-14, 2013}}},\ }\href {\doibase
  10.1142/S0217751X13400186} {\bibfield  {journal} {\bibinfo  {journal} {Int.
  J. Mod. Phys.}\ }\textbf {\bibinfo {volume} {A28}},\ \bibinfo {pages}
  {1340018} (\bibinfo {year} {2013})},\ \Eprint
  {http://arxiv.org/abs/1305.6759} {arXiv:1305.6759 [gr-qc]} \BibitemShut
  {NoStop}%
\bibitem [{\citenamefont {Berti}\ and\ \citenamefont
  {Kokkotas}(2003)}]{Berti:2003zu}%
  \BibitemOpen
  \bibfield  {author} {\bibinfo {author} {\bibfnamefont {E.}~\bibnamefont
  {Berti}}\ and\ \bibinfo {author} {\bibfnamefont {K.~D.}\ \bibnamefont
  {Kokkotas}},\ }\href {\doibase 10.1103/PhysRevD.68.044027} {\bibfield
  {journal} {\bibinfo  {journal} {Phys. Rev.}\ }\textbf {\bibinfo {volume}
  {D68}},\ \bibinfo {pages} {044027} (\bibinfo {year} {2003})},\ \Eprint
  {http://arxiv.org/abs/hep-th/0303029} {arXiv:hep-th/0303029 [hep-th]}
  \BibitemShut {NoStop}%
\bibitem [{\citenamefont {Gleiser}\ and\ \citenamefont
  {Dotti}(2006)}]{Gleiser:2006yz}%
  \BibitemOpen
  \bibfield  {author} {\bibinfo {author} {\bibfnamefont {R.~J.}\ \bibnamefont
  {Gleiser}}\ and\ \bibinfo {author} {\bibfnamefont {G.}~\bibnamefont
  {Dotti}},\ }\href {\doibase 10.1088/0264-9381/23/15/021} {\bibfield
  {journal} {\bibinfo  {journal} {Class. Quant. Grav.}\ }\textbf {\bibinfo
  {volume} {23}},\ \bibinfo {pages} {5063} (\bibinfo {year} {2006})},\ \Eprint
  {http://arxiv.org/abs/gr-qc/0604021} {arXiv:gr-qc/0604021 [gr-qc]}
  \BibitemShut {NoStop}%
\bibitem [{\citenamefont {Cardoso}\ and\ \citenamefont
  {Cavaglia}(2006)}]{Cardoso:2006bv}%
  \BibitemOpen
  \bibfield  {author} {\bibinfo {author} {\bibfnamefont {V.}~\bibnamefont
  {Cardoso}}\ and\ \bibinfo {author} {\bibfnamefont {M.}~\bibnamefont
  {Cavaglia}},\ }\href {\doibase 10.1103/PhysRevD.74.024027} {\bibfield
  {journal} {\bibinfo  {journal} {Phys. Rev.}\ }\textbf {\bibinfo {volume}
  {D74}},\ \bibinfo {pages} {024027} (\bibinfo {year} {2006})},\ \Eprint
  {http://arxiv.org/abs/gr-qc/0604101} {arXiv:gr-qc/0604101 [gr-qc]}
  \BibitemShut {NoStop}%
\bibitem [{\citenamefont {{Xanthopoulos}}(1981)}]{1981RSPSA.378...73X}%
  \BibitemOpen
  \bibfield  {author} {\bibinfo {author} {\bibfnamefont {B.~C.}\ \bibnamefont
  {{Xanthopoulos}}},\ }\href {\doibase 10.1098/rspa.1981.0142} {\bibfield
  {journal} {\bibinfo  {journal} {Proceedings of the Royal Society of London
  Series A}\ }\textbf {\bibinfo {volume} {378}},\ \bibinfo {pages} {73}
  (\bibinfo {year} {1981})}\BibitemShut {NoStop}%
\end{thebibliography}%
\end{document}